\documentclass[fleqn,usenatbib,usedcolumn]{mnras}
\usepackage{mathptmx}

\usepackage[T1]{fontenc}
\usepackage{ae,aecompl}
\usepackage{graphicx}

\usepackage[normalem]{ulem}
\usepackage{cancel} 
\usepackage{graphicx}	
\usepackage{amsmath}	
\usepackage{amssymb}	
\usepackage[usenames,dvipsnames]{color}
\usepackage[british]{babel}
\usepackage{svn-multi}
  \newcommand{\pc}{\,{\rm pc}}
  \newcommand{\kpc}{\,{\rm kpc}}
  
  \newcommand{\kms}{\,{\rm km\,s^{-1}}}
  \newcommand{\Myr}{\,{\rm Myr}}
  \newcommand{\Gyr}{\,{\rm Gyr}}
  \newcommand{\mkG}{\,\umu{\rm G}}
	\newcommand{\cm}{\,{\rm cm}}
	\newcommand{\const}{\text{const}}
	
	\newcommand{\dd}{\mathrm{d}}
  \newcommand{\vect}[1]{\mathbfit{#1}}
	 
	\newcommand{\tc}{\tilde{C}}
	\newcommand{\tb}{\tilde{\beta}}
	\newcommand{\cb}{B}
	\newcommand{\tB}{\widetilde{B}}

\hypersetup{
pdfauthor={I.~Makarenko (orcid.org/0000-0002-0117-5975),
            A.~Shukurov (orcid.org/0000-0001-6200-4304),
            R.~Henderson, 
            L.~F.~S.~Rodrigues (orcid.org/0000-0002-3860-0525),
            P.~Bushby, (orcid.org/0000-0002-4691-6757),
            A.~Fletcher 
           },
pdftitle={Topological signatures of interstellar magnetic fields -- I.
Betti numbers and persistence diagrams},
}
\volume{{\rm in press}}
\begin{document}
\title[Interstellar magnetic field topological signatures]{Topological
	 signatures of interstellar magnetic fields -- I. Betti numbers and persistence diagrams}
\author[I.~Makarenko et al.]{Irina~Makarenko, Anvar~Shukurov,
	Robin~Henderson, Luiz~F.~S.~Rodrigues,
\newauthor
Paul~Bushby, Andrew~Fletcher\thanks{E-mail:
	\href{mailto:irina.makarenko@ncl.ac.uk}{irina.makarenko@ncl.ac.uk},
	\href{mailto:anvar.shukurov@ncl.ac.uk}{anvar.shukurov@ncl.ac.uk},
	\href{mailto:robin.henderson@ncl.ac.uk}{robin.henderson@ncl.ac.uk},
	\href{mailto:luiz.rodrigues@ncl.ac.uk}{luiz.rodrigues@ncl.ac.uk},
	\href{mailto:paul.bushby@ncl.ac.uk}{paul.bushby@ncl.ac.uk},
	\href{mailto:andrew.fletcher@ncl.ac.uk}{andrew.fletcher@ncl.ac.uk}
}
\\
School of Mathematics, Statistics and Physics, Newcastle University,
Newcastle upon Tyne, NE1 7RU, UK
}

\pubyear{2017}
\pagerange{\pageref{firstpage}--\pageref{lastpage}}
\date{Accepted for publication in MNRAS}
\maketitle
\label{firstpage}
\begin{abstract}
The interstellar medium (ISM) is a magnetised system in which transonic or supersonic turbulence
is driven by supernova explosions. This leads to the production of intermittent, filamentary
structures in the ISM gas density, whilst the associated dynamo action also produces intermittent magnetic fields.
The traditional theory of random functions, restricted to second-order statistical
moments (or power spectra), does not adequately describe such systems.
We apply topological data analysis (TDA), sensitive to all statistical
moments and independent of the assumption of Gaussian statistics, to
the gas density fluctuations in a magnetohydrodynamic (MHD) simulation of the multi-phase ISM.
This simulation admits dynamo action, so produces physically realistic magnetic fields.
The topology of the gas distribution, with and without magnetic fields, is
quantified in terms of Betti numbers and persistence diagrams.
Like the more standard correlation analysis, TDA shows that the ISM gas density is sensitive to the presence of magnetic fields. However, TDA gives us important additional information that cannot be obtained
from correlation functions. In particular, the Betti numbers \textit{per
correlation cell} are shown to be physically informative.
Magnetic fields make the ISM more
homogeneous, reducing the abundance of both isolated gas clouds
and cavities, with a stronger effect on the cavities.
Remarkably, the modification of the gas distribution by magnetic fields
is captured by
the Betti numbers even in regions more than $300\pc$ from the midplane,
where the magnetic field is weaker and correlation analysis
fails to detect any signatures of magnetic effects.
\end{abstract}

\section{Introduction}\label{sec:intro}
The interstellar medium (ISM) is turbulent \citep{SE04,ES04}, with the energy injected
by supernova explosions and stellar winds high
enough to maintain a transonic or
supersonic random flow \citep{V-S15}. This makes the compressibility of interstellar
gas
significant. In particular, density structures observable in \ion{H}{i} can
be attributed to converging gas flows aided by self-gravity and thermal instability
\citep{B-PV-SS99,HBV-SKA08,HMLV-S09}. As a result,
the statistical properties of interstellar density fluctuations are non-Gaussian even if the velocity fluctuations
can be approximately described as Gaussian. Deviations from Gaussian statistics for ISM
fluctuations are reflected in the properties of the magnetic field. Apart from the effects of
compressibility, magnetic fields produced by the fluctuation dynamo,
even by a Gaussian random velocity field, are
non-Gaussian, with heavy
power-law tails in the probability distribution of the magnetic field components
\citep{SSSBW17}.
Due to compression in transonic turbulence and
fluctuation dynamo action, interstellar magnetic fields are
\textit{spatially intermittent}, being represented by intense magnetic filaments and
ribbons immersed in a background of weaker, perhaps nearly-Gaussian, magnetic
fluctuations \citep{ZRS90,WBS07}. The term `intermittency' was originally used
to describe spatial and temporal fluctuations in the dissipation rate of turbulence,
but later expanded to include spatial and temporal structures in the turbulent
flow itself, such as filamentary \ion{H}{i} clouds in the ISM and magnetic filaments
and ribbons produced by the fluctuation dynamo
\citep[e.g.,][and references therein]{ZUB15,ZBU16}.

Since the energy density of the interstellar magnetic field is comparable to the
turbulent and thermal energy densities, magnetic intermittency is likely to
significantly affect the statistical properties of ISM turbulence, in particular by making them
non-Gaussian. Second-order statistical moments, such as power spectra and
correlation functions, provide a complete description 
only of a Gaussian random field: for example, the presence of
coherent structures, such as filaments, makes
the random phase approximation of the standard Fourier analysis inapplicable.
However, most diagnostic and interpretation tools in the theory of turbulence
\citep[and of random functions in general --][]{CL13} are designed to work with
Gaussian random flows. Relatively weak deviations from these simple statistical
properties can be captured by higher-order
correlation functions \citep[e.g., ][]{PJNB04}. However, the reliable
estimation of higher-order statistical moments requires taking averages over time- and/or spatial-scales
that grow rapidly with the order of the moment \citep[][]{O77}. This approach is thus of limited value in the case of inhomogeneous turbulence, especially
when the observational or simulated domain is relatively small.

We discuss statistical topological tools sensitive to all statistical
moments of a random field and thus suitable for studies of intermittent 
turbulence. The Minkowski functionals \citep{SB97} and related dimensionless 
measures such as filamentarity and planarity \citep{SSS98,SBMSSS99}
provide convenient morphological descriptors of intermittent turbulent flows
\citep{WBS07,LSD12,MKN07}. However, the morphology of an intermittent random 
field is only one of its aspects. More subtle but no less essential features 
are revealed by topological filtration, which characterises
statistical properties of the extrema
of the random field and connectivity of its isosurfaces
\citep{CarGun2009,ABBSW2010,AdTay2011,Ed2014}. These features are
described in terms of the Betti numbers, $\beta_0$, $\beta_1$ and $\beta_2$; in a space of a
dimension $d$, there are
$d$ Betti numbers. The more familiar Euler characteristic is the
alternating sum of the Betti numbers, $\chi=\sum_{n=0}^{d-1}(-1)^n\beta_n$. Topological data analysis (TDA)
is briefly introduced in Section~\ref{sec:filt}.

Topological data analysis and its applications are still in their infancy:
reliable and efficient algorithms to compute topological characteristics are
still being developed, and the physical significance of
the various topological characteristics of a random field is often elusive.
Nevertheless, some progress has been made in this direction and there are recent examples of applications of
TDA in many areas, including medical imaging \citep{ATW07} and remote sensing \citep{MKY16}.
Examples and a discussion of advanced topological diagnostics of non-Gaussian random fields
are presented in \citet{Hend2017arXiv}. In the context of astrophysics, applications of TDA include the distribution of galaxies in
the cosmological large-scale structure (\citealt{Pranav2017};
see also \citealt{Sousbie2011part1,Sousbie2011part2}) and the \ion{H}{i}
distribution in the Milky Way \citep{Hend2017arXiv}. Genus, a topological measure  closely related to the Euler characteristic, has been used to study density structures in numerical simulations of MHD turbulence \citep{Kowal:2007}, the \ion{H}{i} column density in the Small Magellanic Cloud \citep{Chepurnov:2008} and the polarized synchrotron emission of the Milky Way \citep{Burkhart:2012}. TDA has also been used in solar physics \citep{MKN07,KMU15a,KMU15}
and geophysics \citep{KMKD16}.

Here we apply topological
methods to a magnetohydrodynamic (MHD) simulation of the multi-phase interstellar
medium, driven by random supernova explosions \citep[described in detail by][]{Gent2013_hydro,Gent2013_mag}.
The magnetic field is arguably the least well
understood  component of the ISM, both observationally and theoretically,
but its importance is becoming more evident. However,
progress is slow because the effects of a magnetic field can be subtle
and not easy to discern  \citep[see][and references therein]{EGSFB18}.
We focus upon the ways in which the gas density distribution in the simulated ISM is influenced
by the presence of magnetic fields. Our primary concern is to use topological measures to identify
any differences between magnetic and non-magnetic regions that may not be apparent when traditional methods are used.

In Section~\ref{sec:data}, we describe the physical system and data analysed.
Section~\ref{sec:corr} presents a correlation analysis of the gas density
fluctuations and a discussion of their dependence on the magnetic field
\citep[see also][]{HSSFG17}.
The limitations of traditional approaches are
also discussed. Section~\ref{sec:tda} introduces
topological filtration, Betti numbers and persistence diagrams.
The key results relating to these topological measures are presented in Section~\ref{sec:scal}.
Section~\ref{sec:con} contains further discussion and summarises our conclusions.

\section{The simulated ISM}
\label{sec:data}

Simulations of the diffuse interstellar medium  \citep[e.g.,][]{KBSTN99a,KBST99b,dAB04,Gres08,Pi09,dAABS12,dAB12,HJMLBHKW12,HJMLBHKW12e,Be15,Hen15} include a wide variety of
physical processes and can be treated as physically realistic numerical
experiments. The data used here are obtained from the non-ideal MHD simulations of the
supernova-driven, multiphase ISM of \citet{Gent2013_mag,Gent2013_hydro} that
include  dynamo action, and thus produce physically realistic magnetic fields
\citep[other such models are presented by][]{KBSTN99a,KBST99b,Gres08,Be15}.
The model simulates the ISM in the Solar neighbourhood,
randomly heated and stirred by supernovae (SN),
with external gravity, stratification, differential rotation,
radiative cooling, photoelectric heating and various transport processes.
The local Cartesian frame $\vect{x}=(x,y,z)$ approximates the rotating cylindrical
polar coordinates $(r,\phi,z)$ with the gravity and angular
velocity of rotation oppositely directed and
aligned with the $z$-axis; the azimuthal direction is
identified with the $y$-axis. The simulations use the \textsc{Pencil
Code}\footnote{\url{http://pencil-code.nordita.org/}} \citep{BD02,PencilCode}, a
sixth-order finite difference code for non-ideal MHD. Detailed analyses of the simulations can be found in \citet{Gent2013_mag,Gent2013_hydro},
\citet{EGSFB17,EGSFB18} and \citet{HSSFG17}.

Our analysis is applied to the distribution of the gas number density $n$,
which spans the range $10^{-5} < n< 10^2\cm^{-3}$; the effects of the
magnetic field on $n$ should be more pronounced, and therefore easier to detect, than
those on gas velocity or temperature. We use a computational domain that
extends $1\times1\kpc^2$ horizontally and $2\kpc$ vertically
(symmetric about the galactic mid-plane, placed at $z=0$). The data cube has
$256\times 256\times 544$ uniformly distributed mesh points, providing a
spatial resolution of $4\pc$. The correlation scale of the density fluctuations
is about $50\pc$ at the mid-plane
\citep[][see also Section~\ref{sec:corr}]{HSSFG17}, so there are about 400
correlation cells in each horizontal plane of the computational domain.
Defining $t=0$ to be the time at which a weak seed magnetic field is
introduced into a hydrodynamic system that has already achieved a
statistically steady state, we focus upon 37 snapshots in the range $0.825 \leq t \leq 1.725\Gyr$.
There is a time separation of $\Delta t = 25\Myr$ between each snapshot.
The correlation time of the random velocity field in
these simulations is of the order of $10\Myr$ \citep{HSSFG17}, so the snapshots
are statistically independent to a reasonable accuracy.

\begin{figure}
	\centering
	\includegraphics{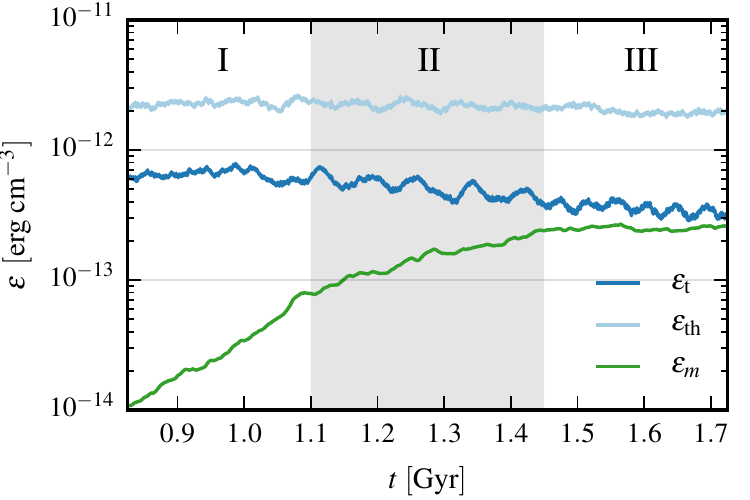}\\
	\caption{The average thermal, turbulent kinetic and magnetic energy
		densities, respectively $\epsilon_{\rm th}$, $\epsilon_{\rm t}$ and
		$\epsilon_{\rm m}$, within the simulation domain. Various stages of the
		magnetic field evolution are marked with numbers at the top of the frame and
		separated with shading in Stage~II. During Stage~I the magnetic field is dynamically insignificant
		so the evolution is essentially hydrodynamic.
		Stage~II is transitional, whereas the magnetic field has
		reached a statistically steady state and is dynamically important in Stage~III.
	}
	\label{fig:enerDens}
\end{figure}

With respect to the magnetic field properties, the simulated ISM evolves through
three main stages during the time period under consideration (Fig.~\ref{fig:enerDens}).
Stage~I is represented by 12 snapshots at
$0.825 \leq t \leq 1.100\Gyr$; here the magnetic field contribution to the total energy
density is negligible and its strength grows exponentially in time, representative of
the kinematic phase of a turbulent dynamo. Snapshots 13--25 cover Stage~II, the
transition from the kinematic dynamo to a statistically steady state of the magnetic
field. Stage~III, represented by the snapshots 26--37 at $1.450\leq t\leq1.725\Gyr$,
is the dynamo-saturated stage where the magnetic
energy is comparable to the kinetic energy of the turbulent flow.

During the whole
evolution, the thermal state of the system remains unchanged (thermal energy density,
$\varepsilon_\text{th}\approx\const$) but the kinetic energy density of the turbulent flow
$\varepsilon_\text{t}$ decreases slightly as the dynamo grows and saturates. At the end of the simulation,
the turbulent kinetic and magnetic energies are comparable;
the mean gas number density within the box is approximately $0.24\cm^{-3}$,
the typical turbulent flow speed is about $14\kms$ and the magnetic field strength is close to
$2.5\mkG$.

\subsection{Gas density fluctuations}\label{sec:reduc}

The gas density $n(\mathbfit{x})$ is represented as the sum of the mean
$\langle n \rangle$ and fluctuating (random) $\delta n$ parts,
both functions of position. The mean density is obtained by
Gaussian smoothing:
\begin{equation}\label{av}
\langle n \rangle(\mathbfit{x}) = \int_V n(\mathbfit{x}\arcmin)\, G_l(\mathbfit{x} - \mathbfit{x}')\,\mathrm{d}^3 \mathbfit{x}'\,,
\qquad \delta n(\mathbfit{x}) = n-\langle n\rangle\,,
\end{equation}
where $G_l(\mathbfit{x}) = \left(2\pi l^2\right)^{-3/2}
			\exp \left[-|\mathbfit{x}|^2/(2l^2) \right]$ is the Gaussian kernel, and $V$ denotes the volume of the computational domain.
The averaging scale for this Gaussian smoothing operation, $l=50\pc$, was obtained by \citet{Gent2013_mag}; this defines the kernel that maximises the difference between the scales of the averaged and fluctuating quantities in the power spectra \citep[see also Section 2.2 of][]{HSSFG17}. The mean value of the
fluctuations, $\langle\delta n\rangle$, is negligible at $z\neq0$ and close
to zero at $z=0$ where the density fluctuations are the strongest.
The deviation of $\langle\delta n\rangle$
from zero is an unavoidable consequence of using Gaussian smoothing as the
averaging procedure; \citet{Ger1992} presents a consistent formalism
for this and similar averaging methods.

We have also tested another averaging procedure where the mean gas density
is defined as the horizontal average,
$\langle n\rangle(z)=\iint_{z=\const} n(\vect{x})\,\dd x\,\dd y$. However, this leads
to physically unacceptable values of the correlation length of the density
fluctuations in excess of $200\pc$, whereas physically
justifiable values (and those observed in the ISM) are about $100\pc$ or less.
It is also worth noting that there is no need for the large-scale density to be perfectly uniform
in $x$ and $y$ and only depend on $z$: the horizontal averaging disregards
this fact \citep[for details, see][]{Gent2013_mag}.

\begin{figure*}
	\centering
	\includegraphics[width=0.35\textwidth]{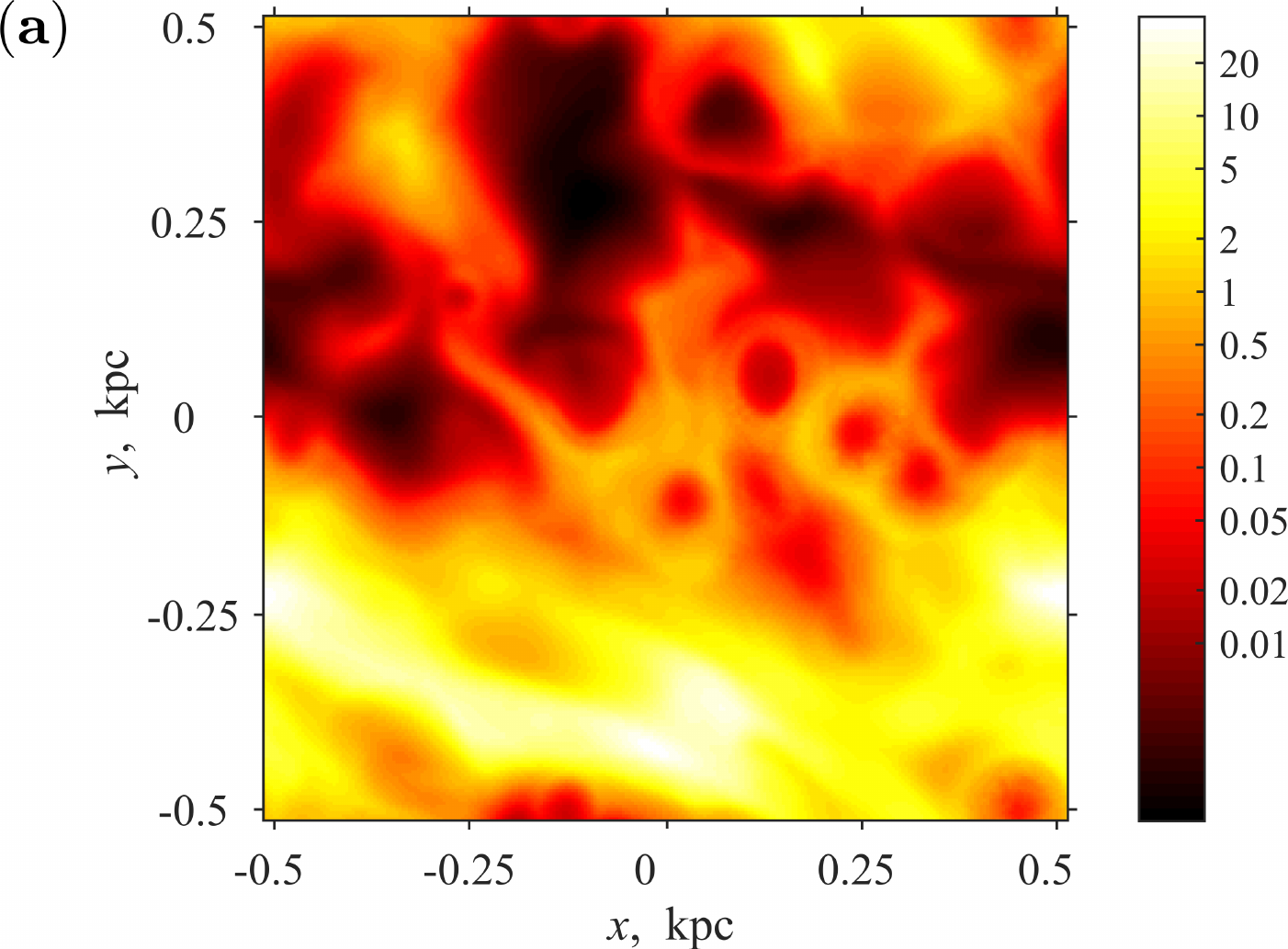} \hspace{5mm}
	\includegraphics[width=0.35\textwidth]{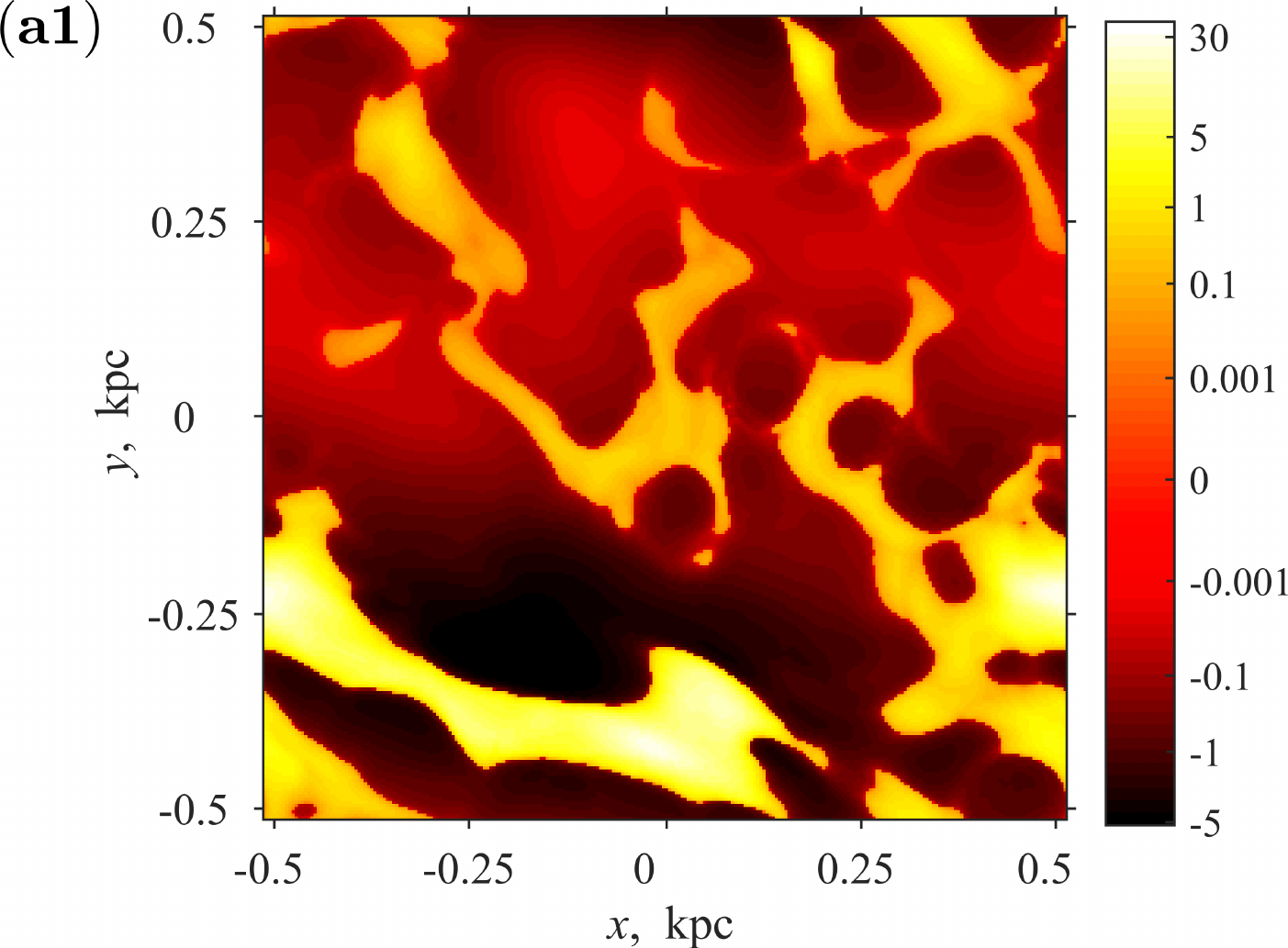}\\	\vspace{3mm}
	\includegraphics[width=0.35\textwidth]{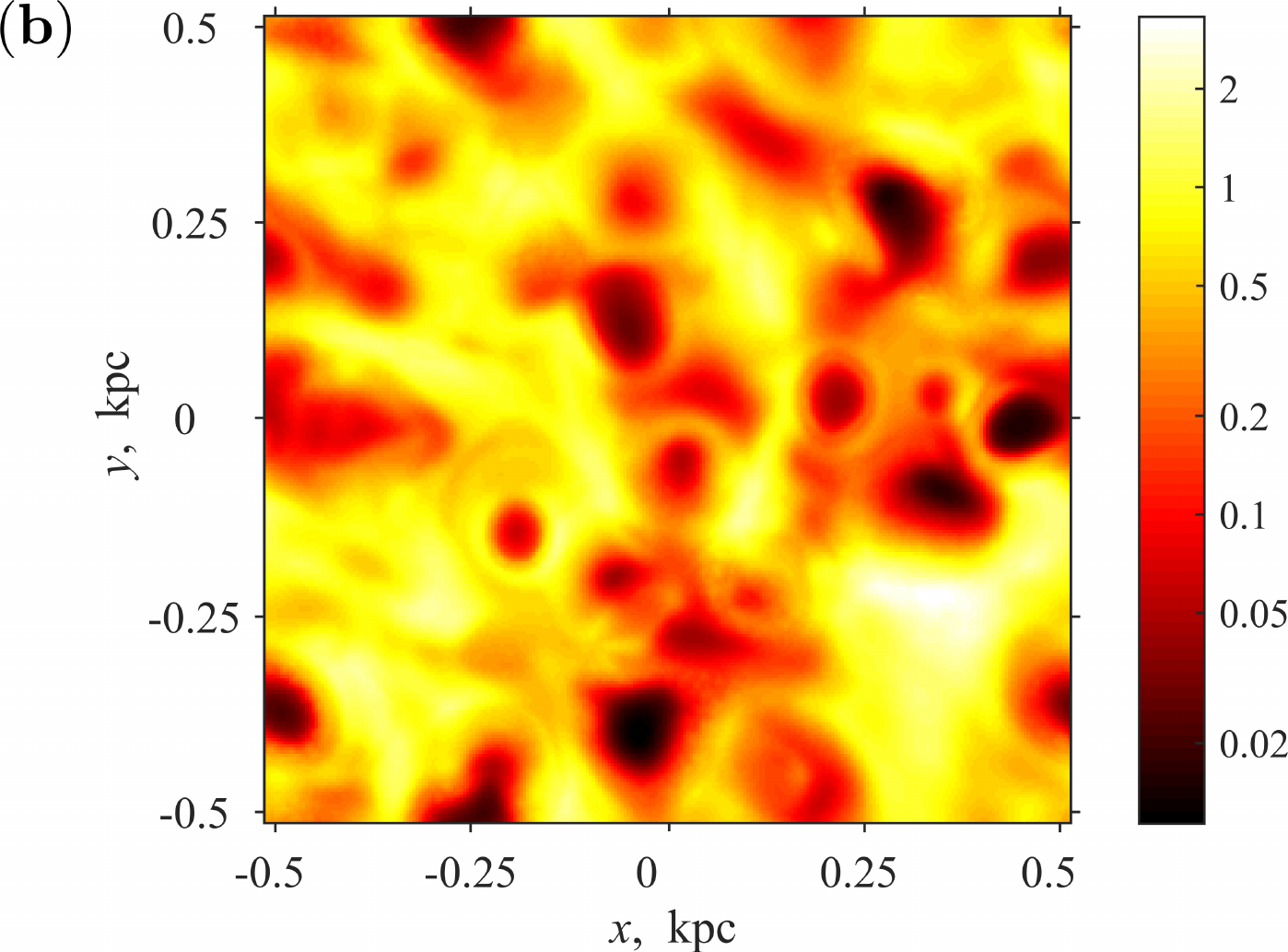} \hspace{5mm}
	\includegraphics[width=0.35\textwidth]{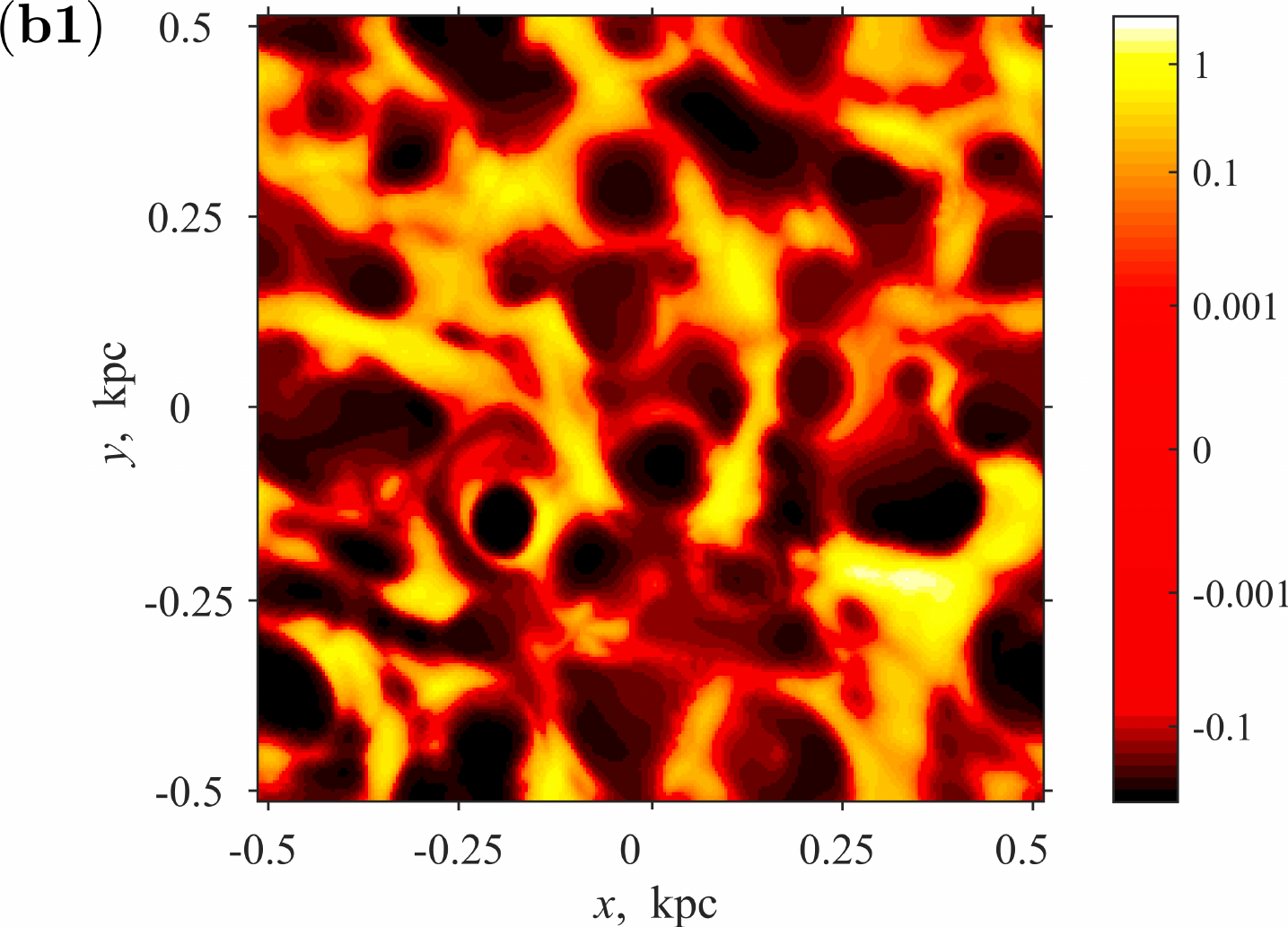}\\	\vspace{3mm}
	\includegraphics[width=0.35\textwidth]{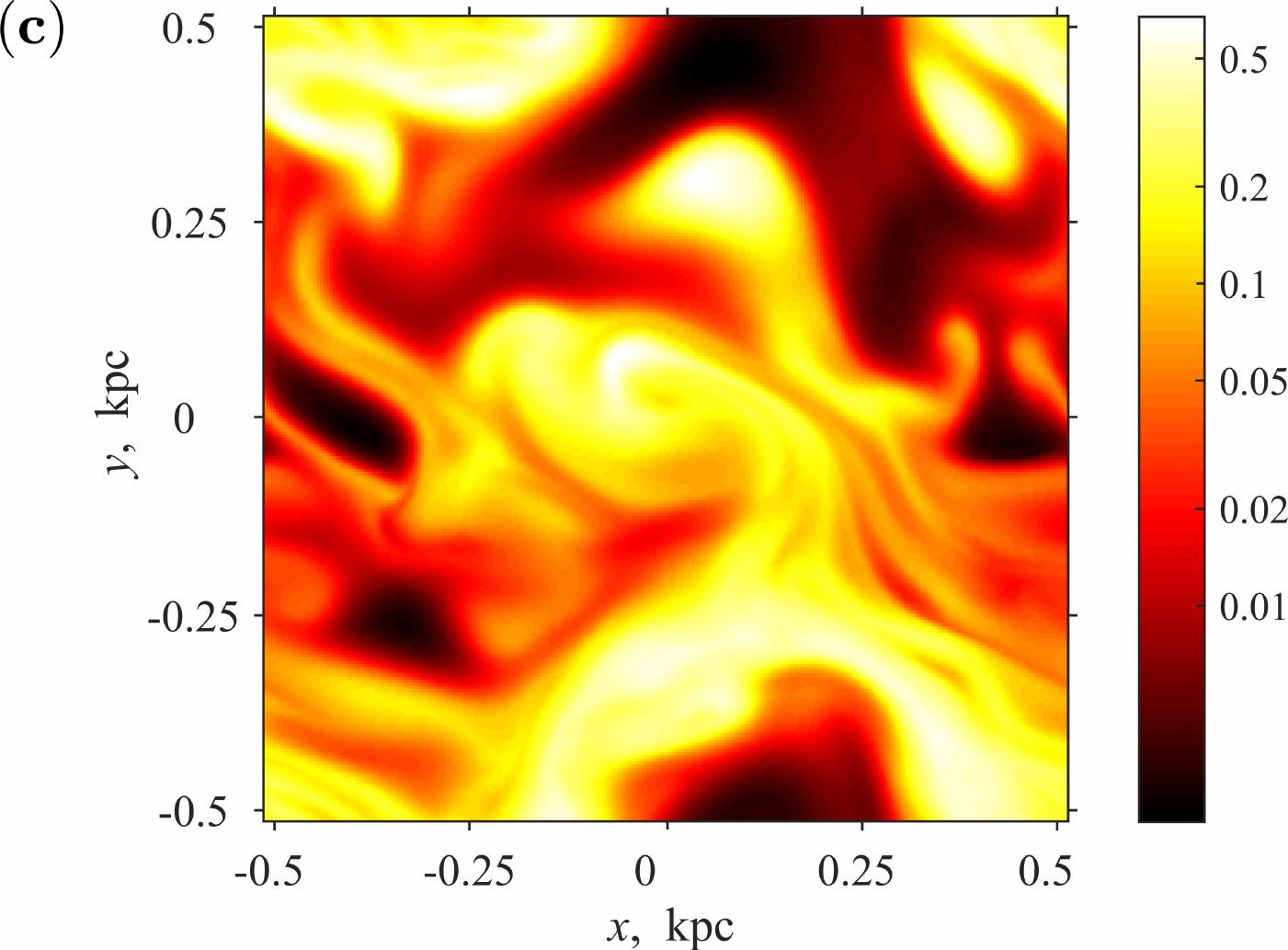} \hspace{5mm}
	\includegraphics[width=0.35\textwidth]{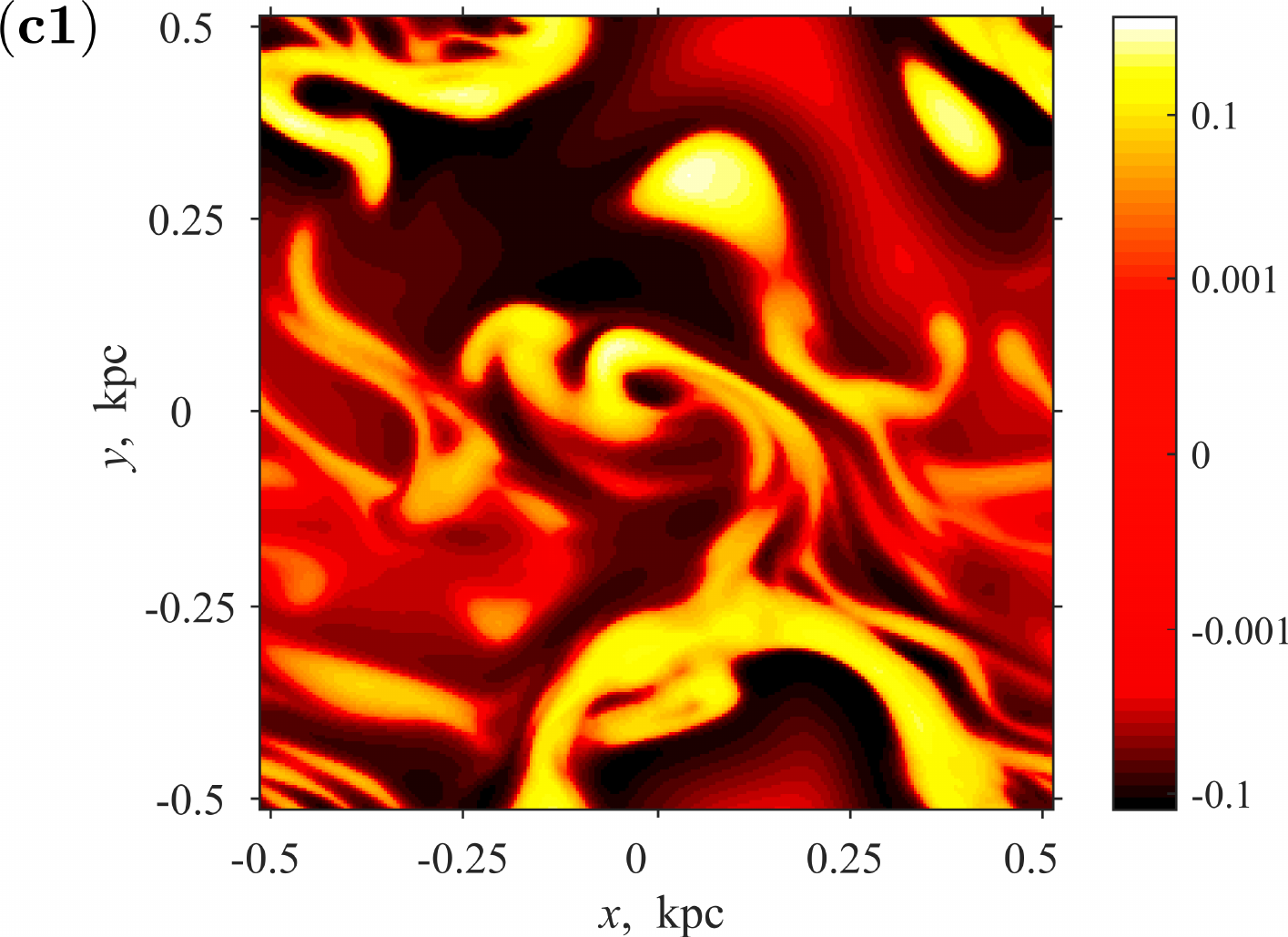}\\	\vspace{3mm}
	\includegraphics[width=0.35\textwidth]{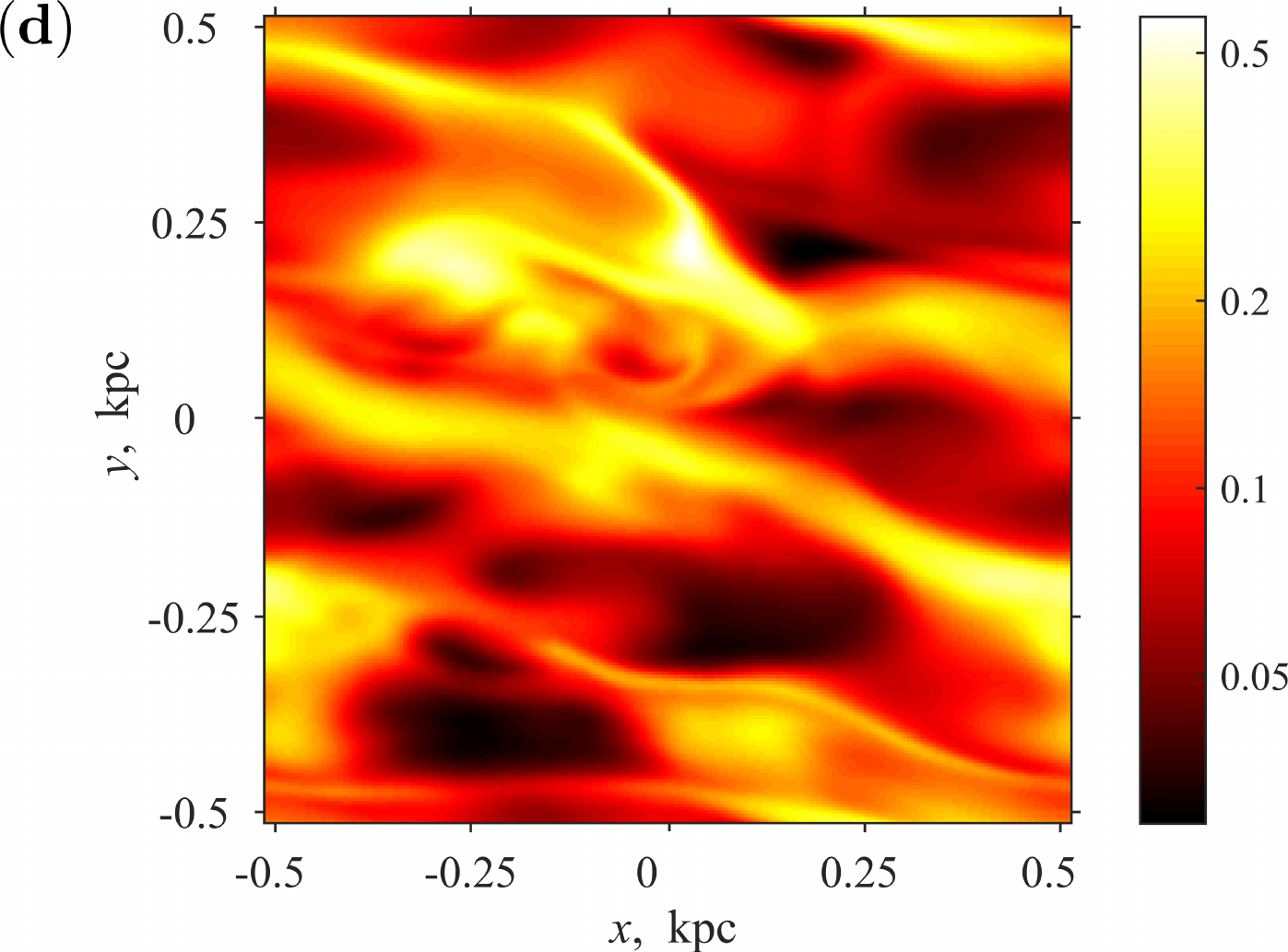} \hspace{5mm}
	\includegraphics[width=0.35\textwidth]{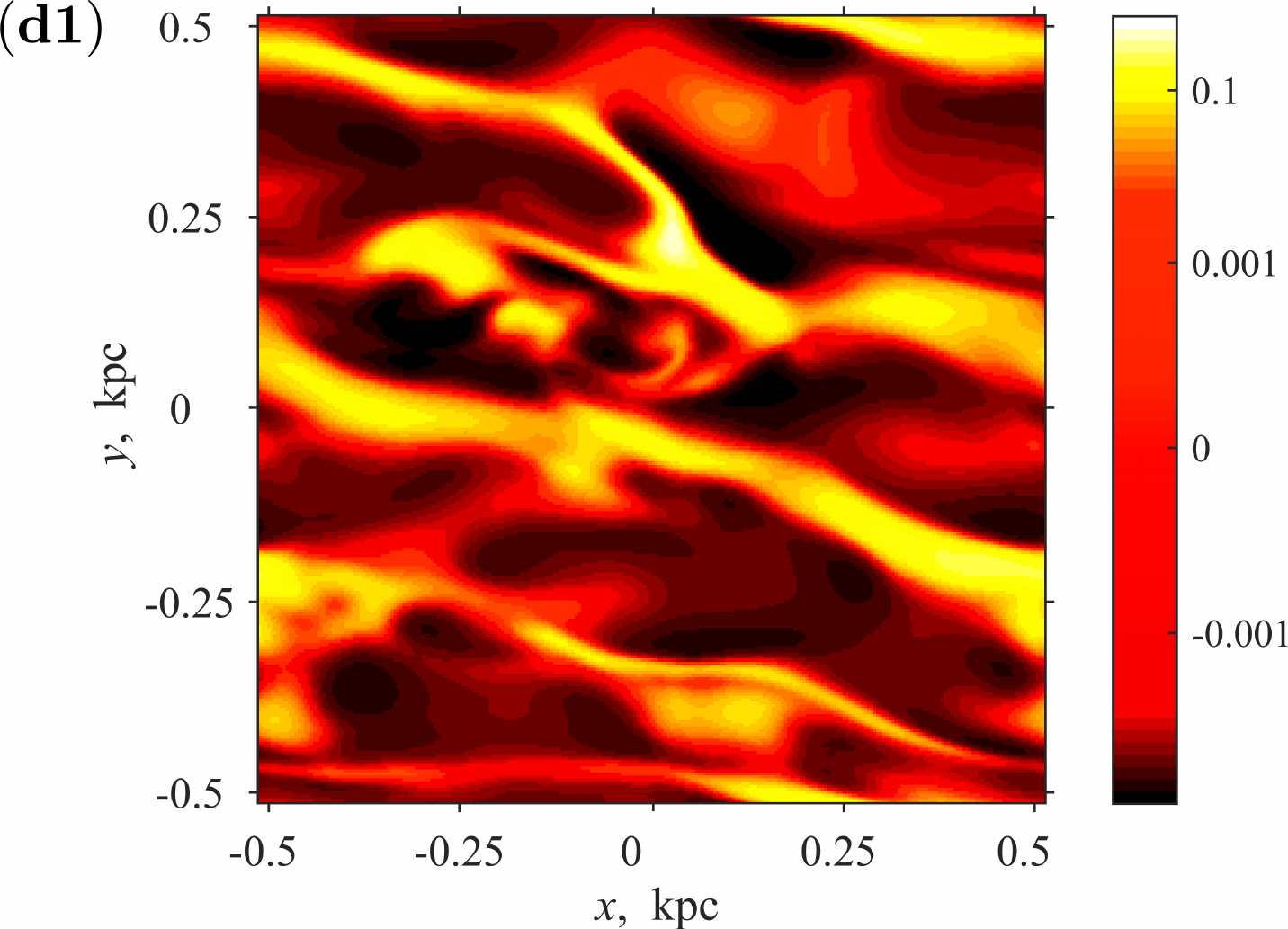}
	\caption{\label{fig:rhoFields}Examples of the gas density distribution $n(x, y)$
		(left column) and density fluctuations $\delta n(x, y)$ (right column), both
		in cm$^{-3}$,
		\textbf{(a)} and \textbf{(a1)}: $z = 0$, $t = 1\Gyr$, 
		\textbf{(b)} and \textbf{(b1)}: $z = 0$, $t = 1.6\Gyr$,
		\textbf{(c)} and \textbf{(c1)}: $z = -0.5\kpc$, $t = 1\Gyr$, and
		\textbf{(d)} and \textbf{(d1)}: $z = -0.5\kpc$, $t = 1.6\Gyr$.
	}
\end{figure*}
\begin{figure}
    \centering
    \includegraphics[width=0.85\columnwidth]{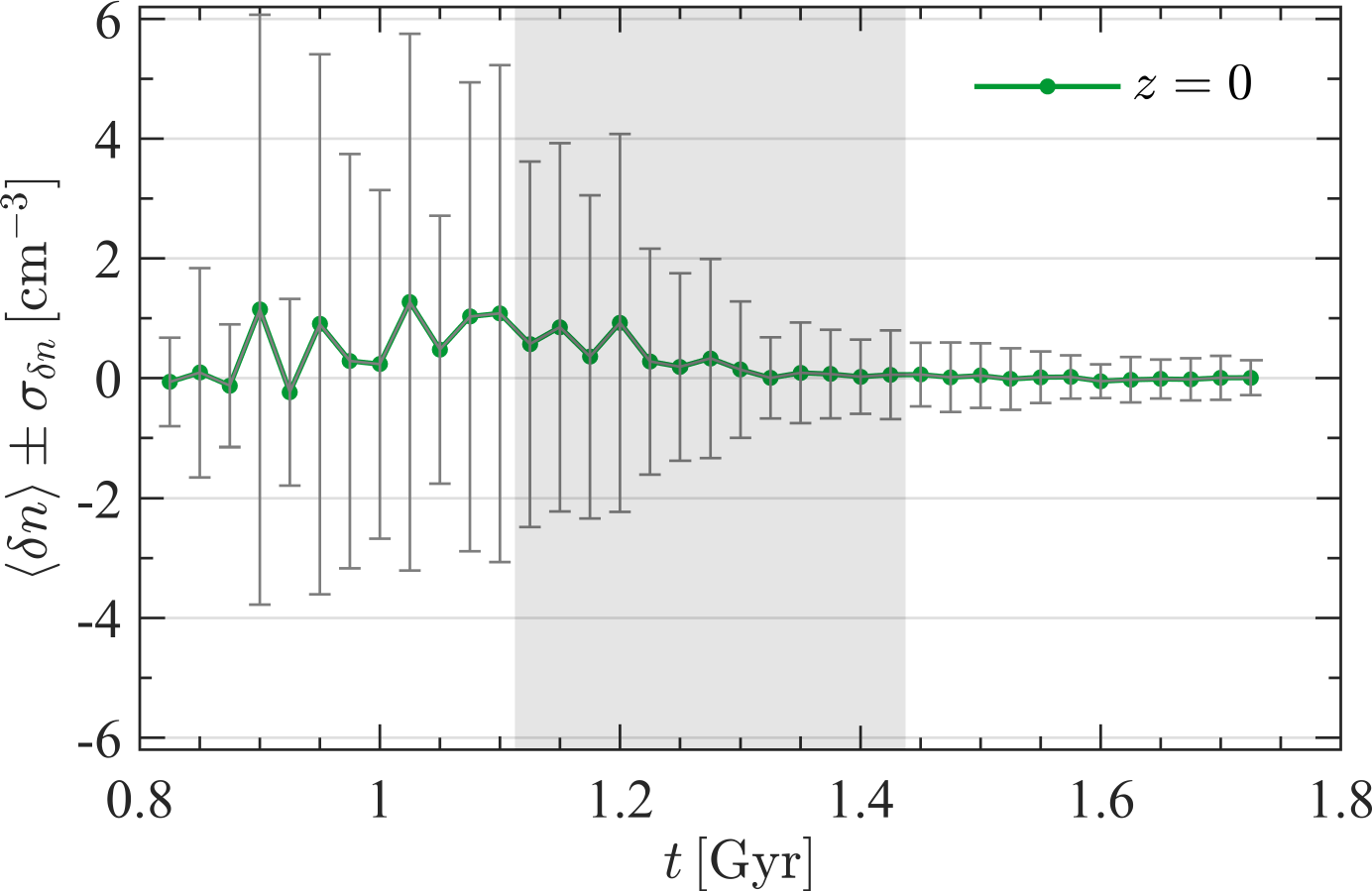}\\
    \includegraphics[width=0.85\columnwidth]{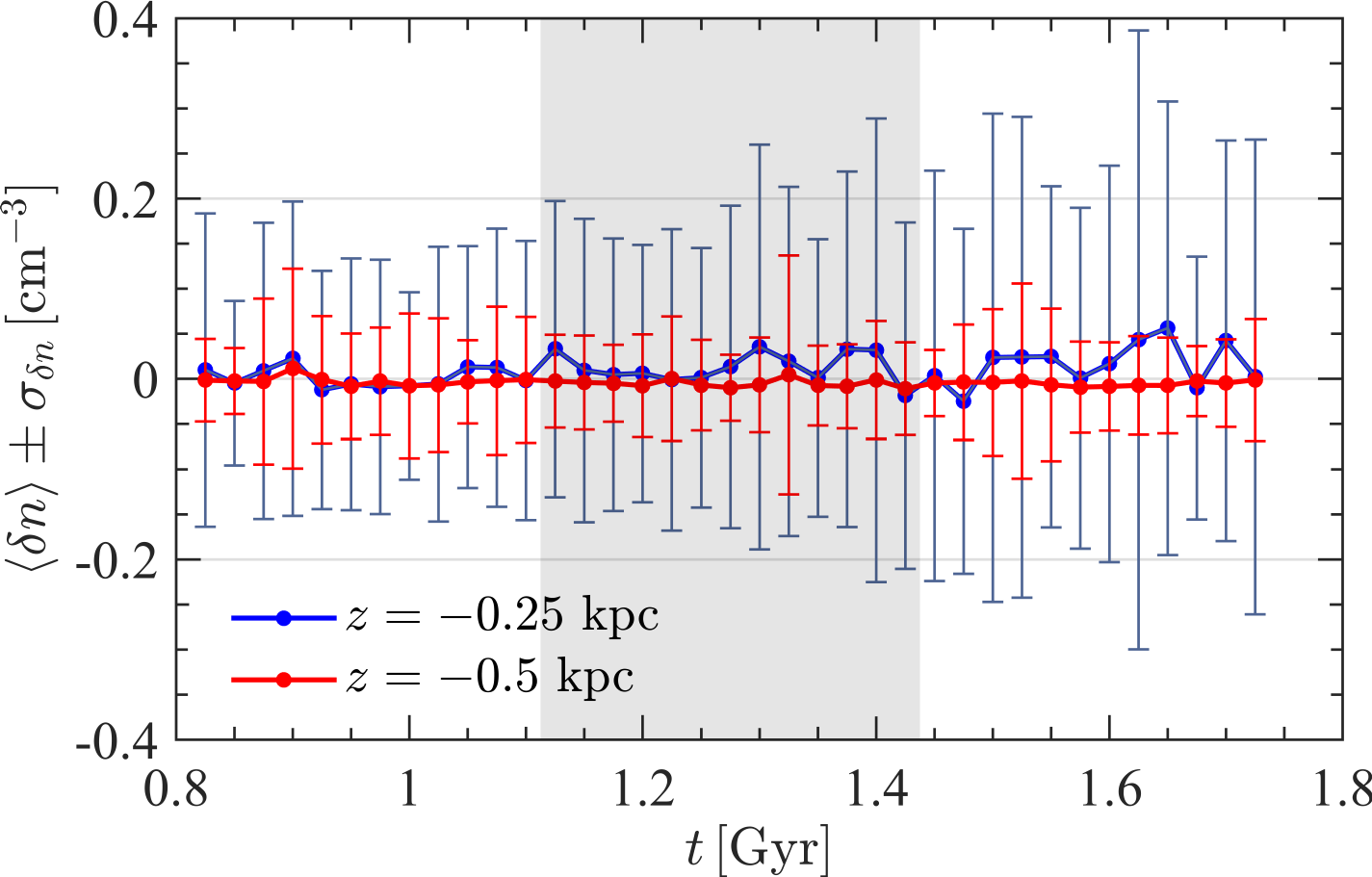}\\
    \includegraphics[width=0.85\columnwidth]{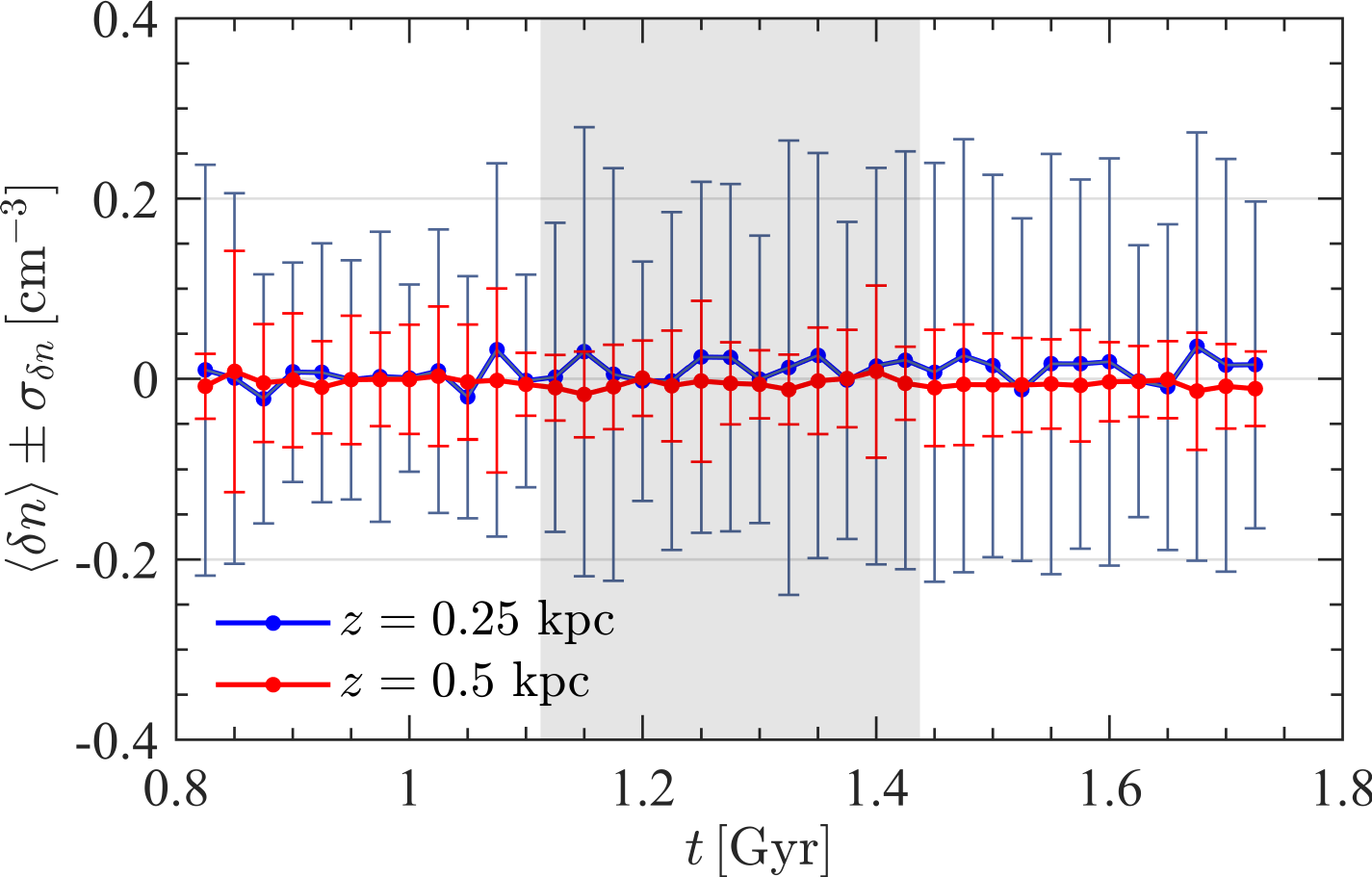}
    \caption{\label{fig:rhoMeanPmStd}Time variation of the mean of the fluctuations
        of total gas density $\langle \delta n \rangle$ and its standard deviation
				$\sigma_{\delta n}$ for the values of $z$ specified in the legend.
				At later times, when the magnetic field is stronger, $\sigma_{\delta n}$  at $z=0$ decreases.}
\end{figure}

The system is stratified, so all physical quantities vary systematically with $z$.
Therefore, we examined horizontal cross-sections of the gas density fluctuations
at fixed $|z|=0,0.25$ and $0.5\kpc$. Examples of the gas density distribution and
its fluctuations are shown in Fig.~\ref{fig:rhoFields}, using snapshots taken from Stages I (kinematic dynamo) and III (fully nonlinear dynamo).
Comparison of Panels (a) and (b), and especially (a1) and (b1) shows that
the strongly magnetised gas is more homogeneous, with
the total, mean and fluctuating gas densities in the ranges
$0\leq n\leq 37\cm^{-3}$, $0\leq\langle n\rangle\leq8\cm^{-3}$ and $-7\leq\delta n\leq31\cm^{-3}$
when the magnetic field is negligible as opposed to
$0\leq n\leq 3.3\cm^{-3}$, $0.2\leq\langle n\rangle\leq1.5\cm^{-3}$
and $-0.7\leq\delta n\leq1.9\cm^{-3}$ when the magnetic field is strong.
The horizontally averaged gas density at $z=0$ changes from $1.8$ to $0.6\cm^{-3}$
from Stage~I to Stage~III.

The supernova rate decreases exponentially with $|z|$, with
a scale height of $90\pc$ (less frequent Type~I supernovae have the scale
height of $325\pc$). The numerous circular structures in Panels (a)--(b1) are supernova
remnants, visible as holes in the gas distribution since gas in their interior is
hot and rarefied. The cellular structure in Panels (b) and (b1), corresponding
to a spongy structure in 3D, is more pronounced in the saturated phase than at earlier times,
as shown in Panels (a) and (a1).
This difference in the density fields, apparently associated with the magnetic field,
can be detected by the naked eye at $z = 0$ but not in Panels (c)--(d1)
that show the gas distribution at $z=-500\pc$. At larger values of $|z|$, the gas is
more homogeneous, with
$0\leq n\leq 0.7\cm^{-3}$, $0\leq\langle n\rangle\leq0.4\cm^{-3}$ and $-0.2\leq\delta n\leq0.5\cm^{-3}$
in Panels (c) and (c1), and
$0.03\leq n\leq 0.57\cm^{-3}$, $0.07\leq\langle n\rangle\leq0.31\cm^{-3}$ and $-0.12\leq\delta n\leq0.31\cm^{-3}$
in Panels (d) and (d1).
By comparing scales, a reduction in the intensity of the fluctuations is evident, and large-scale
gas structures extended at a small angle to the $x$-axis, produced by the
large-scale magnetic field, are clearly visible in Panels (d) and (d1).
Quantification of the small-scale gas structures is the purpose of this paper.

\begin{figure*}
  \centering
  \includegraphics[width=0.4\textwidth]{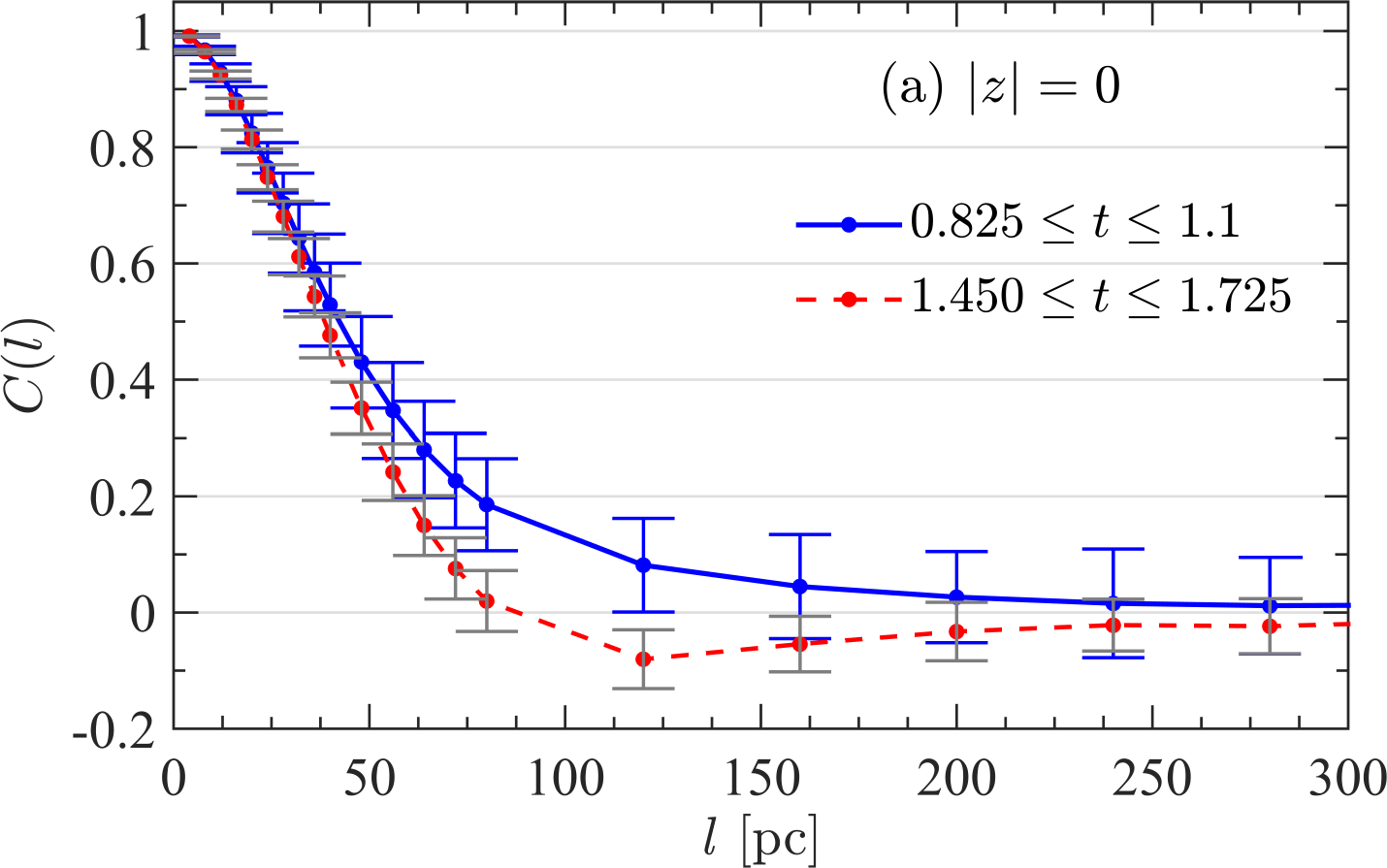}\hspace{4mm}
	\includegraphics[width=0.4\textwidth]{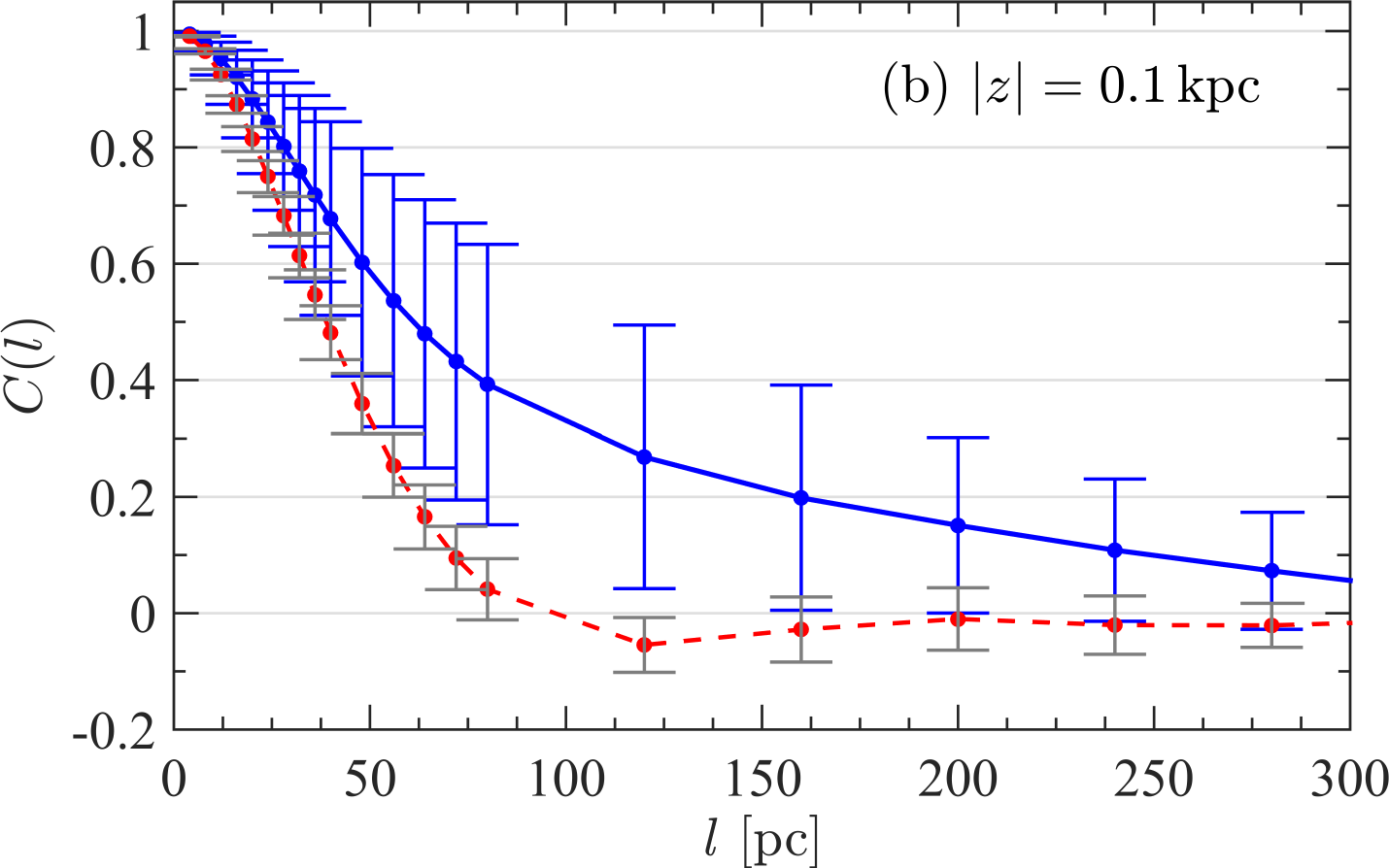}\\
	\vspace{2mm}
	\includegraphics[width=0.4\textwidth]{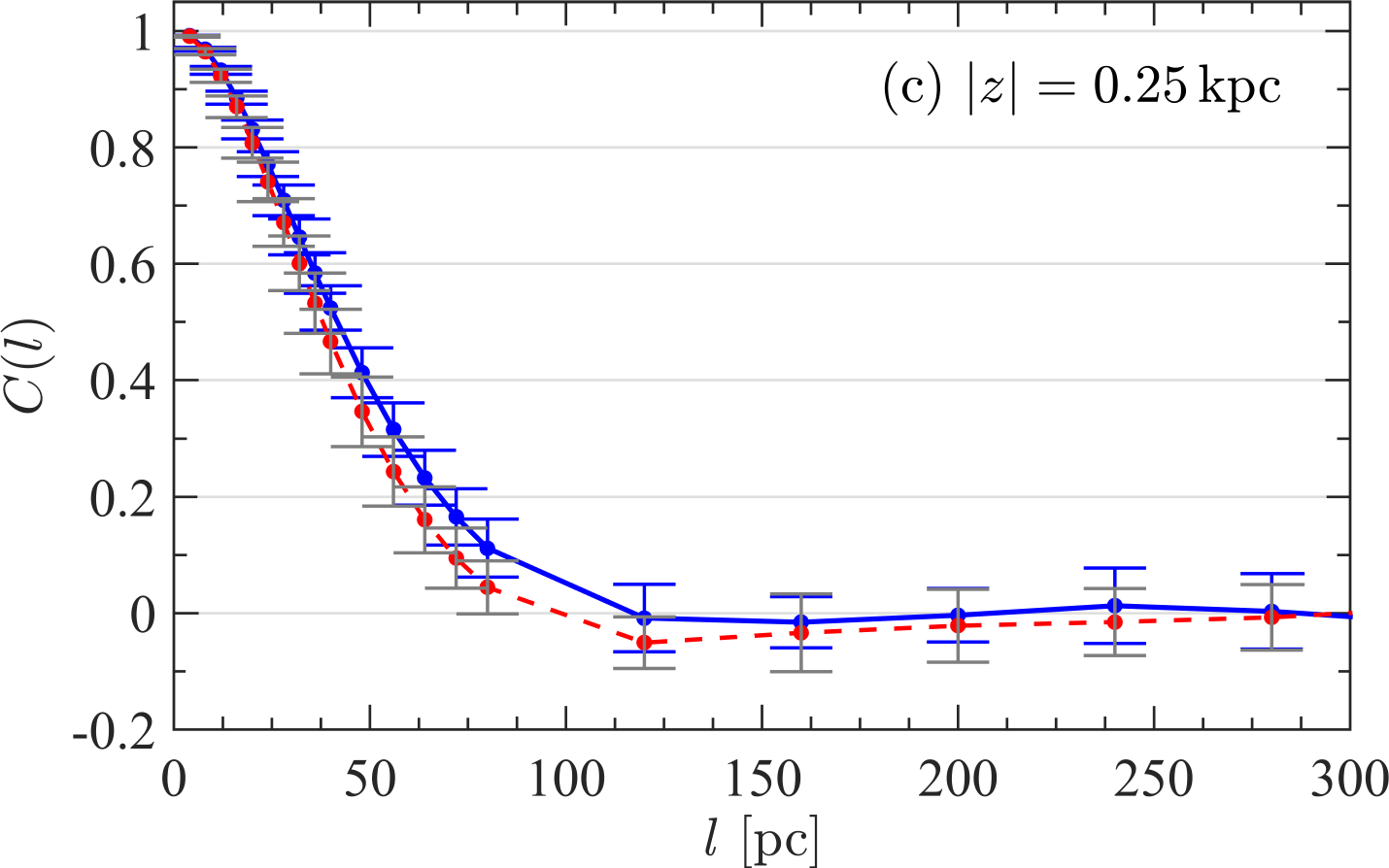}\hspace{4mm}
	\includegraphics[width=0.4\textwidth]{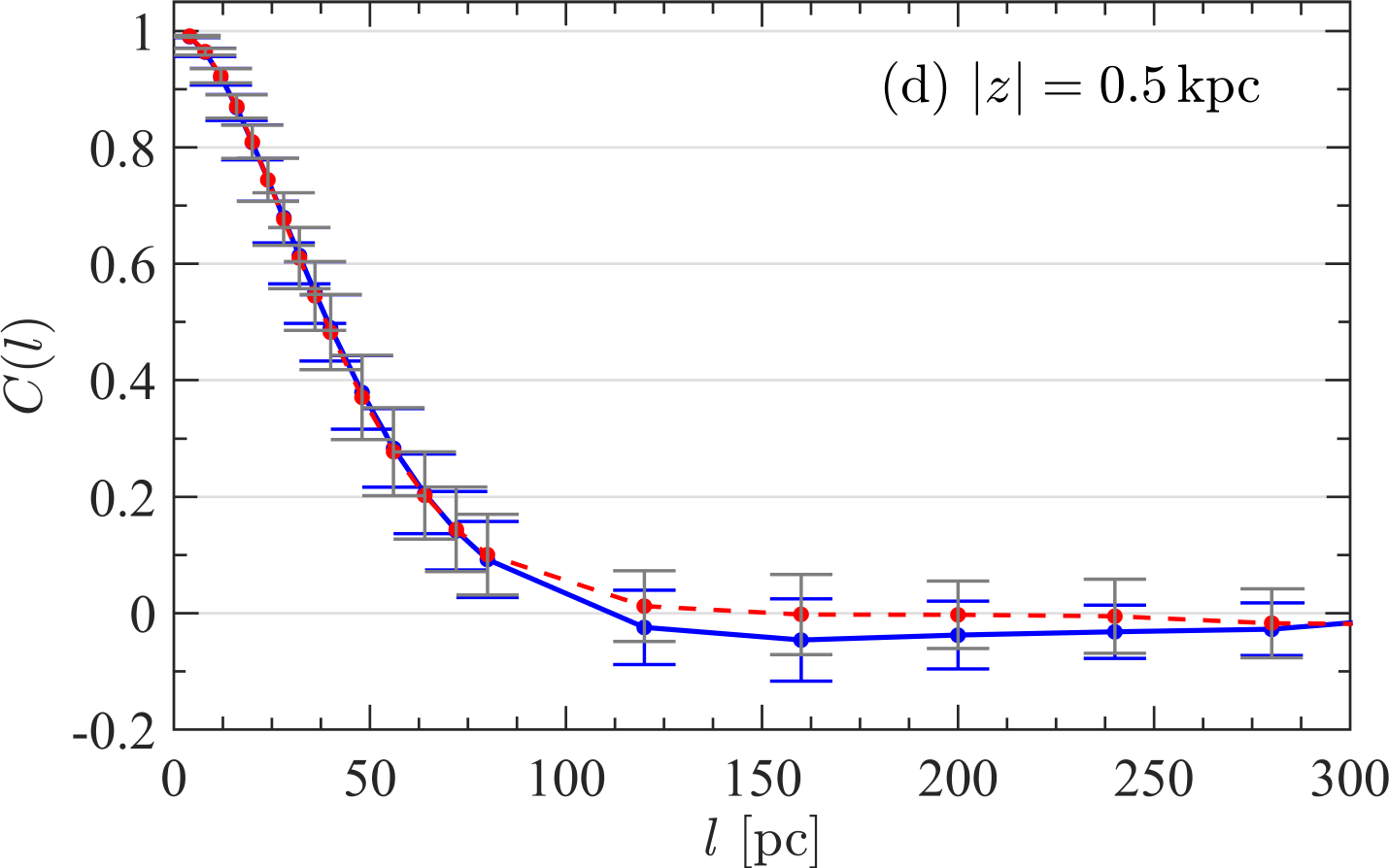}
  \caption{The auto-correlation functions of Eq.~\eqref{Cl},
  	in horizontal planes $z=\const$, for the gas
  	density fluctuations $\delta n$. These have been time-averaged over Stage~I (solid, blue) and
  	Stage~III (dashed, red) at the values of $z$ specified in the legends.
  	The error bars represent the standard deviation around the mean in each stage;
  	the statistical errors of $C(l)$ are negligible.
	}
  \label{fig:acfMean}
\end{figure*}

\begin{figure}
	\centering
	\includegraphics[width=0.85\columnwidth]{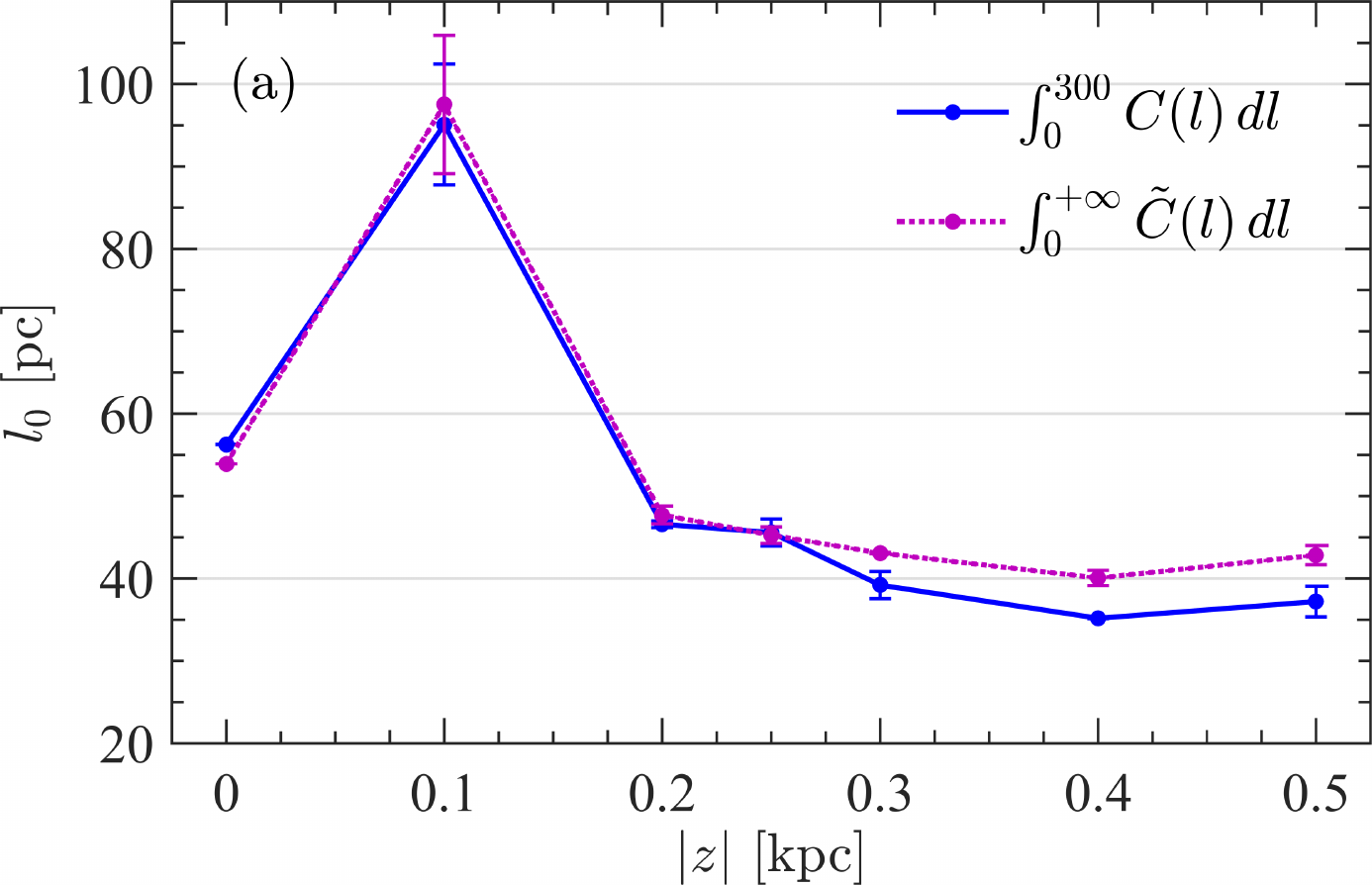}\\ \vspace{3mm}
	\includegraphics[width=0.85\columnwidth]{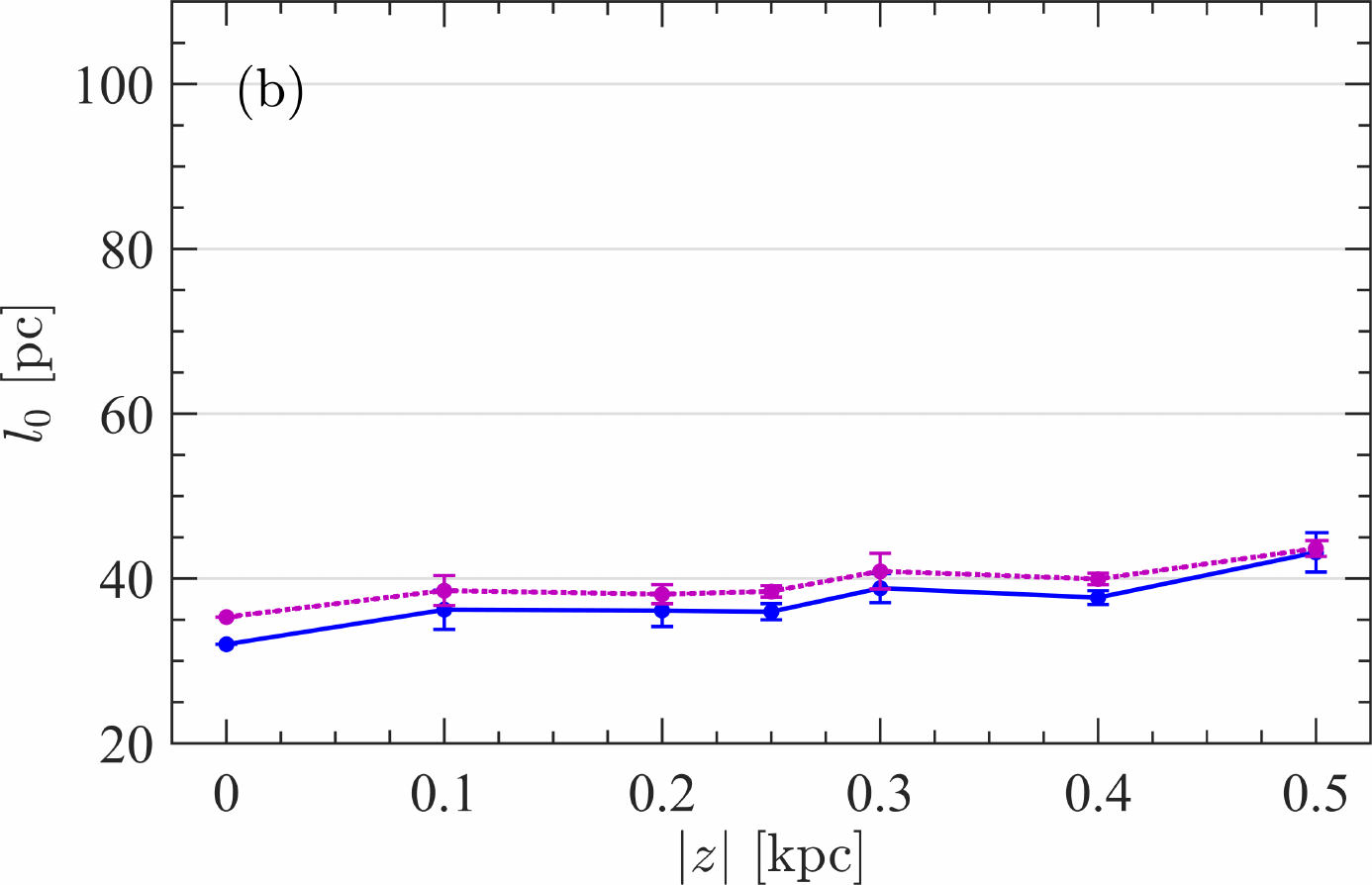}
	\caption{\label{fig:corrL}The correlation length $l_0$ from
		Eq.~\eqref{clen}
		as a function of $z$ in Stage~I
		(upper panel; negligible magnetic field) and Stage~III (lower panel;
		dynamically significant magnetic field). Shown are $l_0$ obtained
		from numerical integration of $C(l)$ over $0\leq l\leq300\pc$
		(solid, blue), analytical integration of $\tc(l)$ over
		$0\leq l<\infty$ (dotted, magenta),
		obtained from the time-averaged autocorrelation functions shown in
		Fig.~\ref{fig:acfMean}.
		}
\end{figure}

Figure~\ref{fig:rhoMeanPmStd} shows variations with time of the mean value of the density fluctuations
$\langle \delta n \rangle$ and its
standard deviation $\sigma_{\delta n}$ at $z = 0$, $\pm0.25$ and $\pm0.5\kpc$.
The fluctuations have a non-vanishing
mean value because the averaging is defined as Gaussian smoothing and it
does not satisfy the Reynolds rules \citep{Ger1992}.
Both the magnitude and the range of density fluctuations are larger
in Stage~I when the magnetic field is weak, and the variation with time is stronger at
the mid-plane. However, the magnetic field reduces the fluctuations very significantly.
This effect is hardly noticeable at $|z|=0.25$ and $0.5\kpc$ even though we will show that it is
readily detectable with the topological data analysis in Section~\ref{sec:scal}.

\section{Correlation analysis} \label{sec:corr}
New methods of analysis rarely invalidate the established ones, and require
careful justification of their advantages. Before
using topological data analysis, we conduct a standard correlation analysis
of the gas density field, not only to extract the information it can provide, but
also to identify its limitations. Correlation analysis of the gas density,
speed and magnetic field in these simulations is provided by \citet{HSSFG17}. Here
we present a more detailed analysis of the gas density fluctuations.

\subsection{Autocorrelation function of gas density}\label{DA}

For a stationary random field $f(\mathbfit{x})$, with $\mathbfit{x}=(x,y,z)$,
the normalised auto-correlation function at a given height $z$ is defined by

\begin{equation}\label{Cl}
C(l)
=\sigma_f^{-2}\iint_{z=\const} \left[f(\vect{x})-\langle f\rangle\right]
			\left[f(\vect{x}+\vect{l})-\langle f\rangle\right]\,\dd x\,\dd y\,,
\end{equation}
where $\sigma_f$ is the standard deviation of $f$, the lag
$\mathbfit{l}=(l_x,l_y,0)$ is confined to the plane $z=\const$ (over which the integral is evaluated), $l=|\mathbfit{l}|$, and
$\langle f\rangle$ is given by Eq.~\eqref{av}.
The auto-correlation function is further averaged over time in each of Stages
I, II and III. Formally, we should write $C(l,z)$ here, to indicate that the auto-correlation function
depends on $z$. However, we only ever consider the $l$ dependence of this function at fixed $z$, so we
have abbreviated this for notational convenience.

The results are shown in Fig.~\ref{fig:acfMean} for Stage~I (solid, blue) and Stage~III
(dashed, red) at $|z| = 0, 100, 250$ and $500\pc$. The gas distribution is
symmetric about the mid-plane $z=0$ on average. However, the size of the computational
domain is relatively small, so that deviations from perfect symmetry
in individual snapshots can be considerable. To verify that this does not affect
our conclusions, we first plotted $C(l)$ separately for $z=250\pc$
and $z=-250\pc$ and similarly for $z=\pm500\pc$. Indeed, the autocorrelation
functions turned out to be symmetric with respect to $z=0$. As $C(l)$ at positive and
negative values of $z$ are close to each other, we show
the averaged correlation functions for $|z|=\const$ in Fig.~\ref{fig:acfMean}.

The difference between the autocorrelation functions obtained with and without
a strong magnetic field (Stages~I and III, respectively) is significant only
at $z=0$
and $\pm0.1\kpc$, shown in Panels (a) and (b) respectively,
but even then the magnetic field does not lead to any characteristic features
that could be used to unambiguously confirm its presence.
For example, a negative tail of $C(l)$ develops at $z=0$ only in Stage~III and could be thought
to be a consequence of the magnetic field. However, at $|z|=500\pc$ a similar (albeit weaker) negative tail
is stronger in Stage~I.
We also found that in Stage~I the autocorrelation function at $|z| = 100\pc$
significantly differs from that at other $z$:
the correlation function in Stage~I decreases with $l$ slower than at other values of $|z|$,
with larger scatter between the snapshots reflected in longer error bars,
and the change between Stages~I and III is stronger.
This is apparently related to the fact
that supernova activity reduces significantly beyond $|z|\simeq100\pc$. However,
the autocorrelation function has a similar form at all distances from the
mid-plane in Stage~III, presumably because magnetic pressure
redistributes gas along $|z|$, reducing the significance of the supernova layer.

As discussed by \citet{EGSFB18}, the magnetic field is strongest at
$|z|=200\text{--}300\pc$, so its effects on the gas distribution can be expected
to be most pronounced there. However, the autocorrelation functions do not
show any signs of this.

\subsection{Correlation length}\label{CL}

The correlation length is defined as
\begin{equation}\label{clen}
l_0 = \int_0^{\infty} C(l)\,\dd l\,.
\end{equation}
Calculation of $l_0$ from experimental or numerical data that is restricted to a
relatively narrow range of $l$ requires some caution.
For example, integration of $C(l)=\exp(-l/l_0)$ to $l=L$ leads to a relative
error of $\exp(-L/l_0)$ in the correlation length, about five per cent for
$L=3l_0$. The problem can be aggravated by statistical errors in $C(l)$.
As suggested by \citet{HSSFG17}, as well as integrating $C(l)$ numerically
within the range $0\leq l\leq300\pc$, we approximated the derived autocorrelation functions as
\begin{equation}\label{Ct}
\tc(l) = \cos\left(al^2+bl\right)\,\exp\left[-l^2/(2 c^2)\right]\,,
\end{equation}
and obtained $a$, $b$ and $c$ from least-squares fits to the autocorrelation
functions of Section~\ref{DA}. The unweighted fits were used, and the fit quality was verified;
the $\chi^2$ was test satisfied for all of the fits.
This analytic approximation was then integrated
over $0\leq l<\infty$. The term $\cos\left(al^2+bl\right)$ is required to allow for the
negative values of $C(l)$ at $|z|\ga100\pc$ (see Fig.~\ref{fig:acfMean}).
Another property of a correlation function is its
curvature at $l=0$, usually characterised in terms of the Taylor micro-scale,
$\lambda$, defined via
\citep[][]{TL72}
\begin{equation}\label{TM}
C(l)\approx 1-(l/\lambda)^2
\qquad\text{for } l/\lambda\ll1\,.
\end{equation}
\noindent However this a more difficult quantity to measure than the correlation length, because
it is determined by correlations at small-scales, which can be influenced by the grid resolution.
For similar reasons, differences in the small-scale structures in these simulations are difficult
to characterise using only a correlation analysis.

\begin{table*}
	\centering
	\caption{\label{tab:notations}
	Summary of notation.}
	\begin{tabular}{cll}
		\hline
		Symbol & \centering{Meaning} & Reference\\
		\hline
		$l_0$ & Correlation length & Section 3\\
		$\lambda$ & Taylor microscale & Section 3\\
		$h$ & The level of an isocontour or isosurface & Section 4\\
		$\beta_0$ & A number of components of a sublevel set at a level $h$ & Section 4\\
		$\beta_1$ & A number of holes of a sublevel set at a level $h$ & Section 4\\
		$\cb_0$ & The total number of components in a filtration, equal to the number of points in the $\beta_0$ persistence diagram &
		Section 4 \\
		$\cb_1$ & The total number of holes in a filtration, equal to the number of points in the $\beta_1$ persistence diagram &
		Section 4\\
		$\tb_0$ & A number of components of a sublevel set at a level $h$ per area of $l_0^2$ & Section 4 \\
		$\tb_1$ & A number of holes of a sublevel set at a level $h$ per area of $l_0^2$ & Section 4\\
		$\tB_0$, $\tB_1$ & The total number of components and holes in a filtration per correlation cell & Section 5\\
		$\chi$ & Euler characteristic, in 2D: $\beta_0 - \beta_1$; per correlation length $\tB_0 - \tB_1$  &  Section 5\\
		\hline
	\end{tabular}
\end{table*}

Figure~\ref{fig:corrL}
shows the $z$-dependence of the correlation length,
obtained as described in the figure caption.
The estimates of $l_0$ at positive and negative values of
$z$ are very close to each other, so we present the average values.
It is clear from the comparison of the upper and lower panels of
Fig.~\ref{fig:corrL} that magnetic fields significantly affect the
statistics of the gas distribution in the region $|z|\la300\pc$,
with significant differences in the corresponding correlation lengths between Stages I and III.
Further from the mid-plane, the correlation lengths do not change appreciably with time.

\section{Topological data analysis}\label{sec:tda}

Unlike correlation analysis, topological data analysis is not restricted to finite-order
statistical moments of a random field.
Its aim, achieved through
\textit{topological filtration}, is to isolate significant
properties of a random field that can be used to simplify it and thus to
make it amenable to analysis, comparison and statistical inference.
Rigorous definitions of the Betti numbers and related concepts can be found
in
\citet{ABBSW2010,EH10,AdTay2011} and \citet{Ed2014}
while \citet{Park2013} and
\citet{Pranav2017} provide useful and less formal expositions. Here we briefly
present the basics at an intuitive level.

Betti numbers $\beta_n$
characterise the topological structures that form
a random field $f(\vect{x})$. Firstly, an isosurface at $f(\vect{x})=h$, where $h$ is a constant, of a 3D random field
is defined
The topology of the isosurface is then
characterised
in terms of isolated
components, loops and closed shells (known as \textit{cycles} of dimension 0,
1 and 2, respectively). The Betti numbers quantify this structure:
$\beta_0$ is the number of components, $\beta_1$ is the
number of loops that cannot be reduced to a point by a continuous deformation.
The Betti numbers are topological invariants, i.e. they are not affected  by translations,
rotations and continuous deformations of the random field.
In a typical astrophysical application, the components
	represent matter clumps or clouds, the cycles are closed chains of
matter and the shells surround regions of reduced density (voids).

Representation of the random field in terms of the Betti numbers, and their
variation as the level $h$ of the isosurface changes, is called the
\textit{filtration} of the random field: it isolates topologically significant
features of the field, and thus facilitates their analysis and comparison.
For a given $h$, the \textit{superlevel} and \textit{sublevel} sets are defined to be 
the sets of points $\vect{x}$ where $f(\vect{x})>h$ or $f(\vect{x})<h$
respectively.
Topological filtration is therefore a collection of topologically
distinct superlevel or sublevel sets of a random field, similar to those shown in
Figs.~\ref{fig:filt}a and \ref{fig:filt}b. The topology of such
a set only changes when the level $h$ passes through a critical point of the
random field. Therefore, a filtration of the random field only contains
topologically significant information about it. The topology of the superlevel
or sublevel sets is characterised in terms of the Betti numbers, $\beta_n$.

In this paper we will also use the number of components or holes for the whole filtration
(i.e. the total numbers obtained for all levels $h$), which
is equal to the total number of components and holes at all levels $h$.
These two quantities, related to the Betti numbers $\beta_0$ and $\beta_1$,
are denoted as $\cb_0$ and $\cb_1$, respectively.
Those topological features of the isosurfaces
that occur continuously over a wide range of
isosurface levels are said to be
\textit{persistent} and considered to be the most important.
In the next section we discuss in more detail the idea of topological filtration
and Betti numbers and then illustrate them using a specific example in
Section~\ref{IE}. Some readers may find it useful to read Section~\ref{IE}
in conjunction with or even before Section~\ref{sec:filt}.
For later convenience, the key notation is presented in Table~\ref{tab:notations};
the section where	each notation appears in the text for the first time is shown in the right column.

\subsection{Topological filtration and Betti numbers}\label{sec:filt}

The above description refers to a three-dimensional random field. In two
dimensions, only the Betti numbers $\beta_0$  and $\beta_1$ are defined. To
illustrate the technique, consider a continuous random function $f(x, y)$
defined in a finite domain, and its isocontour of a level $h$, i.e. a curve
in the $(x,y)$-plane where $f=h$.
The set of points $(x, y)$ where $f>h$ is therefore the superlevel set;
its complement, $f<h$,  is the sublevel set.
To filter out insignificant noise but allow the important elements of $f$
to pass through the filter, we vary $h$ from smaller to larger values and
record the number of components (clouds) and holes
in each isocontour together with the values of $h$ where components and
holes appear and disappear. The isocontours can also be scanned
down
from larger to smaller values of $h$: this does not affect the results.
The transition from the sublevel sets to superlevel sets swaps the persistent
diagrams of $\beta_0$ and $\beta_1$ (in 2D). Here we use the sublevel sets,
i.e. scan $f(x,y)$ up from smaller to larger values.

\begin{figure*} 
	\centering
	\includegraphics[width=0.3\textwidth]{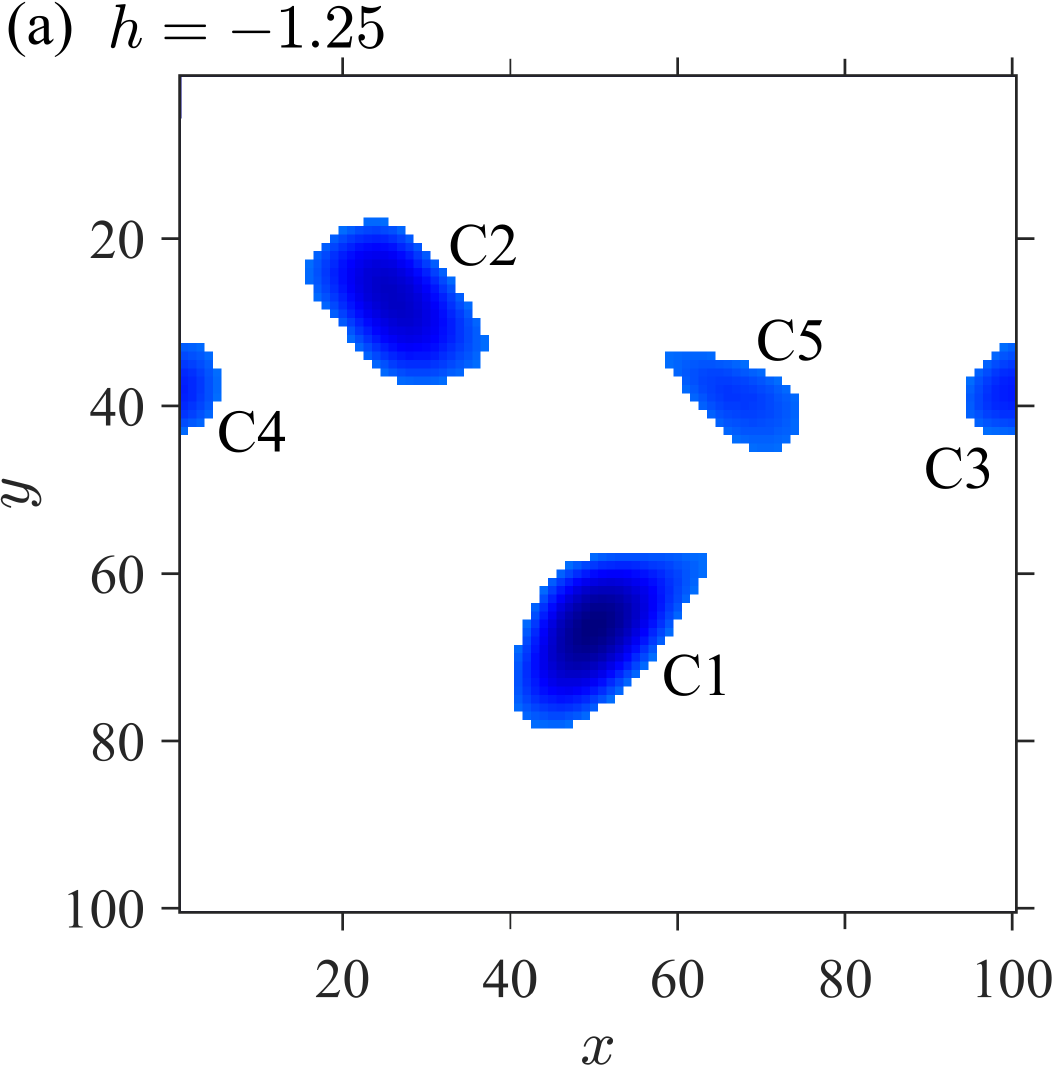}\hspace{4mm}
	\includegraphics[width=0.3\textwidth]{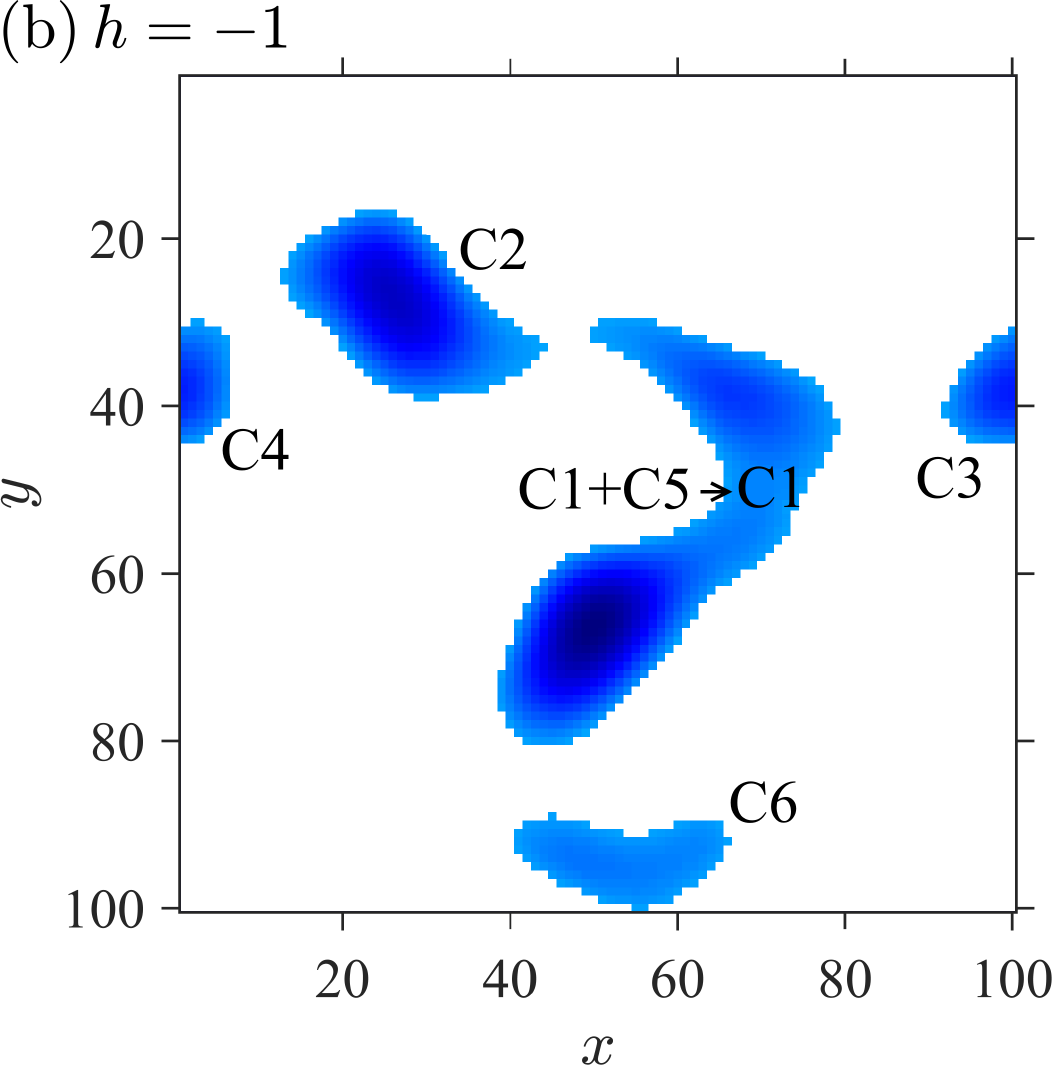}\hspace{2mm}\hfill
	\includegraphics[width=0.3\textwidth]{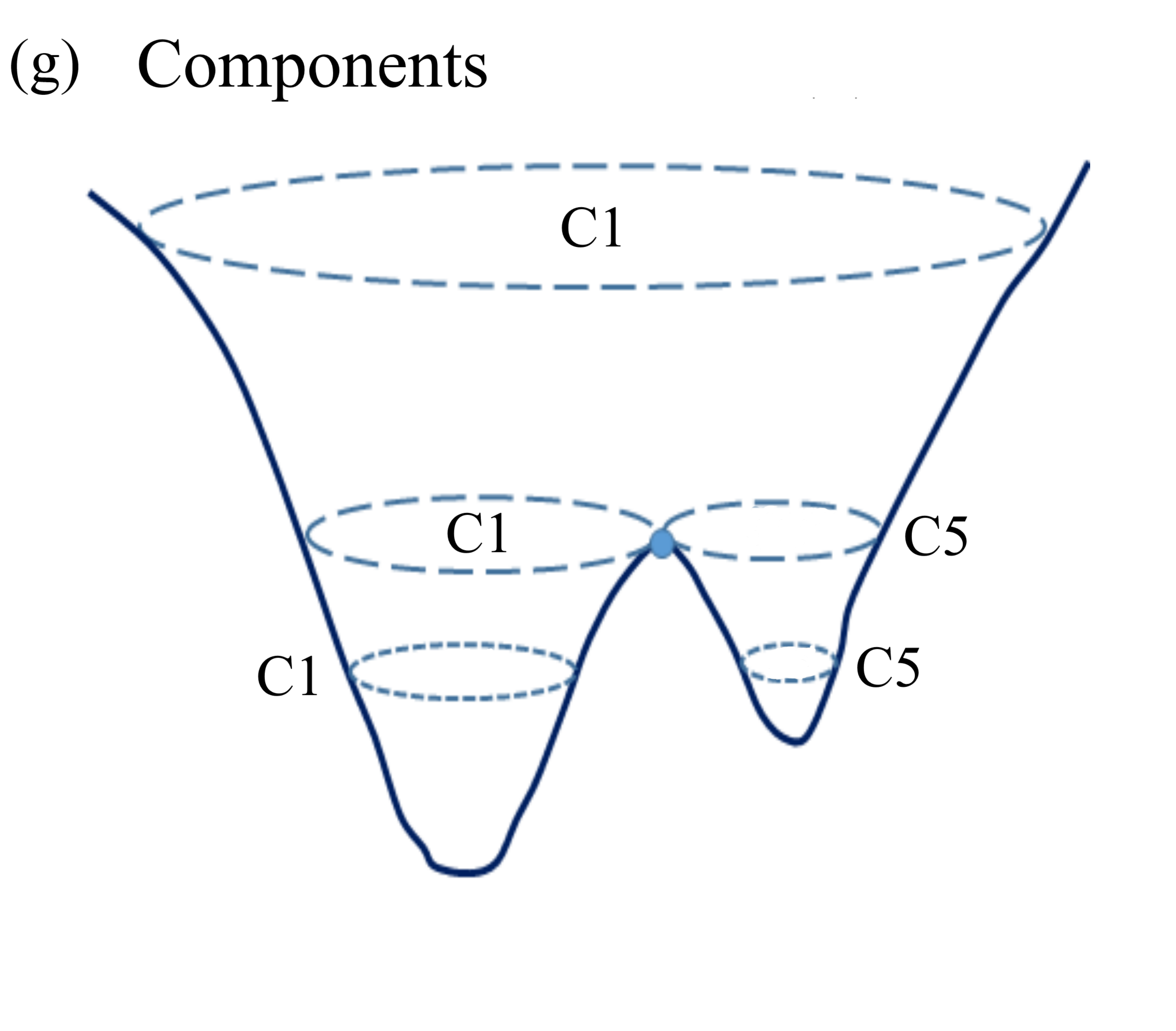}\vspace{1mm}\\%
	\includegraphics[width=0.3\textwidth]{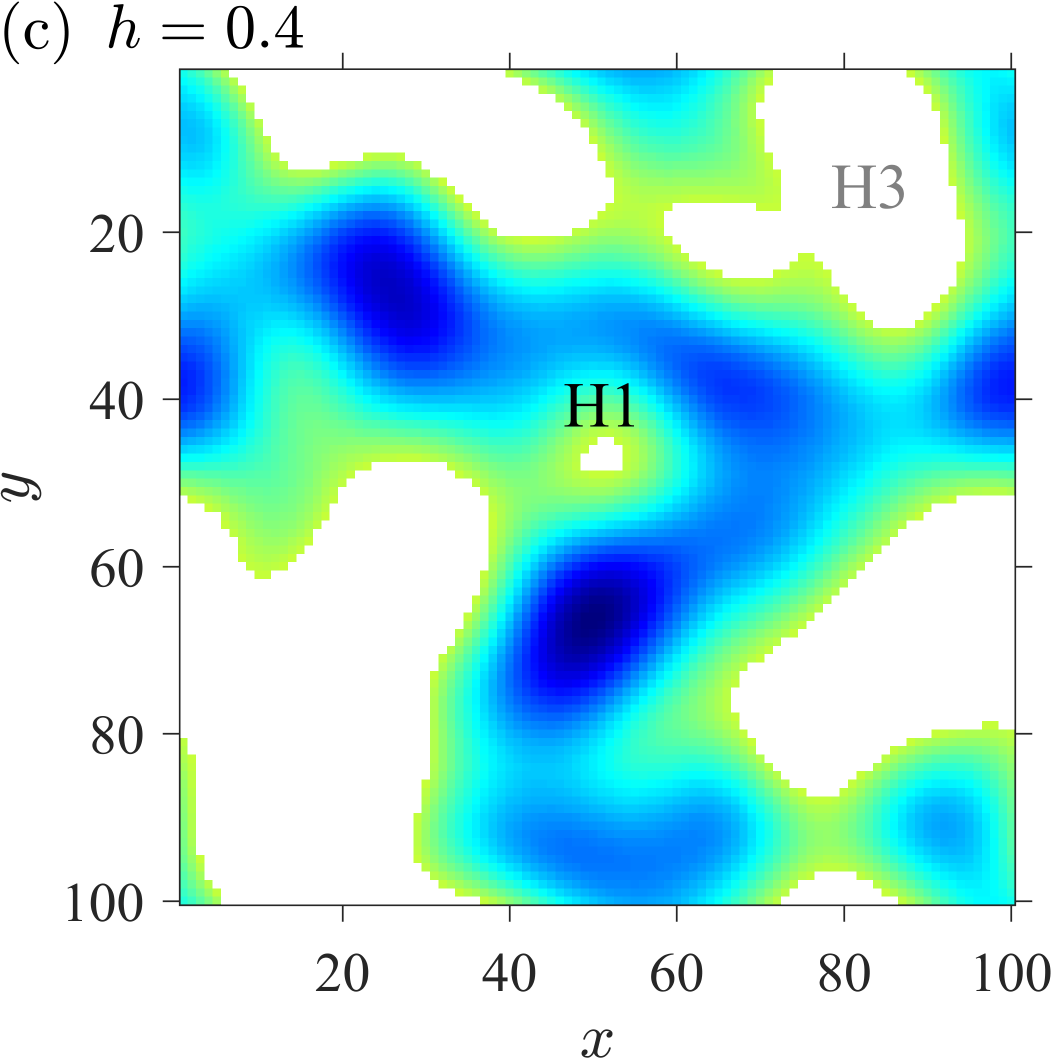}\hspace{4mm}
	\includegraphics[width=0.3\textwidth]{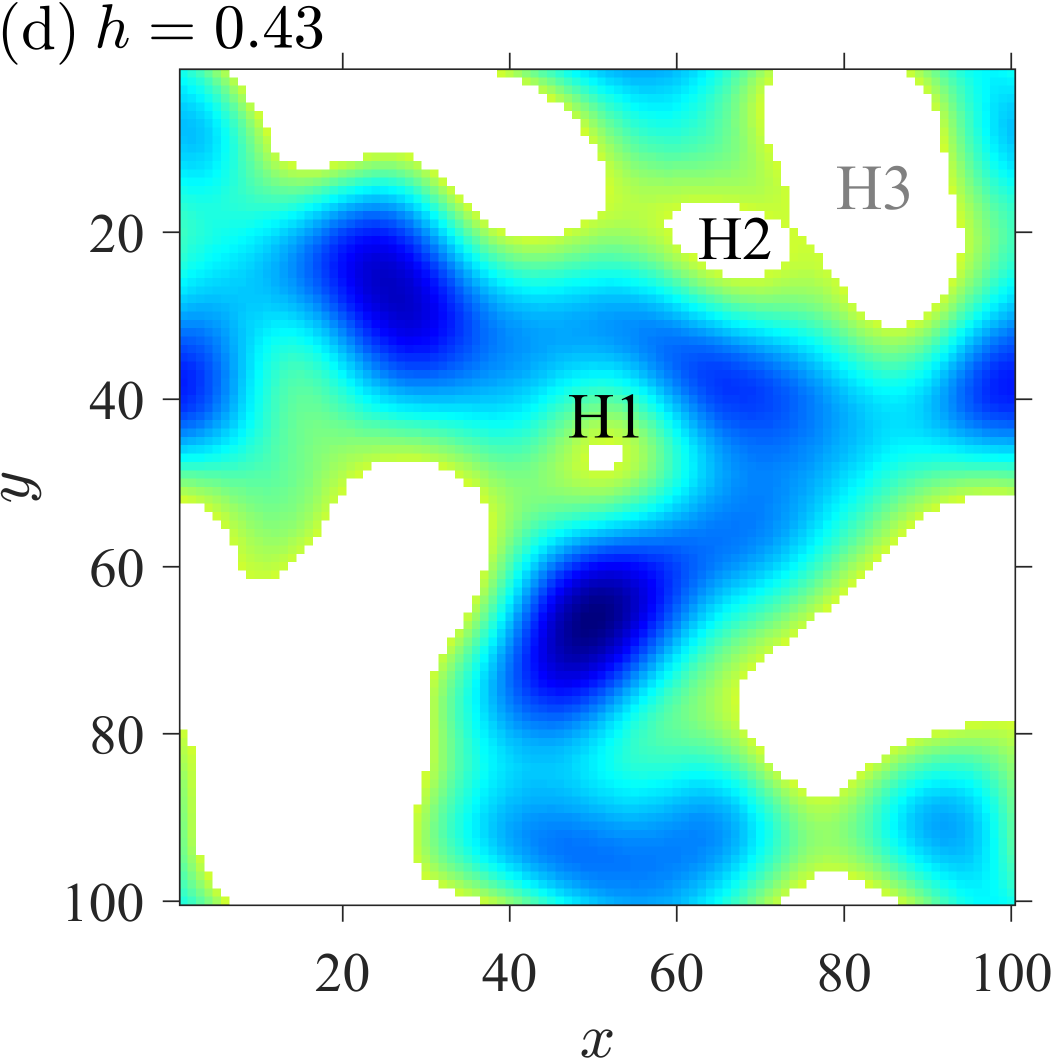}\hspace{2mm}\hfill
	\includegraphics[width=0.3\textwidth]{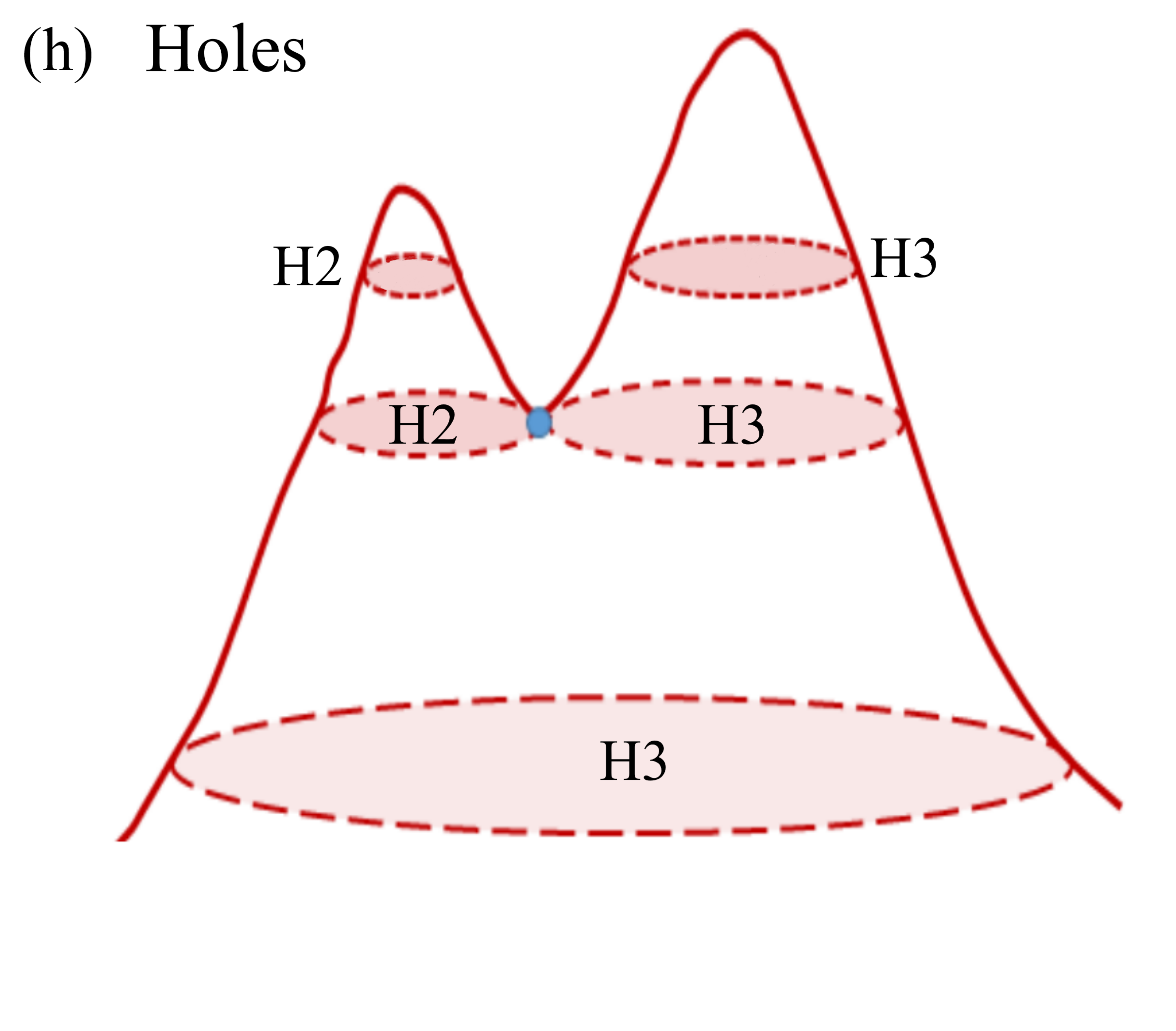}\vspace{1mm}\\
	\includegraphics[width=0.3\textwidth]{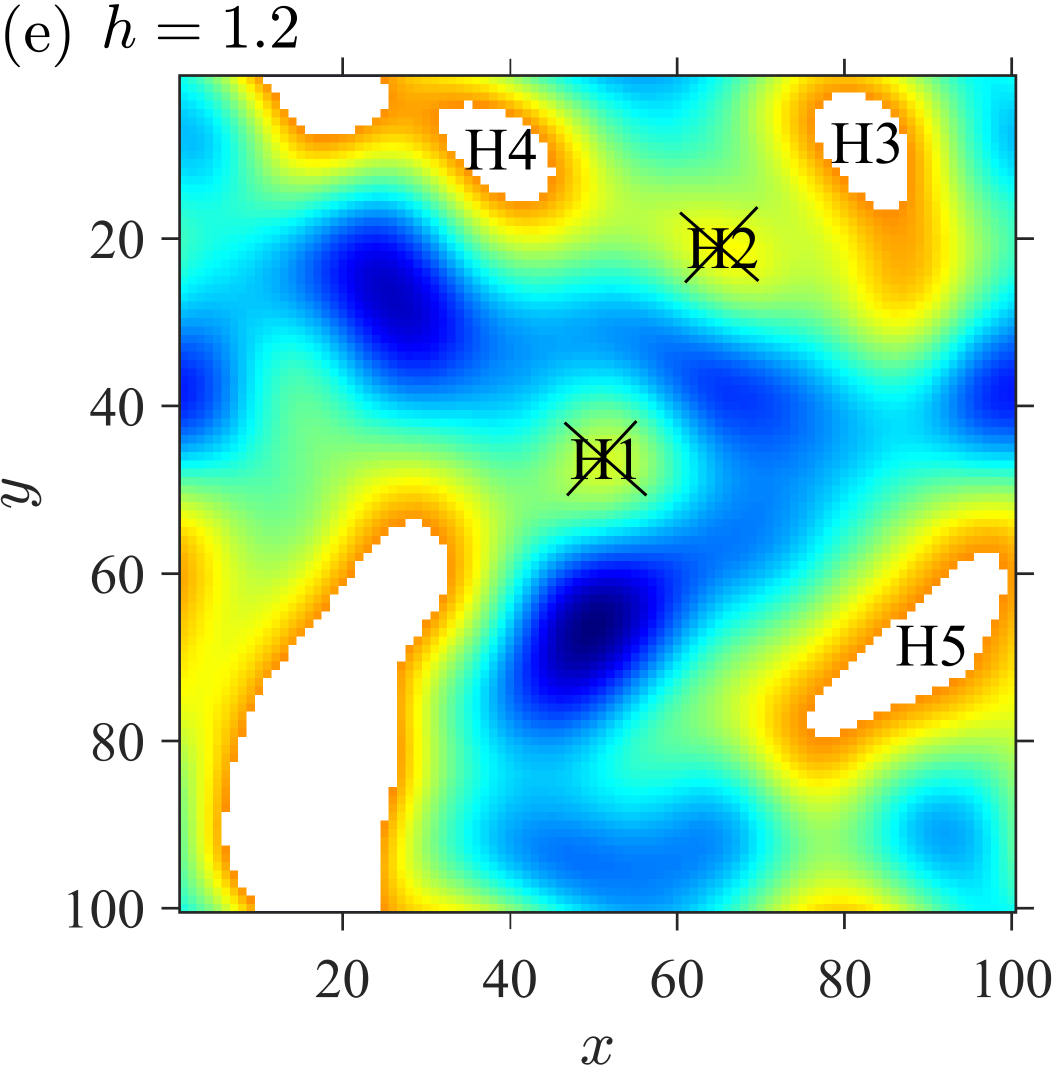}\hspace{3.8mm}
	\includegraphics[width=0.35\textwidth]{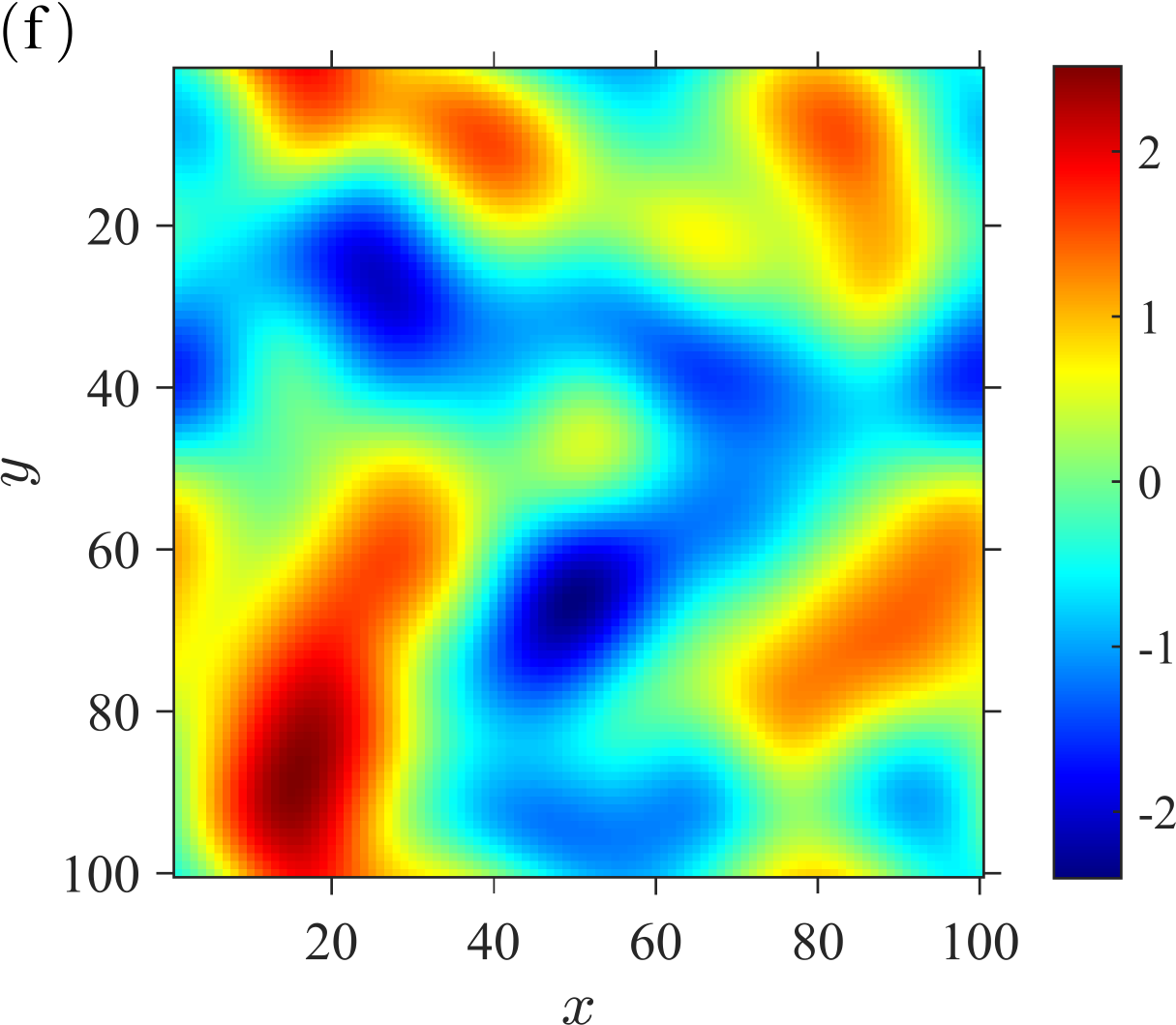}\hfill
	\caption{\label{fig:filt}
	Topological filtration of a continuous, smooth, 2D Gaussian scalar field, 
	$f(x,y)$, which has a vanishing mean value, unit standard deviation, 
	and an auto-correlation function $C(l)=\exp\left[-l^2/(2L^2) \right]$, 
	where $L\approx15$. The colour table, which is defined by the
	colour bar shown near Panel (f), shows the spatial variations of this 
	scalar field. Sublevel sets, i.e. regions where $f(x,y)<h$, are shown
	for increasing values of $h$:
         \textbf{(a)}~$h=-1.25$, \textbf{(b)}~$h=-1$ \textbf{(c)}~$h = 0.4$, 
	\textbf{(d)}~$h = 0.43$  and \textbf{(e)}~$h = 1.2$.
	The components (C) and holes (H) are labelled in
	Panels~(a), (b), (c), (d) and (e) as described in the text. 
	Panels~\textbf{(g)} and \textbf{(h)} show how the components merge and
	the holes split as the level $h$ changes. Panel \textbf{(f)} presents the full 
	range of $f(x,y)$.}
	\end{figure*}

When the level $h$ reaches the lowest value of $f$ in the domain,
the first component emerges as illustrated in Fig.~\ref{fig:filt}.
Each local minimum of $f$ adds a new component
when it is reached by the increasing isocontour level $h$.
Two (or more) components can merge, when $h$ increases, if they are connected by a
gorge (where a saddle point should be located); then the labelling convention
is that the
\textit{younger} component (i.e., that formed at a larger $h$) dies whereas
the older one continues to exist. Two components merge when the isocontour
contains a saddle point, see Fig.~\ref{fig:filt}, panel (g); three components merge, when $h$ passes through a
monkey saddle (a degenerate critical point with a local minimum along one direction
and inflection point in another, as opposed to the ordinary saddle
with a minimum in one direction an maximum in another).
Three (or more) components can merge, when there are three (or more)
saddle points at the same level $h$.

The components can merge to form a loop whose interior is a hole; each hole at a level $h$ surrounds
a local maximum of $f$ that will occur at some higher level.
Holes are born when a loop is formed and die when $h$ passes through the corresponding local maximum of $f$.
Holes can split up at a saddle point, in which case the labelling convention is
to ascribe the original birth time to the hole with the larger eventual local maximum, 
and deem the current level to be the birth time of the second hole.
Fig.~\ref{fig:filt}, panel (h), illustrates this.

\begin{figure*}
	\centering
	\includegraphics[width=0.4\textwidth]{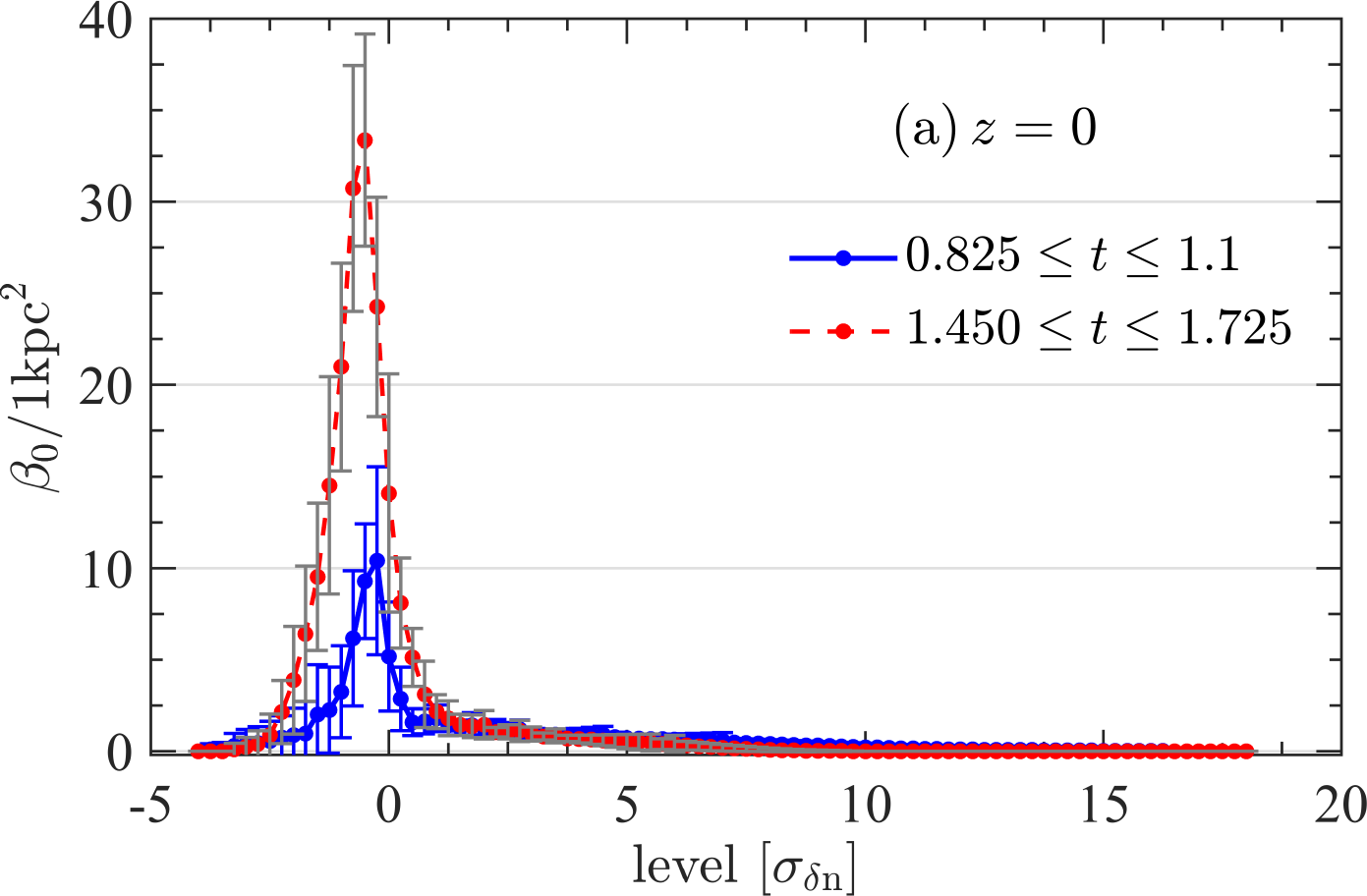} \hspace{3mm}
	\includegraphics[width=0.4\textwidth]{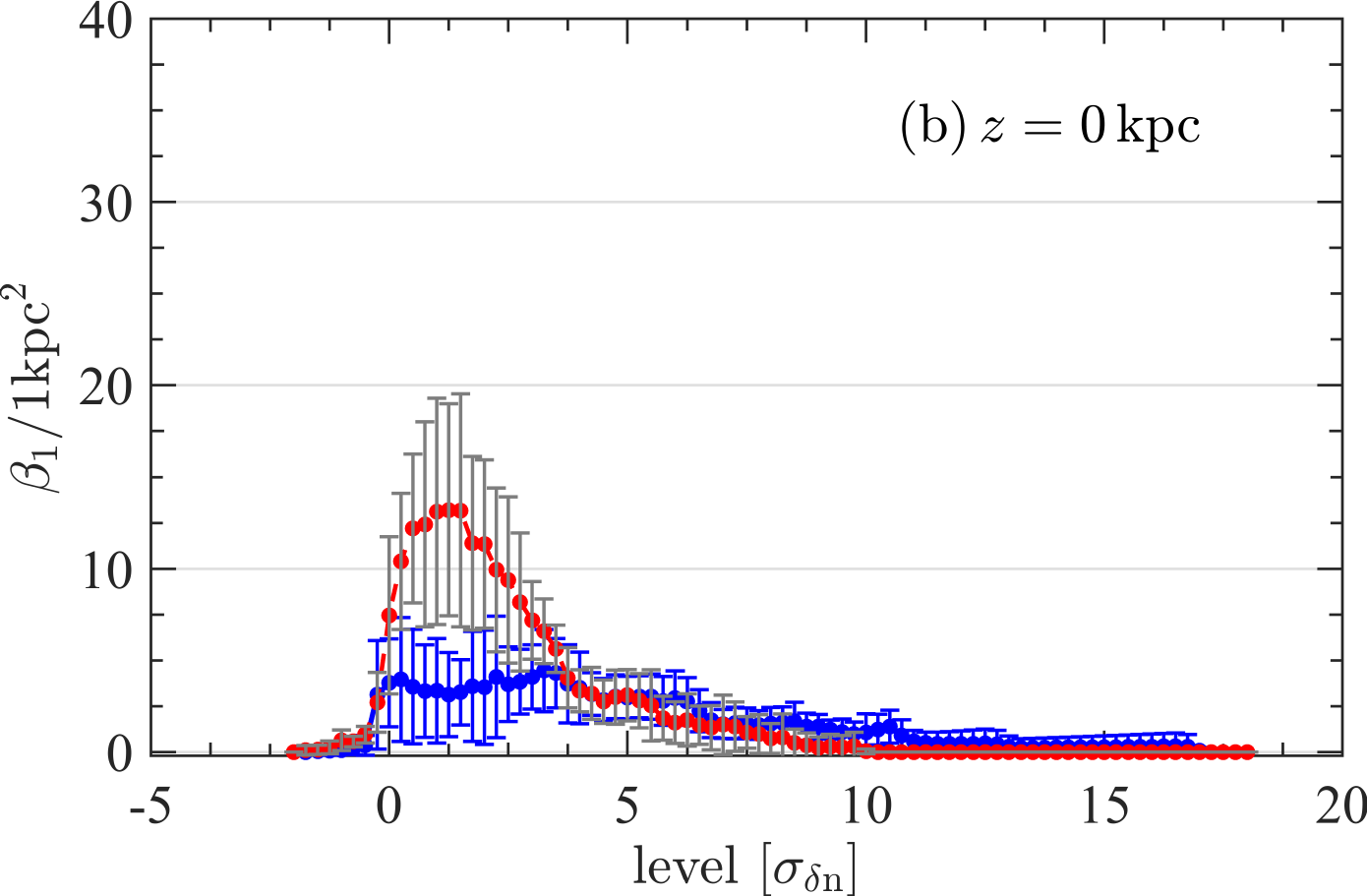}\\	\vspace{3mm}
	\includegraphics[width=0.4\textwidth]{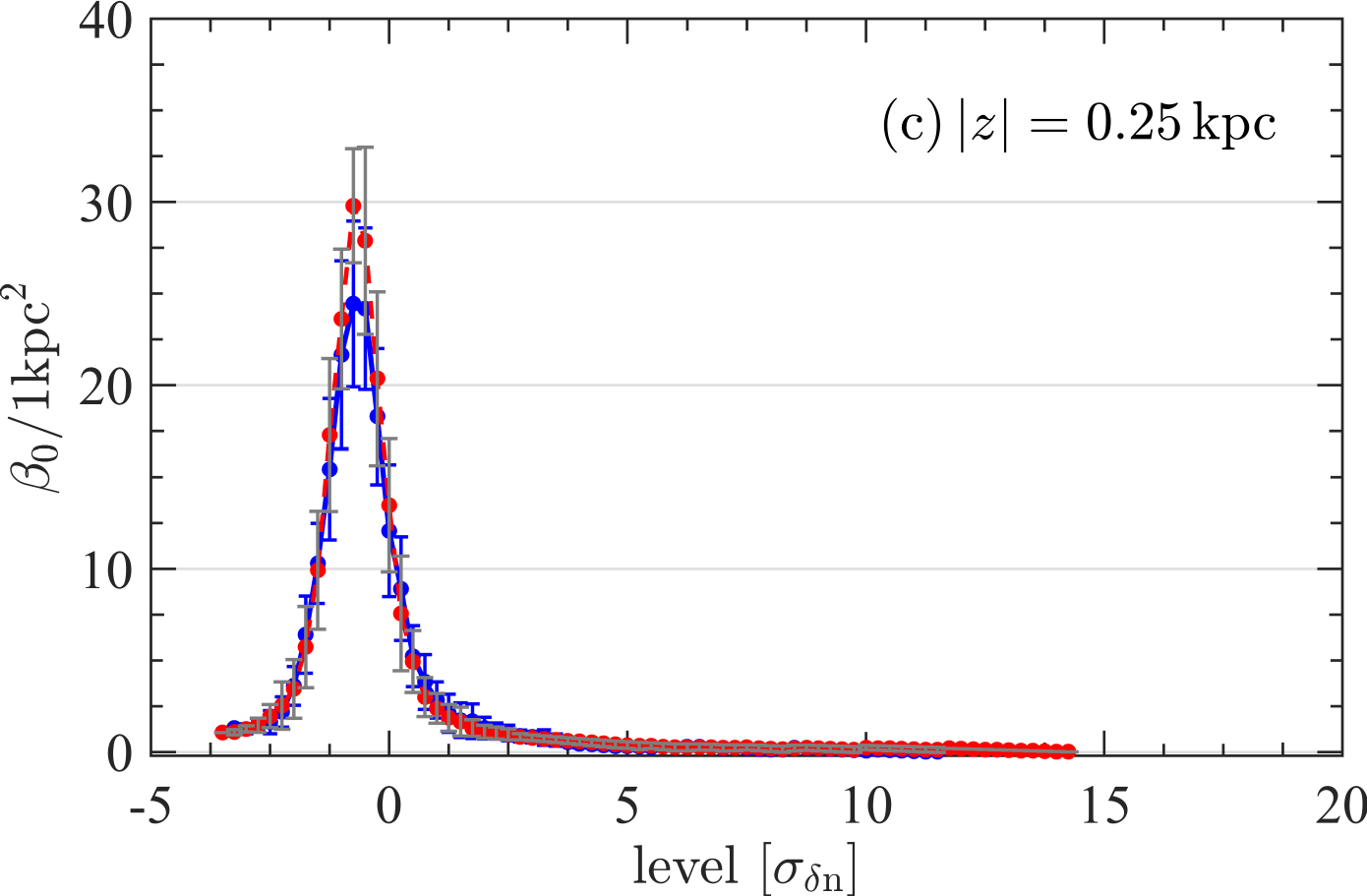}\hspace{3mm}
	\includegraphics[width=0.4\textwidth]{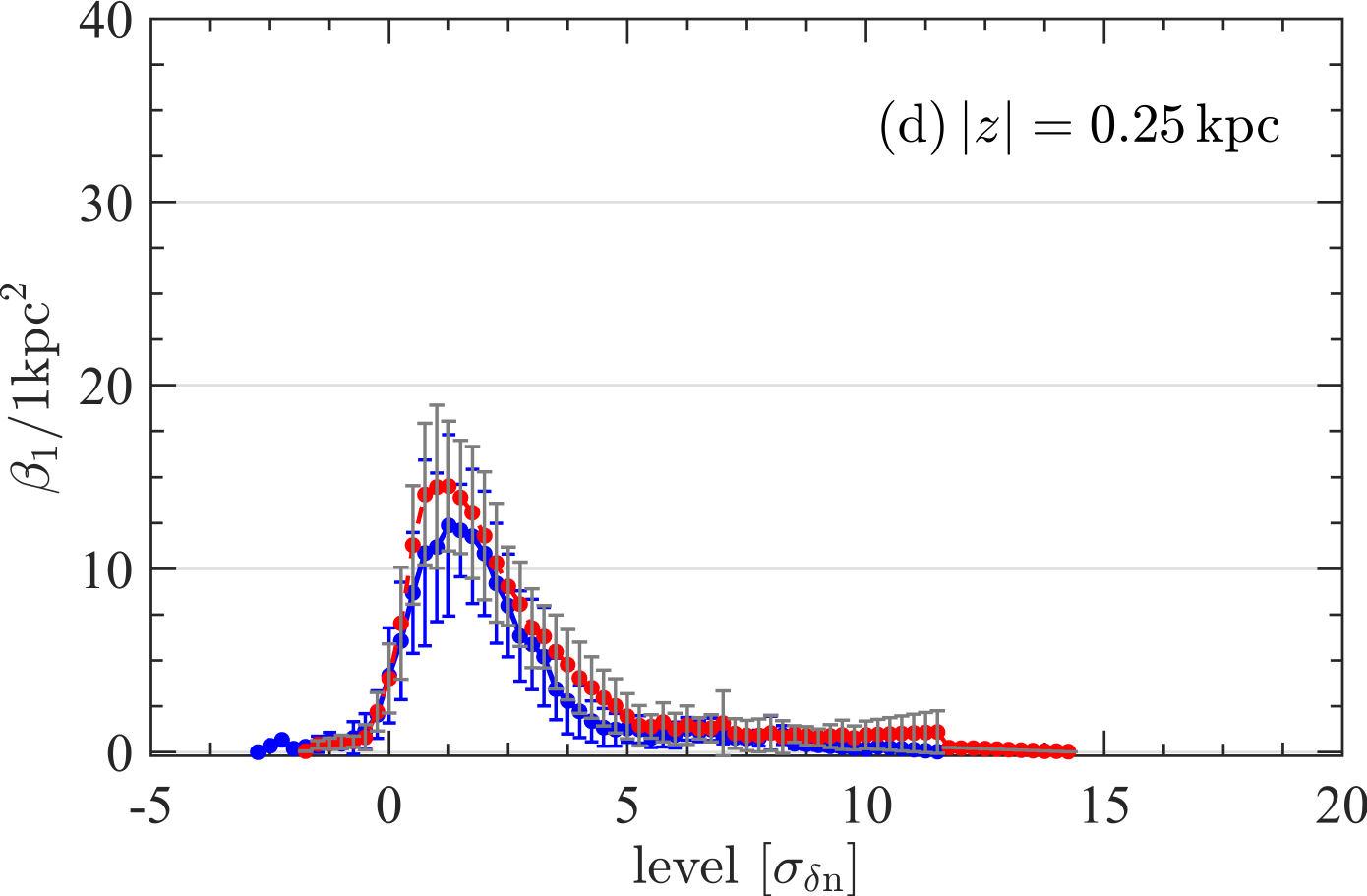}\\
	\includegraphics[width=0.4\textwidth]{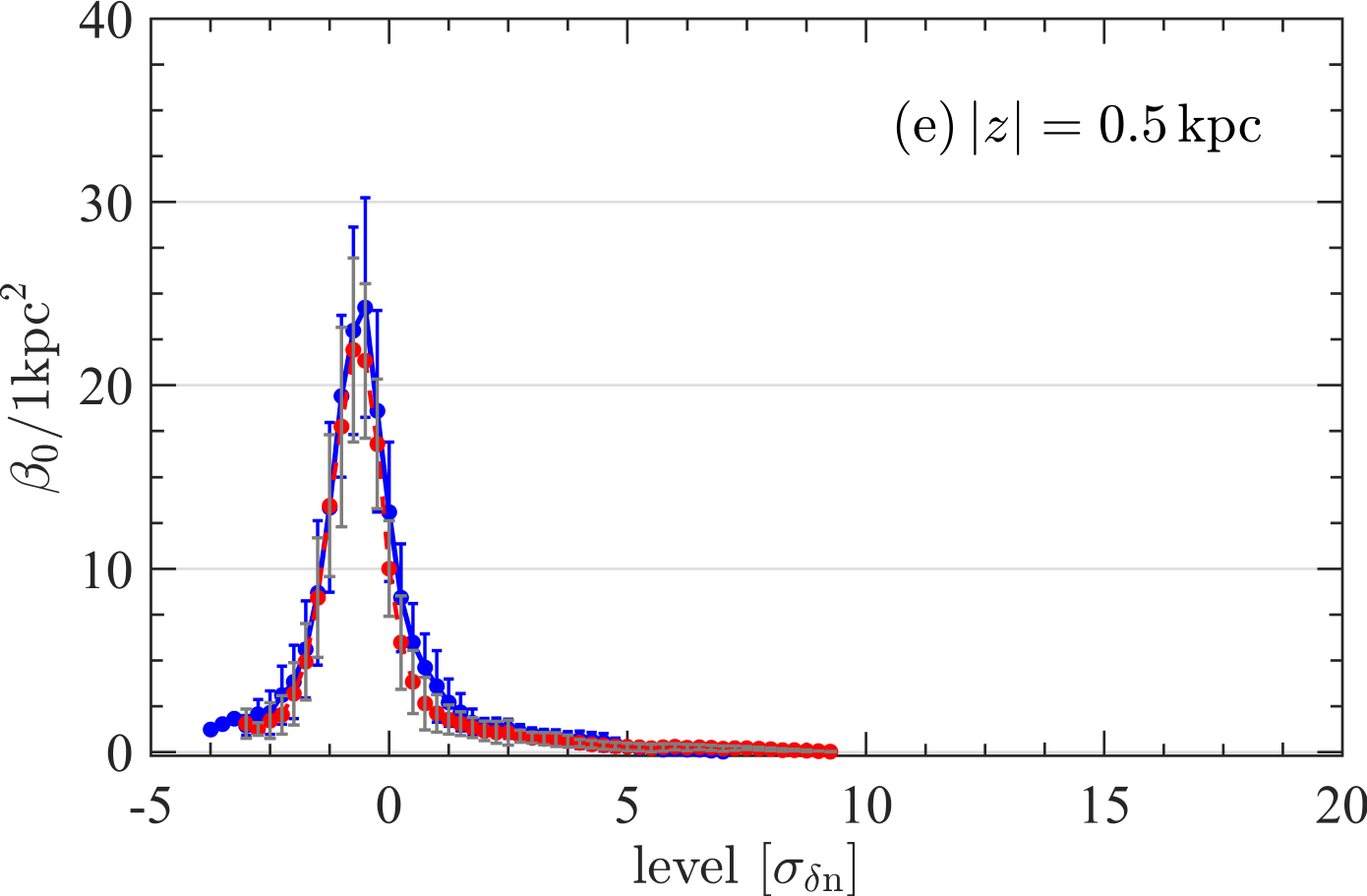}\hspace{3mm}
	\includegraphics[width=0.4\textwidth]{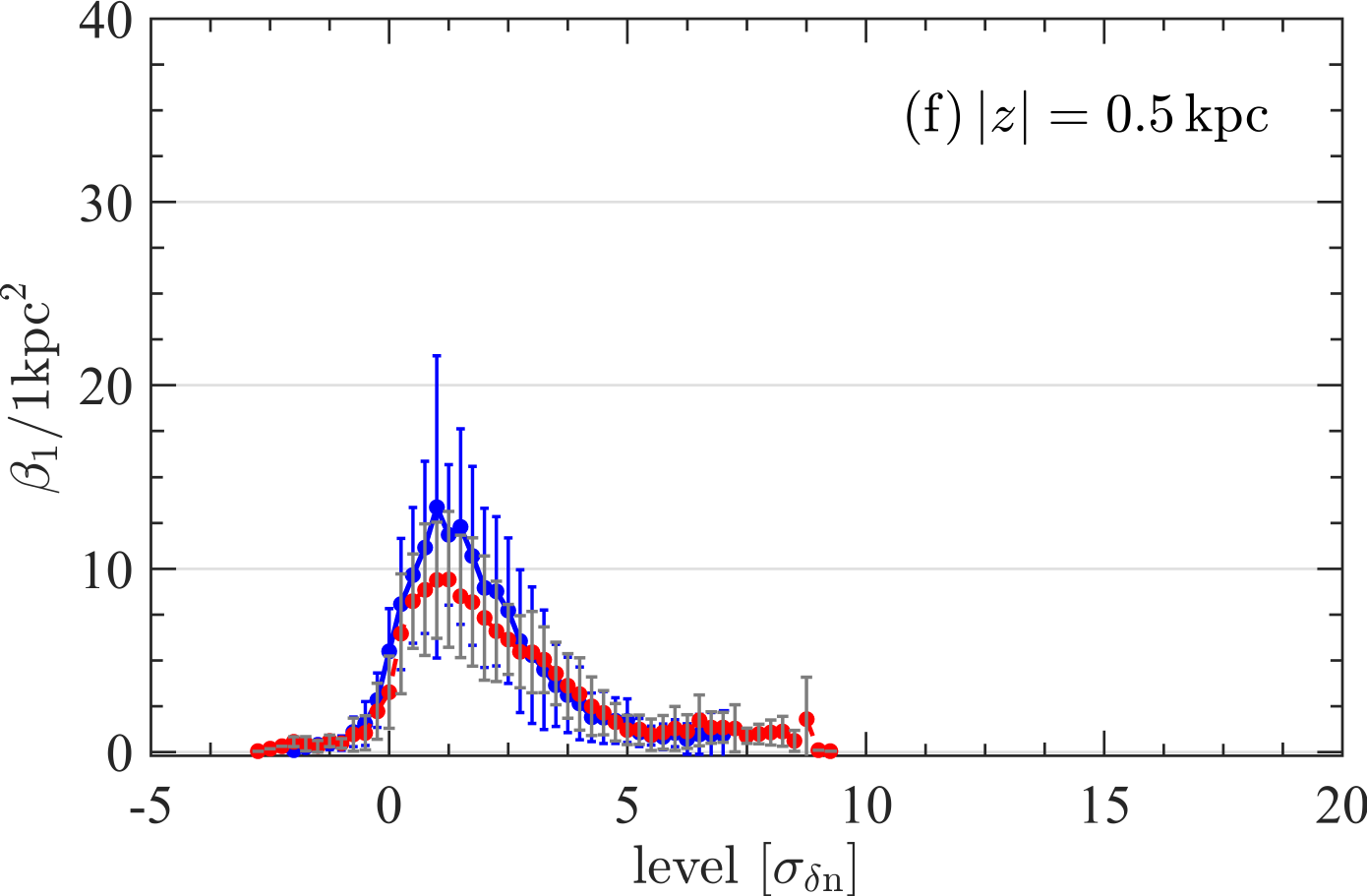}\\
	\caption{\label{curves} The dependence of the Betti numbers per unit area,
		$\beta_0$ in the left-hand column and $\beta_1$ on the right,
		of the gas density fluctuations at a level $h$ (specified in the units
		of $\sigma_{\delta n}$, the standard deviation of the density
		fluctuations). These are plotted at various distances from the mid-plane $z=0$ as given in
		the legends, in Stage~I (blue) and Stage~III (red). Error bars
		represent the standard deviation of the scatter of data points
		between the snapshots involved in the time averaging. The difference
		between Stages~I and III is significant only at $z=0$.}
\end{figure*}

It is then clear that the birth and death of components and holes is intrinsically related
to the nature of, and connections between, the stationary points of the random field
(its extrema and saddles) and to the values of $f$ at those points. Betti numbers contain
rich information about the random function.
Eventually, as $h$ approaches the absolute maximum of $f$ in the domain, 
only one, most persistent component remains (and no holes). Therefore 
$\beta_0=1$ and $\beta_1=0$ at levels $h$ exceeding the absolute maximum of $f$ in the domain.
The `lifetime', or persistence of a component or a hole is characterised by
the range of $h$ where it exists. Selecting only those features that are
more persistent, one distils a simplified (and therefore, better manageable)
topological portrait of the random field.

\subsection{Illustrative example}\label{IE}

The topological filtration of the scalar Gaussian random field
that is illustrated in Fig.~\ref{fig:filt} has already been 
referred to in the previous discussion. Here we discuss this filtration in more
detail, describing how the Betti numbers can be calculated.
Panel~(f) represents a realisation of a 2D Gaussian random field
$f(x, y)$ whose absolute minimum,  $f =-2.4$, occurs at $(x,y)=(50,35)$.
Thus, the first component C1 is born at $h = -2.4$ and there are no holes
at this level: we have $\beta_0=1$ and $\beta_1=0$ at $h=-2.4$.
As $h$ increases, more components are born. At $h = -1.25$, Panel~(a), there are
five components C1--C5 and no holes: $\beta_0=5$ and $\beta_1=0$.
At $h = -1$, Panel~(b), component C6
has appeared, whereas C1 and C5 have merged via a saddle point between them,
passed through at a smaller $h$; the surviving component is labelled
C1 whereas C5 has died as it was born later than C1.
There are no holes at $h=-1$: $\beta_0=5$ and $\beta_1=0$.
At a higher level $h=0.4$, shown in Panel~(c),
most of the components are merged into a big island, and there is one more small island in the
left bottom corner of the panel; moreover, one hole H1 appeared: $\beta_0=2$, $\beta_1=1$.	
We have indicated on this sketch where H3 will appear as the level is raised (it is not
yet defined as a hole because it is not yet bounded).
At a level $h=0.43$, Panel~(d),	the number of components is the same as 
in the previous panel, $\beta_0=2$. One more hole H2 has appeared, $\beta_1=2$.
At a higher level $h=1.2$, Panel~(e), all components have merged into a single island, $\beta_0=1$.
There are three holes labelled H3--H5, each around a local maximum of $f$, so $\beta_1=3$.
Note again that some holes (bordering on the field frame) have not yet completely 
formed and are not
taken into account as holes. There are two such cases in panel (e).
The holes H1 and H2, that surround the maxima lower than $f=1.2$,
have already died (contracted).

\subsection{Application to the ISM simulation}

These ideas can be applied directly to the numerical simulation. 
The Betti numbers were computed using the algorithm suggested in \citet{EH10}.
The description of the algorithm can be also found in \citet{MMMKM14}.
To represent the result of topological filtration of the field
one can plot the number of components $\beta_0$ and the number of holes $\beta_1$ as a
function of level $h$. Then, for a 2D field we obtain two curves, one for the components and
one for the holes, showing changes that take place in the field structure.
The Betti numbers obtained from scanning the gas density fluctuations
in planes $z=\const$ from smaller to larger values of $\delta n$ are
	shown in Fig.~\ref{curves} as functions of the isocontour level $h$.
	Apart from at the mid-plane, $z=0$, the difference between Stages~I and III
	is insignificant and can hardly be used for diagnostic purposes. As we
	discuss in Section~\ref{sec:scal}, a more detailed analysis is required
	to reveal the differences.

\subsection{Persistence diagrams}\label{PD}

Returning to our illustrative example, having specified the structure and connections of the
stationary points of $f$ in
terms of the Betti numbers, $\beta_0$ and $\beta_1$, at various levels $h$,
we turn to the persistence
of each structural feature. Let $u$ be a level at which a component or a hole
appears (is born) and $v$ a level where it disappears (dies).
Apart from the list of features at each level $h$, the filtration
described above results in the lists of $(u, v)$ pairs for both the components
and the holes. The lifetime of each topological feature is defined as
$v-u$. The longer the lifetime, the more
prominent (persistent) is the feature.
Since we are interested in the $h$-dependent connections
between the features, which do not change if $f$ is continuously deformed
(and thus represent topological invariants), the positions of the features
in the $(x,y)$-plane are of no interest.

The pairs $(u, v)$ can be represented as points in the $(u,v)$-plane to obtain
a \textit{persistence diagram}.
Then, filtration of a 2D field results in two persistence diagrams, one for
each Betti number.
As mentioned above, the total number of components and holes 
obtained from the filtration as the level $h$ changes from the lowest minimum to the highest 
maximum of the field,
are referred to as $\cb_0$ and $\cb_1$ in this paper (see Table~\ref{tab:notations}).

\begin{figure*}
	\centering
	\includegraphics[width=0.3\textwidth]{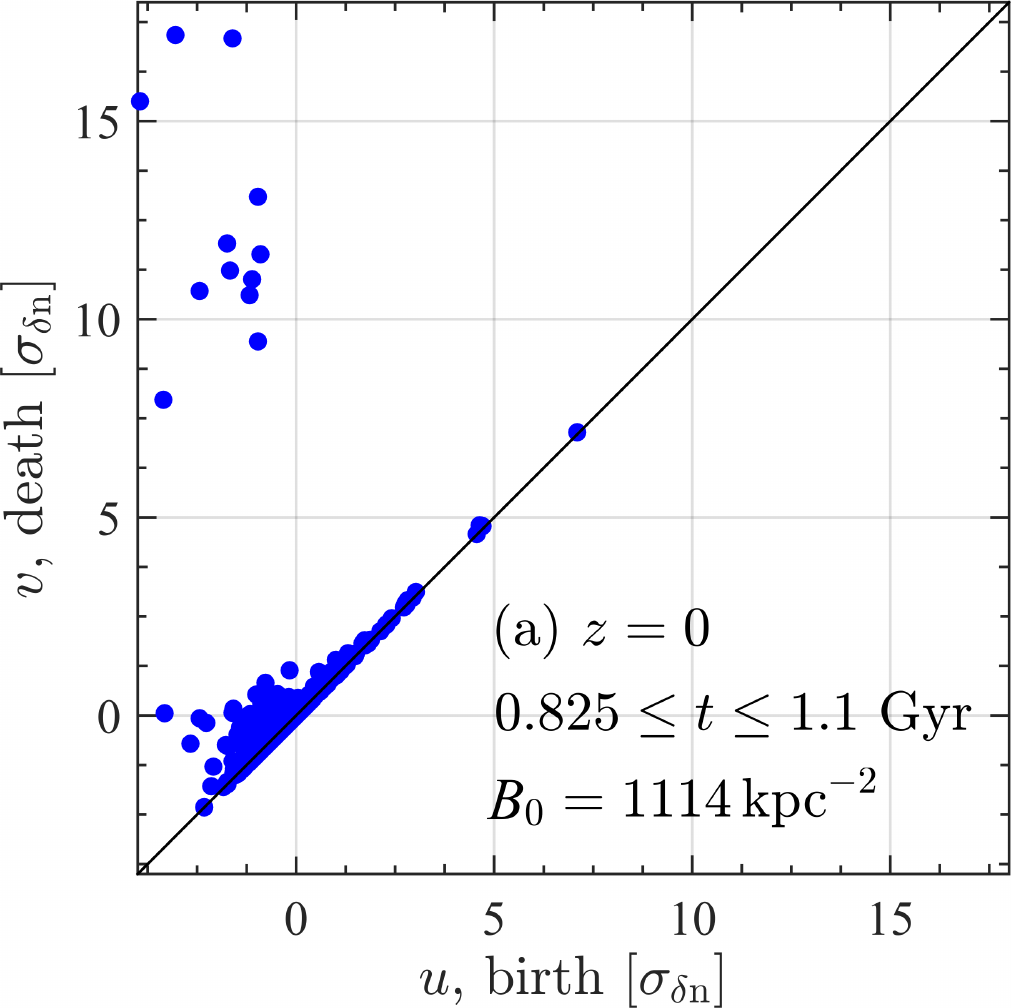} \hspace{3mm}
	\includegraphics[width=0.3\textwidth]{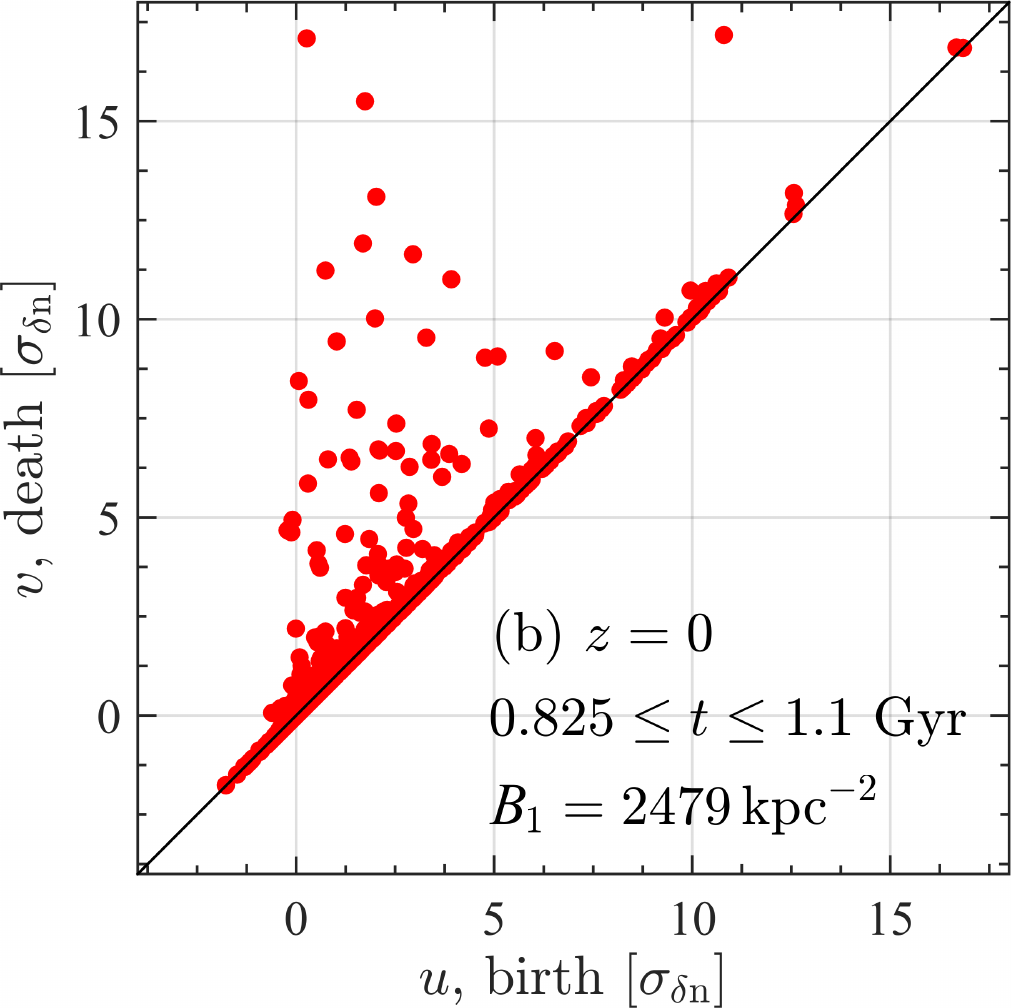}\\	\vspace{3mm}
	\includegraphics[width=0.3\textwidth]{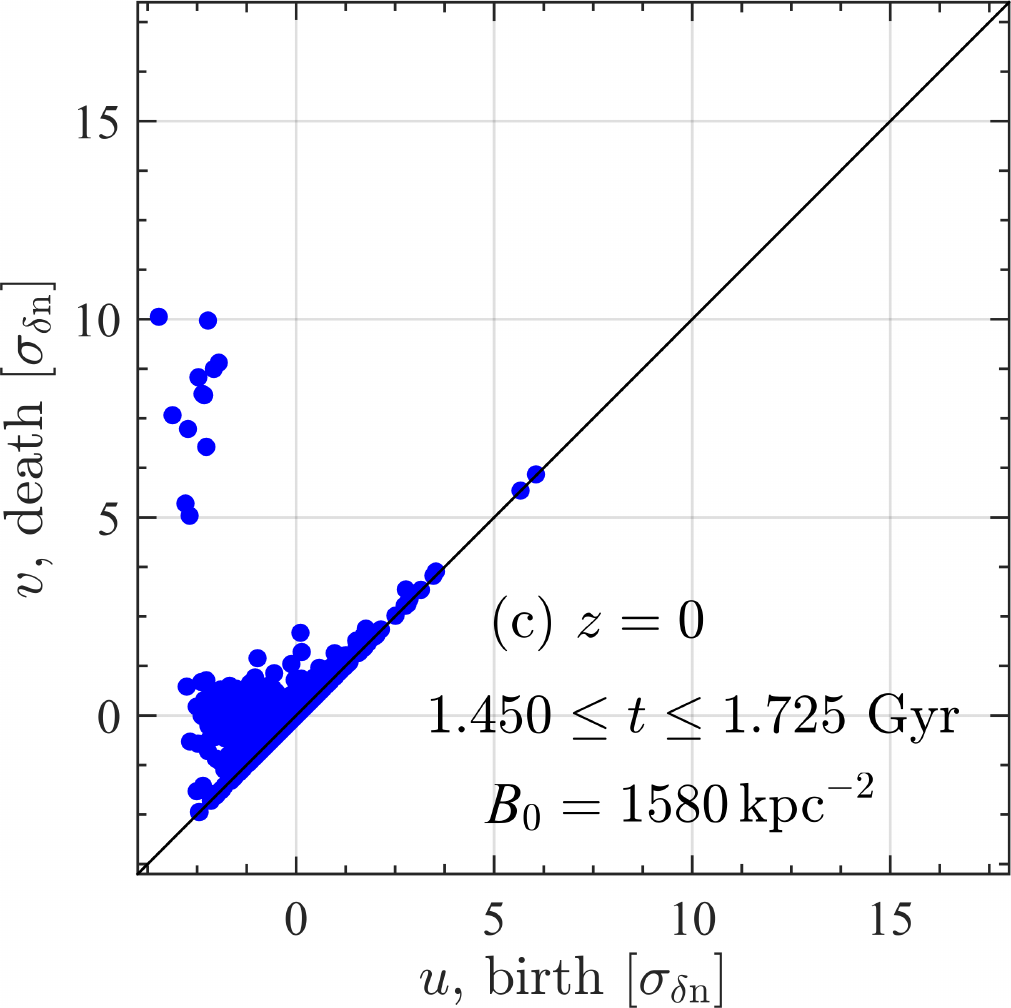}\hspace{3mm}
	\includegraphics[width=0.3\textwidth]{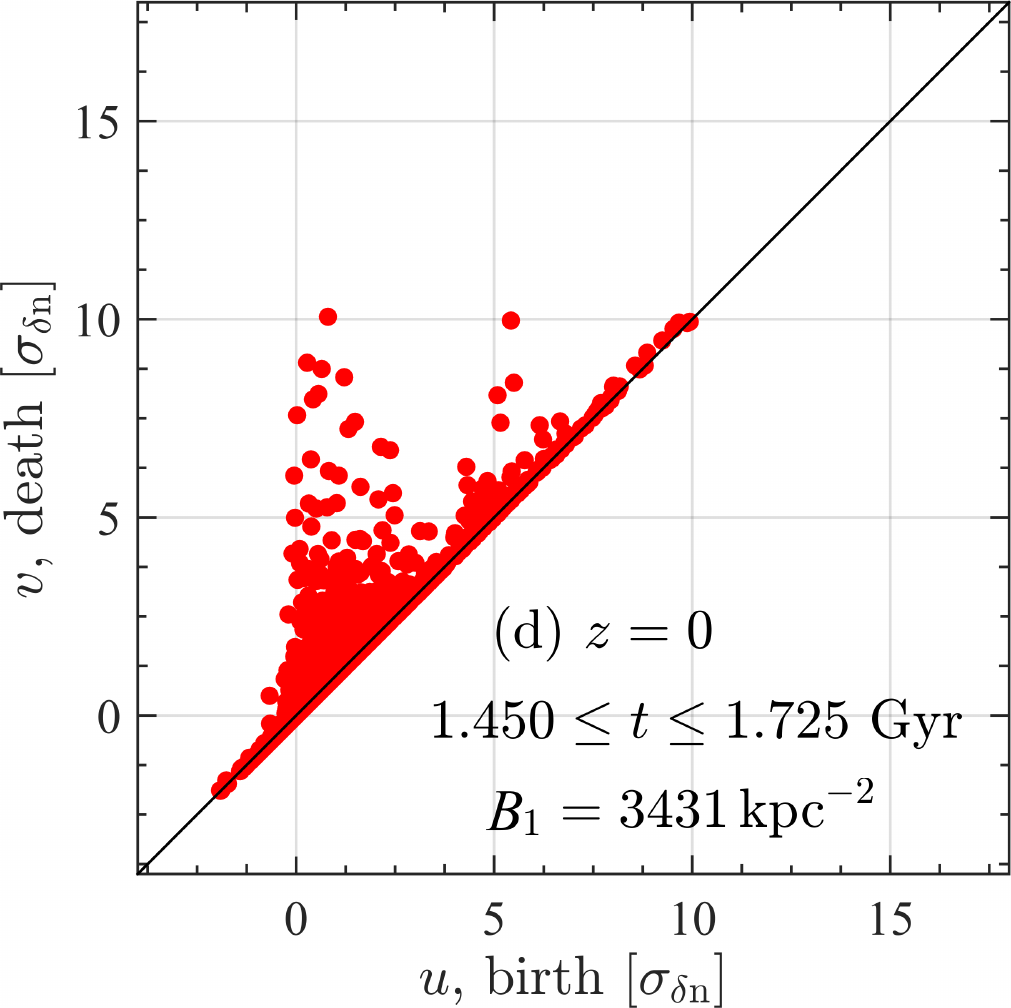}\\
	\caption{\label{fig:cpd}
	  Persistence diagrams of the gas density fluctuations from all snapshots at $z=0$ 
		for the components (left-hand panels, blue) and holes (right-hand panels, red) 
		in \textbf{(a)--(b)}~Stage~I and \textbf{(c)--(d)}~Stage~III. 
		Twelve persistence diagrams for each of Stage~I and Stage~III have been
		combined for each Betti number.
	 }
	\end{figure*}

Figure~\ref{fig:cpd} shows persistence diagrams for
the gas density fluctuations $\delta n$ in all snapshots from Stage~I
(upper row) and Stage~III (lower row) at $z=0$. Points at the largest
distance from the diagonal represent the most stable, persistent features
that have larger span in $\delta n$.
The diagrams for both $\cb_0$ and $\cb_1$ are more compact in both
$u$ and $v$ in Stage~III reflecting a narrower range
of the gas density fluctuations, i.e. a more homogeneous medium.
In both stages, there is a relatively small number of  very persistent components
(points with large $v$ in the $\cb_0$ diagram), whereas the persistence
of holes ($\cb_1$) is less extreme. On the other hand, the number of
structures, either components ($\cb_0$) or holes ($\cb_1$),
is larger by about 40 per cent in Stage~III. This suggests a more
structured gas distribution.
Points at the bottom left corners of each diagram correspond to
weak fluctuations at the lowest values of $\delta n$. Most components
have $u,v<0$, indicative of numerous local minima in the gas density
distribution since the filtration proceeds from smaller to larger
values of $\delta n$. On the other hand, holes ($\cb_1$) occur
mostly at $u,v>0$, representing cavities in a relatively dense gas.

\subsection{Normalisation of Betti numbers}\label{MBN}
For a statistically stationary random field in 2D (or 3D), the magnitudes of the
Betti numbers are proportional to the area (or volume) sampled,
and they are therefore often presented per unit area (or volume).
\citet{Park2013}, \citet{Pranav2015}, \citet{Pranav2017}, \citet{Sousbie2011part1} and \citet{Sousbie2011part2}
discuss Betti numbers per cubic pixel of Mpc$^3$. With such a
normalisation, the result would change if the random field was rescaled
spatially, $\vect{x}\to a\vect{x}$ with $a=\const$,
despite the fact that its statistical properties remained
essentially unchanged. Of course, the normalisation volume should change
correspondingly, but this would not happen when the normalisation volume
	is chosen arbitrarily or when the observational or computational region
	is fixed in size. To allow for such trivial differences in random
	fields, the Betti numbers should be normalised to a physically
	significant volume (or area in 2D),
	leading to a dimensionless quantity.

	For a random field, an obvious inherent spatial scale is
the correlation length $l_0$, so the normalisation volume or area can be
chosen as  $l_0^3$ or $l_0^2$ in 3D or 2D, respectively (in the isotropic case).
To provide a relevant context, we note that the number of
	local maxima or minima of a 2D Gaussian, isotropic random field per the
	correlation cell area $l_0^2$ follows from \citet{L-H57,L-H57a} as
	\begin{equation}\label{Nmax}
		N_0=\frac{l_0^2}{6\pi\sqrt3}
		\left.\frac{\dd^4C(l)/\dd l^4}{-\dd^2C(l)/\dd l^2}\right|_{l=0} =\frac{\sqrt3}{12}\approx0.14\, ,
	\end{equation}
	for the correlation function $C(l)=\exp[-\pi l^2/(4l_0^2)]$
	\citep[for 	a discussion in terms of the autocorrelation function, see
	section 30 of][]{Sv66}. It is also worth noting that
	the probabilistic properties of the local extrema of a Gaussian random field
	are controlled by the form of the autocorrelation function at small
	values of $l$, so the Taylor micro-scale $\lambda$ (see Eq.~\ref{TM}) is another natural normalisation lengthscale
	to consider. We tried using $\lambda^2$ for normalisation but found it less discriminating than $l_0^2$, probably due to the difficulties in its accurate determination, as mentioned in Sect.~\ref{CL}.

\begin{figure}
	\centering
	\includegraphics[width=0.4\columnwidth]{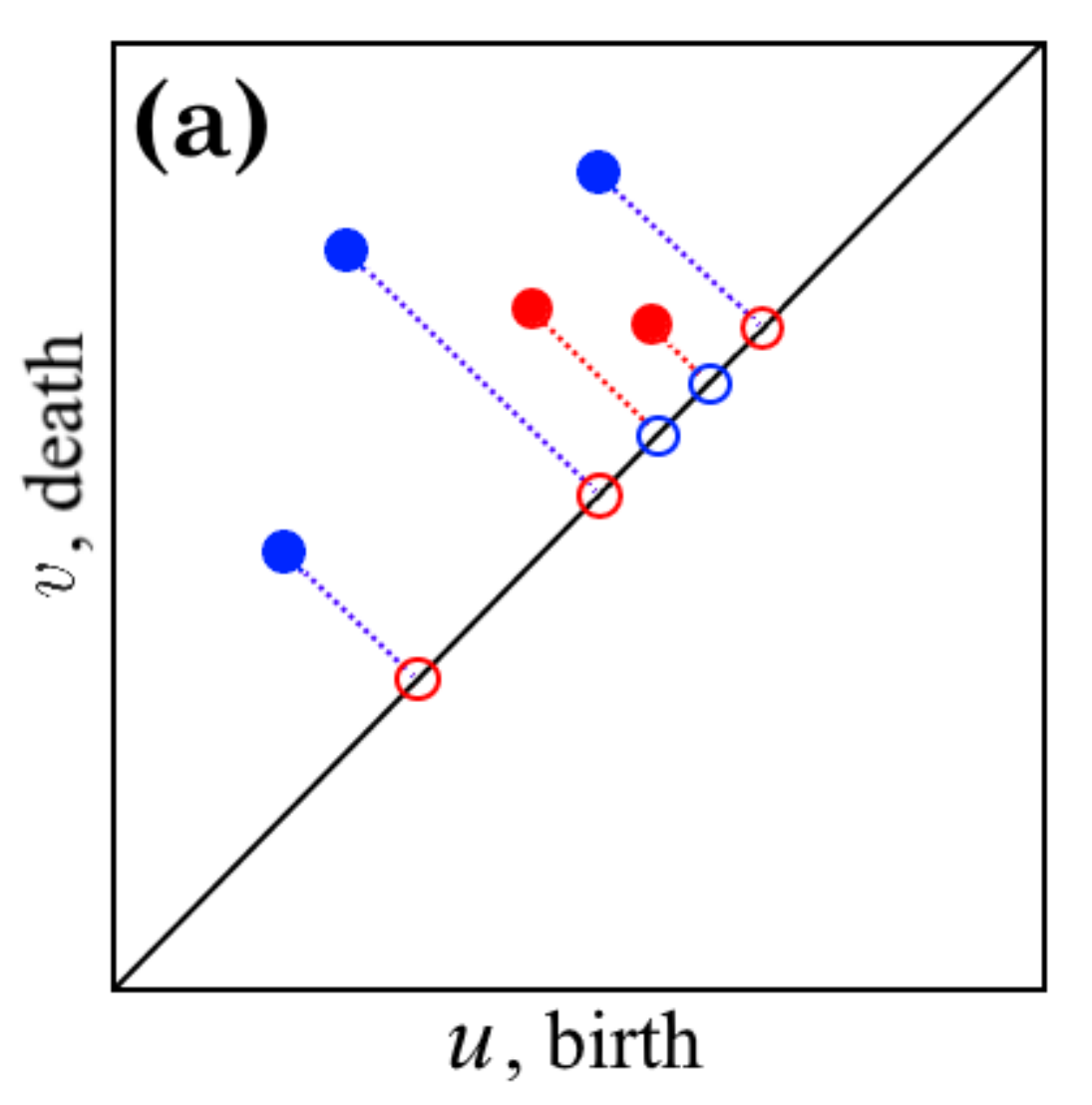}
	\hspace{1mm}
	\includegraphics[width=0.4\columnwidth]{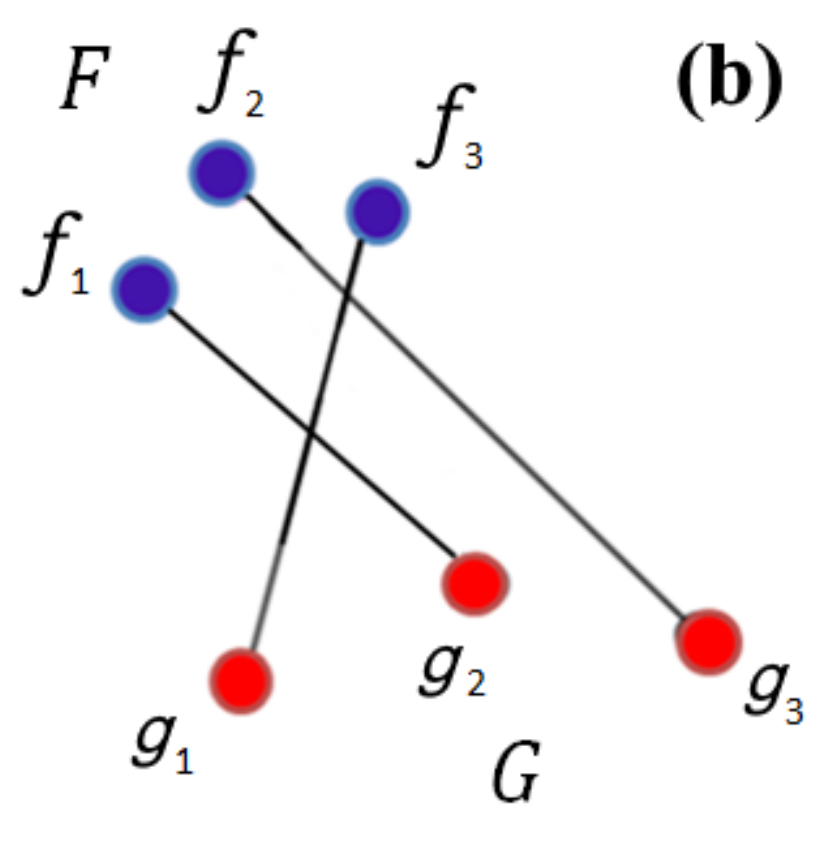}
	\caption{
		Illustration of the definition of the bottleneck distance, \textbf{(a)}: 
		adding points on a diagonal of the persistence diagram, \textbf{(b)}: 
		one of the possible matchings between two sets of	points.
		}
	\label{fig:match}
\end{figure}

\subsection{The bottleneck distance between persistence diagrams}\label{sec:bd}

Topological filtration produces persistence diagrams for the
Betti numbers, i.e. sets of points in the $(u,v)$-plane
shown in Fig.~\ref{fig:cpd}.
Comparison of random fields then requires a quantification of differences
between such clouds of points,
i.e. the introduction of a measure in the space of persistence diagrams.
One such measure is the bottleneck distance \mbox{\citep{EH10,Ed2014}}
often used to compare persistence diagrams in the exploration and development of
various TDA techniques.
We first present a formal definition of the bottleneck distance and
then explain it with an example.

Consider two persistence diagrams
$\mathbb{F}$ and $\mathbb{G}$, i.e. two sets of points in the $(u,v)$-plane.
The bottleneck distance between two diagrams is
$$
D(\mathbb{F},\mathbb{G})= \inf_{\mu:\,\mathbb{F}\to\mathbb{G}}\sup_{\mathbfit{x}\in\mathbb{F}}\left\| \mathbfit{x}-\mu(\mathbfit{x})\right\|_\infty,
$$
where $\mu:\,\mathbb{F}\to\mathbb{G}$ is a bijection, the infimum is over all bijections from $\mathbb{F}$ to $\mathbb{G}$, and
a distance between points $\mathbfit{x}=(u_\mathbb{F}, v_\mathbb{F})$ and $\mathbfit{y}=(u_\mathbb{G}, v_\mathbb{G})$ is measured as 
$\left\| \mathbfit{x}-\mathbfit{y}\right\|_\infty = \max\{ |u_\mathbb{F}-u_\mathbb{G}|, |v_\mathbb{F}-v_\mathbb{G}|\}$.

To clarify the definition of the bottleneck distance, consider how it is evaluated in practice
for two persistence diagrams, $\mathbb{F}$ and $\mathbb{G}$, i.e. two sets of points 
in the $(u,v)$-plane shown in 
solid blue and red in Fig.~\ref{fig:match}.
The bottleneck distance between them can be measured only if they contain the same number of points.
Since two persistence diagrams can have different number of points, the diagrams are augmented as follows.
We find the orthogonal projections of all points of $\mathbb{F}$ on the diagonal 
$u=v$ (red circles) and add them to the diagram $\mathbb{G}$ to obtain the set $G$ shown 
in Fig.~\ref{fig:match}a. Then we add the projections of all points of $\mathbb{G}$ (blue circles)
to the diagram $\mathbb{F}$ to obtain the set $F$. Now $F$ and $G$ have the same size.

Consider all possible one-to-one matchings (or correspondences, bijections) 
between the sets $F$ and $G$. Suppose that there are $k$ such matchings. 
For example, for the configuration of points in Fig.~\ref{fig:match}b, 
there are 6 such matchings ($k = 6$):
\[
\begin{tabular}{@{}ccc@{}}
	($\mathbfit{f}_1$ $\mathbfit{g}_1$), & ($\mathbfit{f}_1$ 
	$\mathbfit{g}_1$), & ($\mathbfit{f}_1$ $\mathbfit{g}_3$),\\
	($\mathbfit{f}_2$ $\mathbfit{g}_2$), & ($\mathbfit{f}_2$ 
	$\mathbfit{g}_3$), & ($\mathbfit{f}_2$ $\mathbfit{g}_2$),\\
	($\mathbfit{f}_3$ $\mathbfit{g}_3$), & ($\mathbfit{f}_3$ 
	$\mathbfit{g}_2$), & ($\mathbfit{f}_3$ $\mathbfit{g}_1$),\vspace{4mm}\\
	($\mathbfit{f}_1$ $\mathbfit{g}_3$), & ($\mathbfit{f}_1$ 
	$\mathbfit{g}_2$), & ($\mathbfit{f}_1$ $\mathbfit{g}_2$),\\
	($\mathbfit{f}_2$ $\mathbfit{g}_1$), & ($\mathbfit{f}_2$ 
	$\mathbfit{g}_1$), & ($\mathbfit{f}_2$ $\mathbfit{g}_3$),\\
	($\mathbfit{f}_1$ $\mathbfit{g}_2$), & ($\mathbfit{f}_3$ 
	$\mathbfit{g}_3$), & ($\mathbfit{f}_3$ $\mathbfit{g}_1$).\\
\end{tabular}
\]
Each pair of points (denoted $\mathbfit{f}_j$ and $\mathbfit{g}_j$) in a given matching has a cost
defined as a geometric distance between them in the $(u,v)$-plane, denoted
$|\mathbfit{f}_j,\, \mathbfit{g}_j| = 	[(u_{fj}-u_{gj})^2+(v_{fj}-v_{gj})^2]^{1/2}$.
If both points are at the diagonal, the cost of this pair is zero. 
The largest cost in a given matching is then introduced as
\[
d_k = \max_{j}{|\mathbfit{f}_j,\, \mathbfit{g}_j|},
\qquad \mathbfit{f}_j \in F, \qquad \mathbfit{g}_j \in G,
\]
where $k$ enumerates individual matchings of the sets $F$ and $G$.
Having quantified all possible matchings between $F$ and $G$,
the largest cost in each matching is recorded and its minimum value among all matchings is called
the bottleneck distance between $F$ and $G$:
\[
D(F, G) = \min_k d_k.
\]

\noindent Details of algorithms for computation of the bottleneck distance 
can be found in \citet{Kerber2016}. We used the software package \textsc{dipha} at
https://github.com/DIPHA/dipha.

\begin{figure*}
	\centering
	\includegraphics[width=0.4\textwidth]{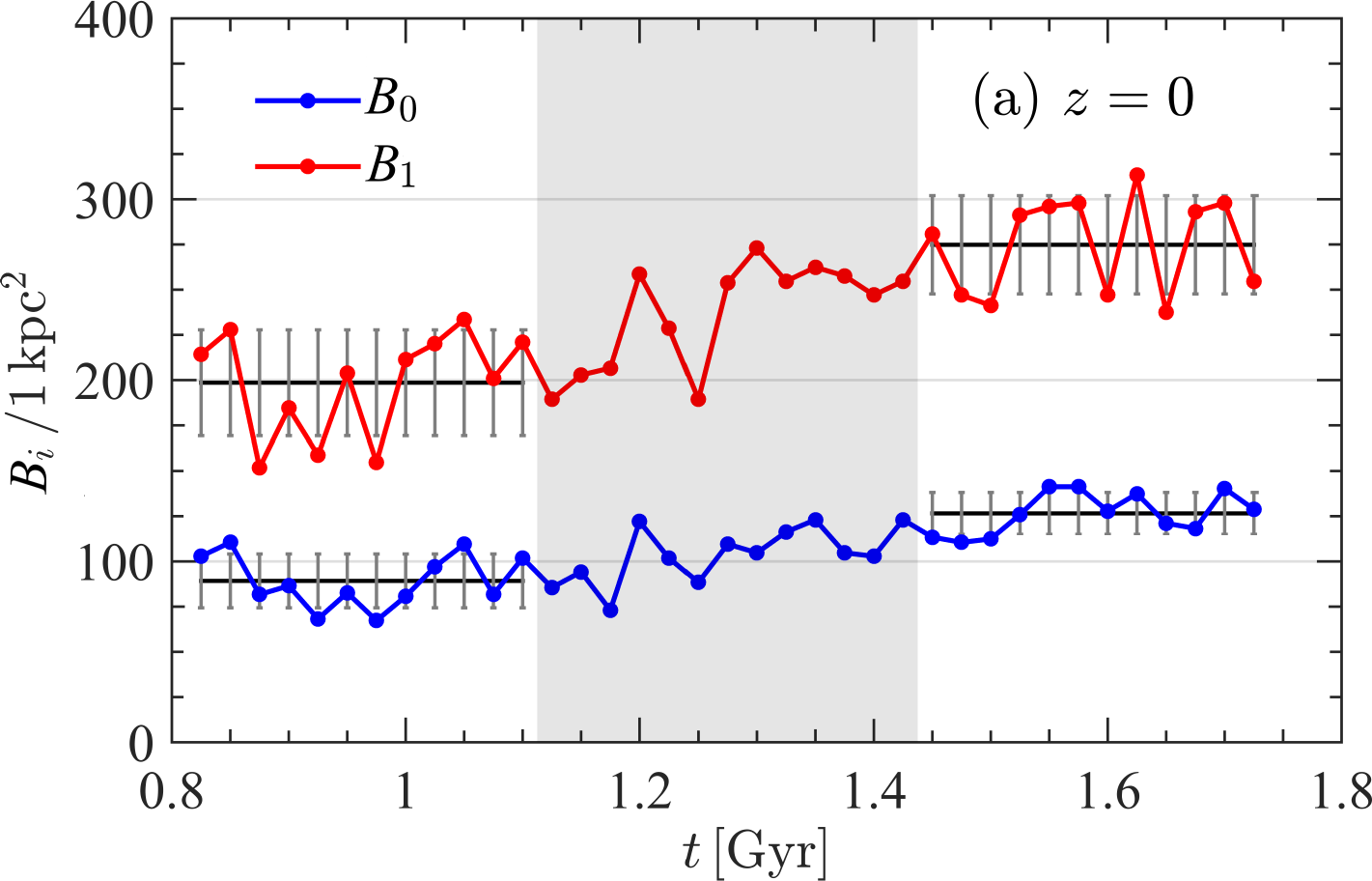}\\
	\vspace{2mm}
	\includegraphics[width=0.4\textwidth]{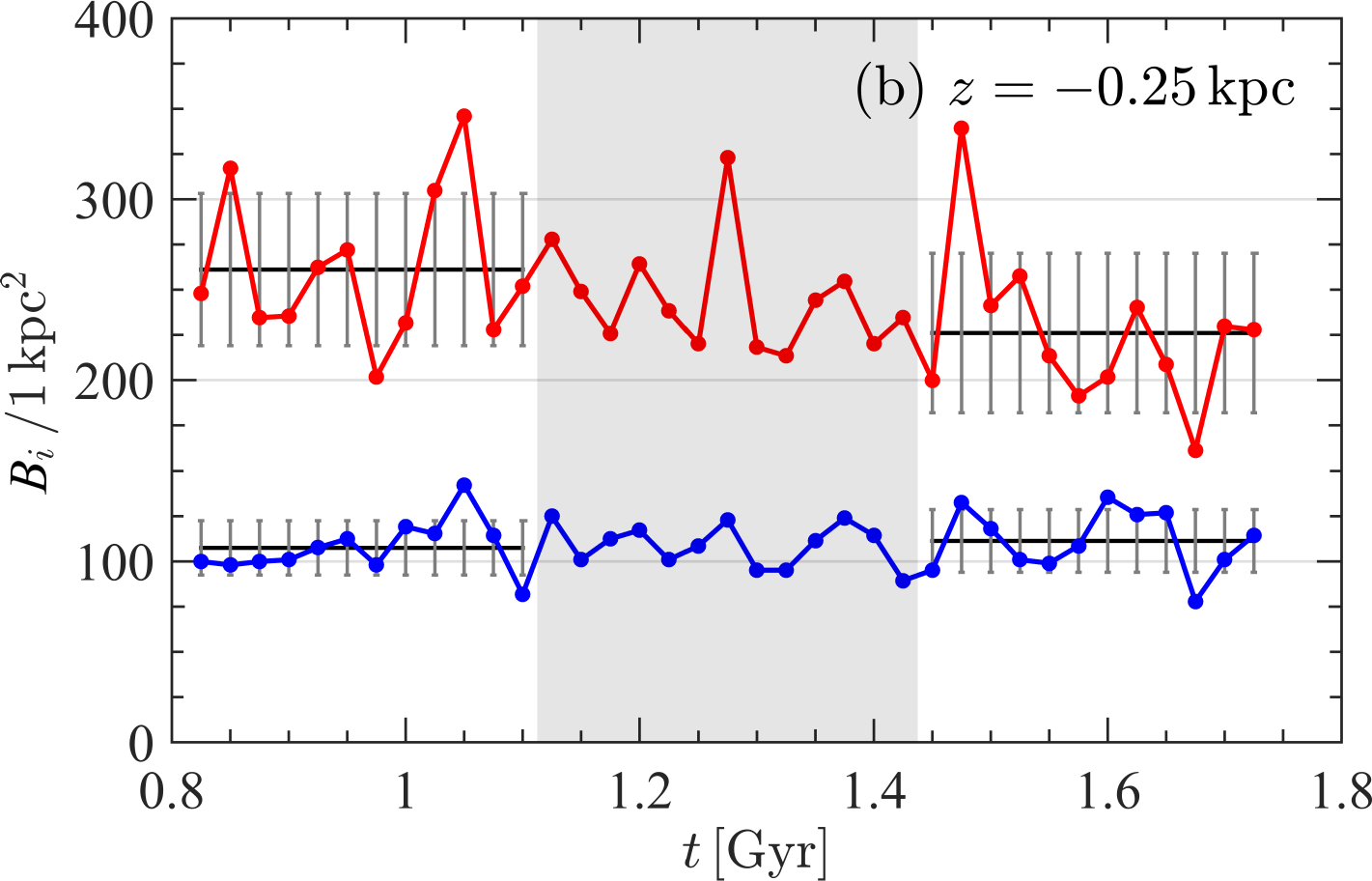} \hspace{4mm}
	\includegraphics[width=0.4\textwidth]{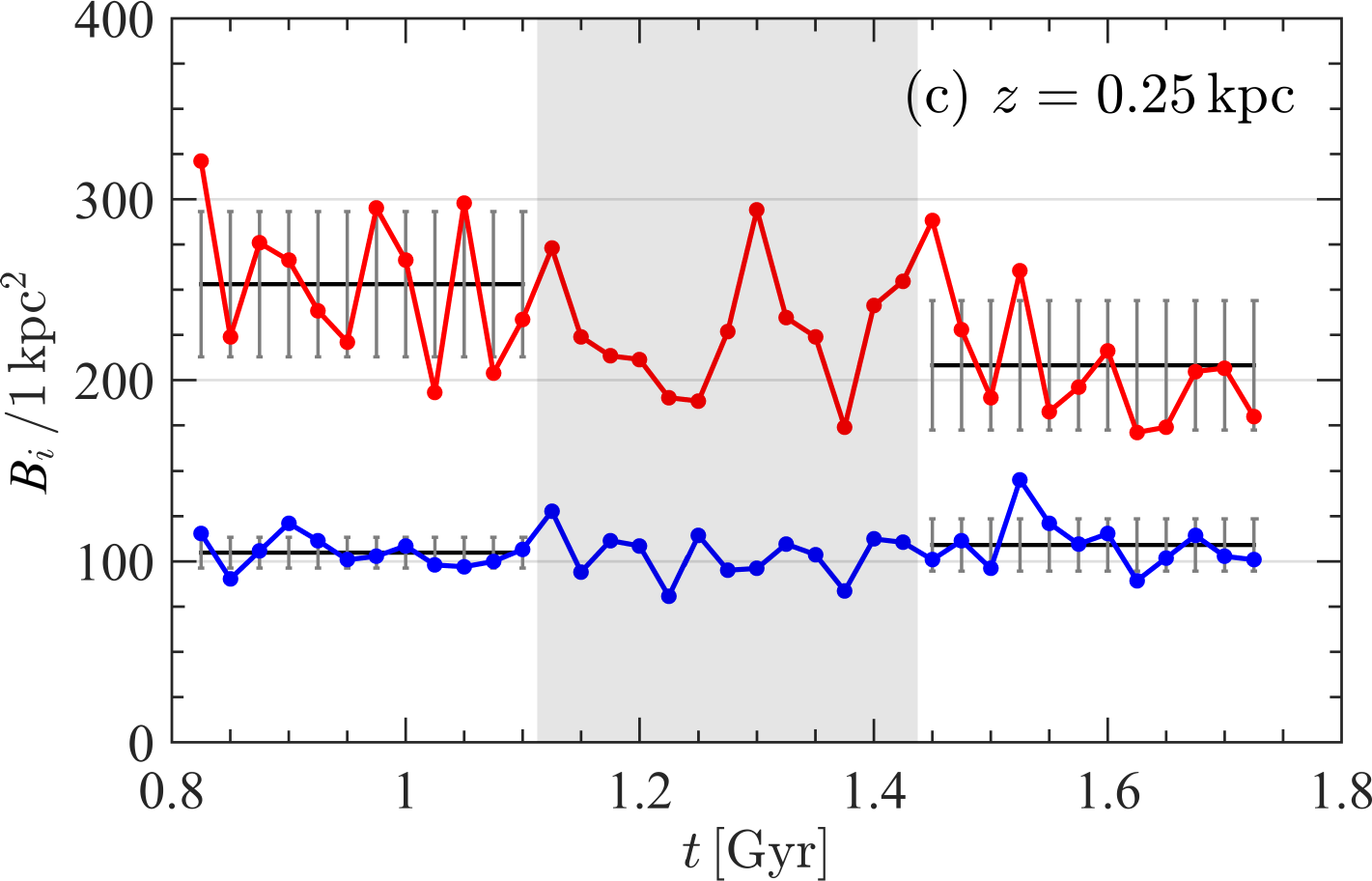}\\
	\vspace{2mm}
	\hspace{1mm}
	\includegraphics[width=0.4\textwidth]{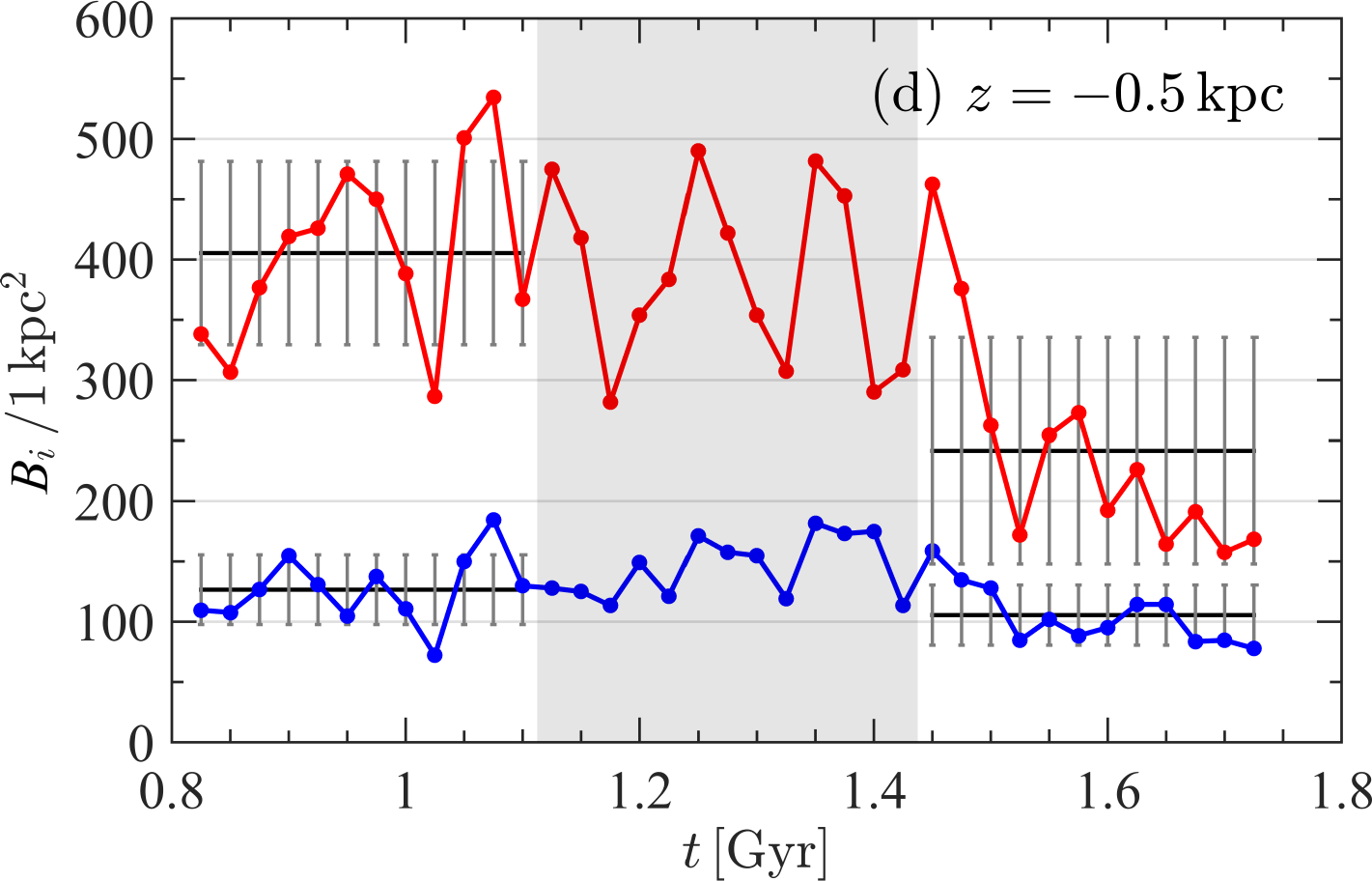} \hspace{6mm}
	\includegraphics[width=0.4\textwidth]{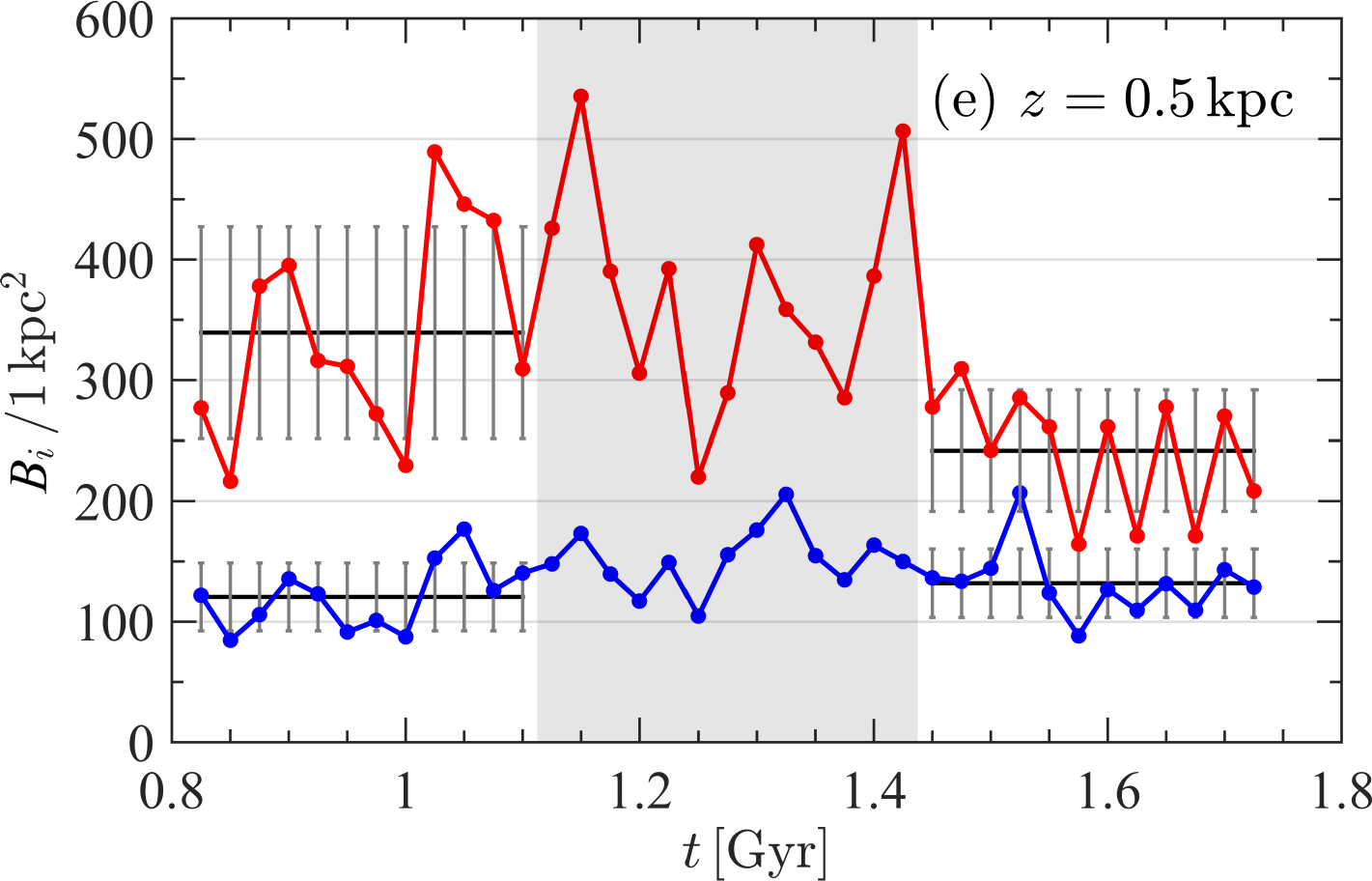}
	\caption{\label{fig:bettiViaTime}Evolution of the 
		Betti numbers $\cb_0$ (blue) and $\cb_1$ (red) per $1\kpc^2$ at various 
		distances $z$ from the mid-plane.
		The shaded area corresponds to the transition between the
		kinematic dynamo regime of Stage~I (with negligible magnetic field)
		to the magnetically steady state of Stage~III (where the magnetic field contribution is dynamically significant). Error bars represent the
		standard deviation of individual values of $\cb_0$ and $\cb_1$ 
		within a stage.}
\end{figure*}

\section{Betti numbers of the gas density fluctuations}\label{sec:scal}
To focus on topological properties of the fluctuations in the gas density distribution,
it is useful to \textit{standardise} them, i.e. to reduce them
to zero mean and renormalise to the unit standard deviation. In the rest of the
text, we therefore use the following standardised density fluctuations,
\begin{equation}\label{tn}
\delta\tilde{n}=\frac{\delta n-\langle n\rangle}{\sigma_{\delta n}}\,,
\end{equation}
where $\sigma_{\delta n}$ is the standard deviation of $\delta n$.

Figure~\ref{fig:bettiViaTime} and Table~\ref{tab:bettiPerKpcT} show the
Betti numbers $\cb_0$ and $\cb_1$ 
per $1\kpc^2$ as a function of time.
Despite significant asymmetry of the gas distribution around $z=0$ in
individual snapshots, the Betti numbers are rather symmetric, and thus
capture the overall, time-averaged symmetry of the system with respect to
$z=0$. It also clear that the Betti numbers reflect a change in the gas
distribution apparently associated with the magnetic field. Since $\cb_1>\cb_0$
in all cases, the gas structure is cellular (`spongy'), with numerous holes,
apparently 
supernova remnants filled with dilute gas.

\begin{table}
	\centering
	\caption{\label{tab:bettiPerKpcT}The 
	Betti numbers $\cb_0$ and $\cb_1$ (per $1\kpc^2$)
	of the gas density fluctuations averaged in Stages~I and III at various
	values of $z$, the distance to the mid-plane of the computational domain.
	The standard deviations have been calculated between individual snapshots.}
	\begin{tabular}{cccccc}
		\hline
		& \multicolumn{2}{c}{$\cb_0\pm \sigma_{\cb_0}$ $[\text{kpc}^{-2}]$}  &&
		\multicolumn{2}{c}{$\cb_1\pm \sigma_{\cb_1}$ $[\text{kpc}^{-2}]$}\\
		\cline{2-3}\cline{5-6}
		$z\ [\text{pc}]$ 	& Stage I 	& Stage III 	&& Stage I 		& Stage III \\
		\hline
$\phantom{-}0$&$\phantom{0}90\pm15$ &$127\pm12$ &&$199\pm30$ 	&$275\pm27$\\
		$-250$ 		&$108\pm15$ &$111\pm17$ 	&&$261\pm42$ 	&$226\pm44$\\
$\phantom{-}250$&$105\pm9\phantom{0}$&$109\pm15$&&$253\pm40$ 	&$208\pm36$\\
		$-500$  	&$127\pm29$ &$106\pm25$ 	&&$405\pm76$ 	&$242\pm94$\\
$\phantom{-}500$  	&$121\pm28$ &$132\pm28$ 	&&$339\pm88$ 	&$242\pm50$\\
		\hline
	\end{tabular}
\end{table}

\begin{figure*}
  \centering
  \includegraphics[width=0.4\textwidth]{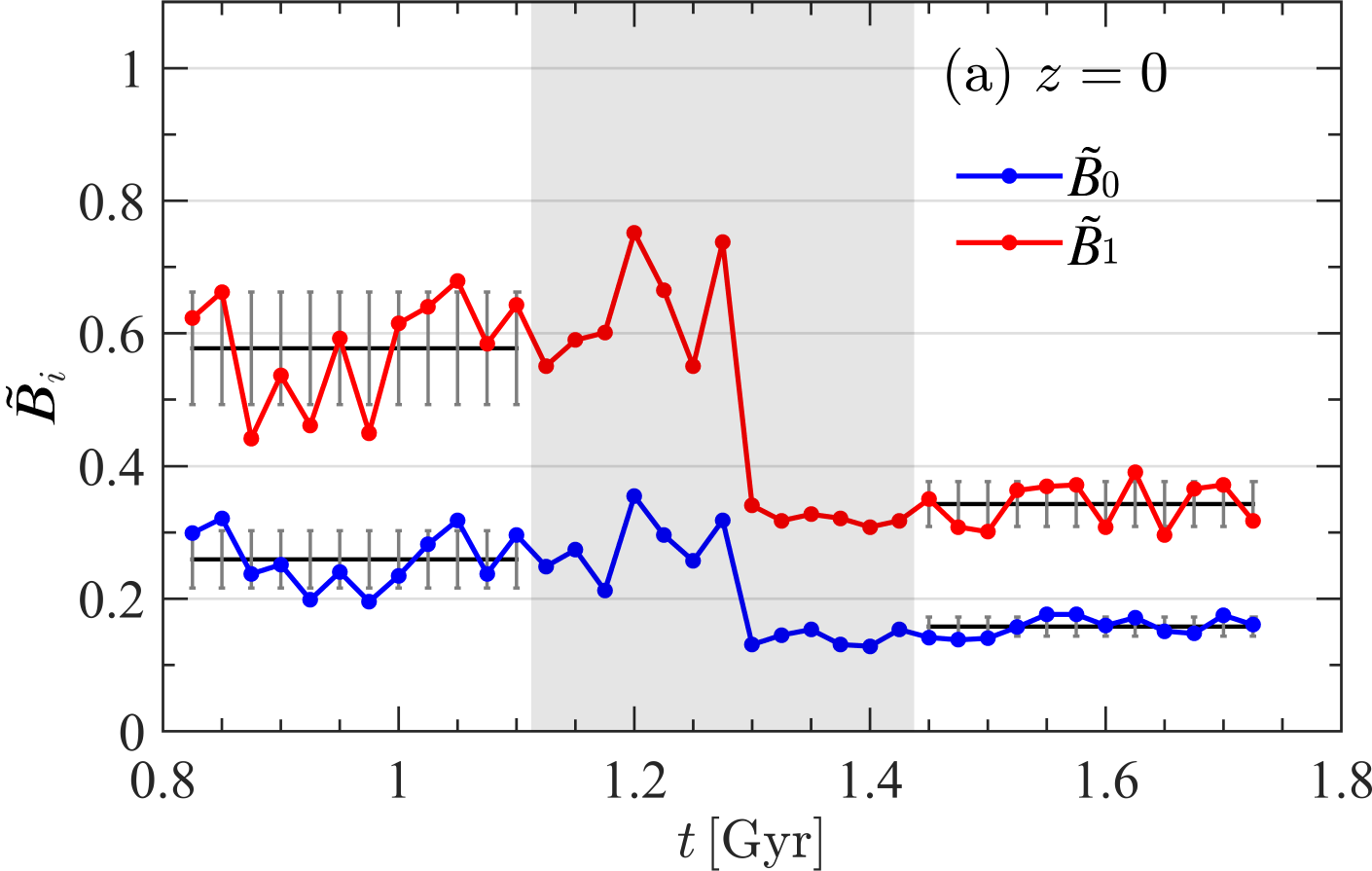}\\
	\vspace{2mm}
	\includegraphics[width=0.4\textwidth]{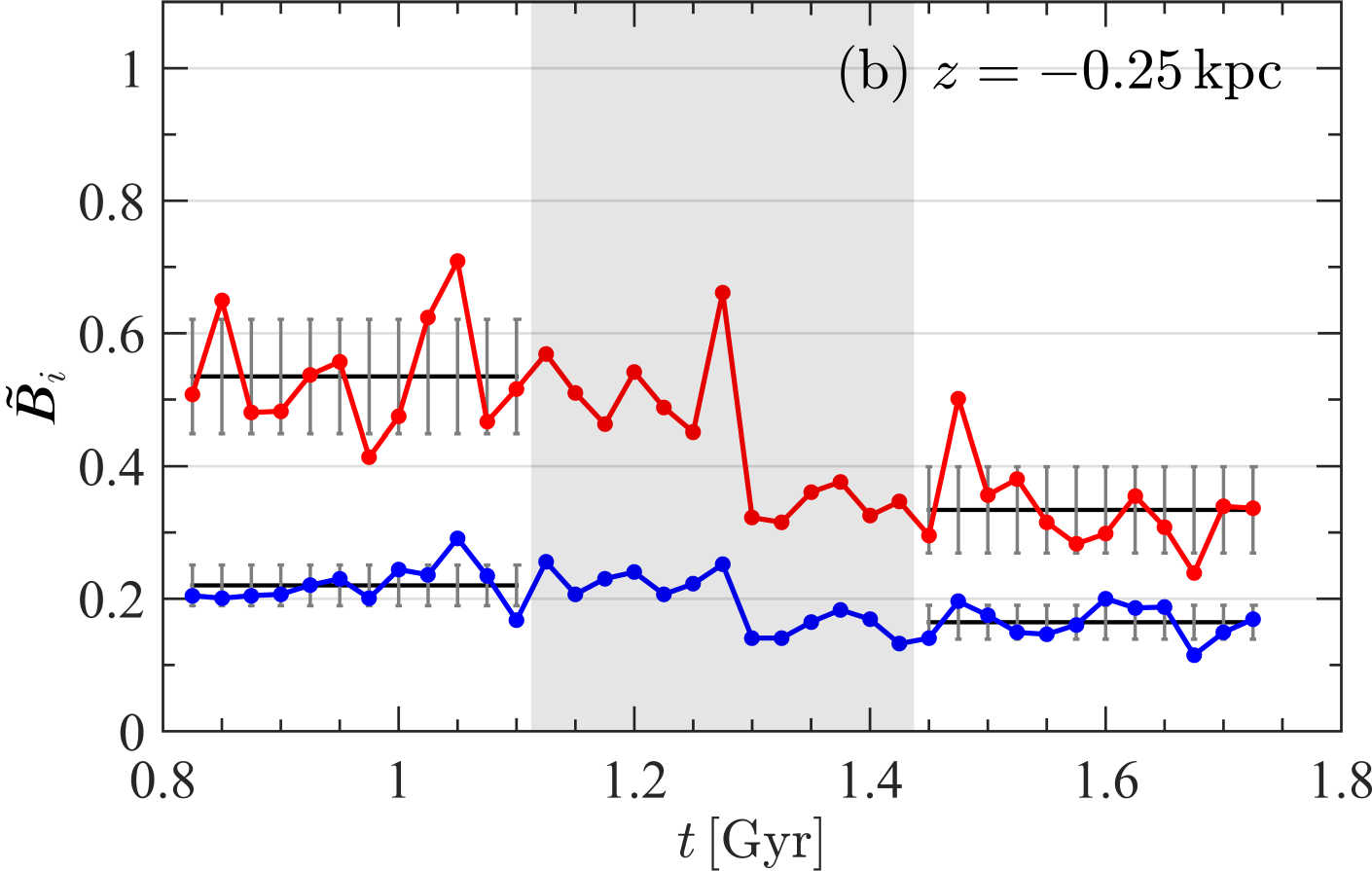} \hspace{4mm}
	\includegraphics[width=0.4\textwidth]{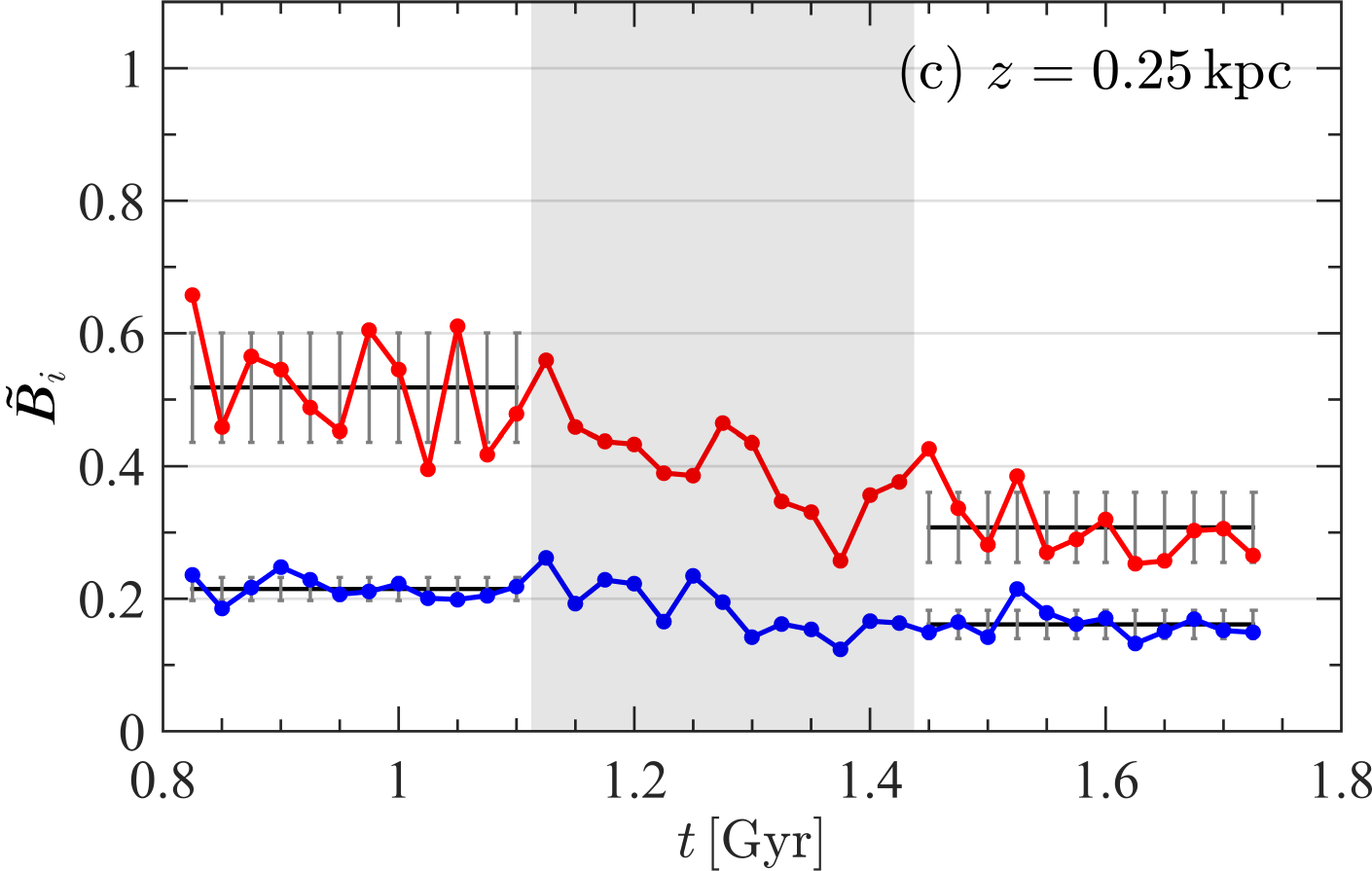}\\
	\vspace{2mm}
	\hspace{1mm}
  \includegraphics[width=0.4\textwidth]{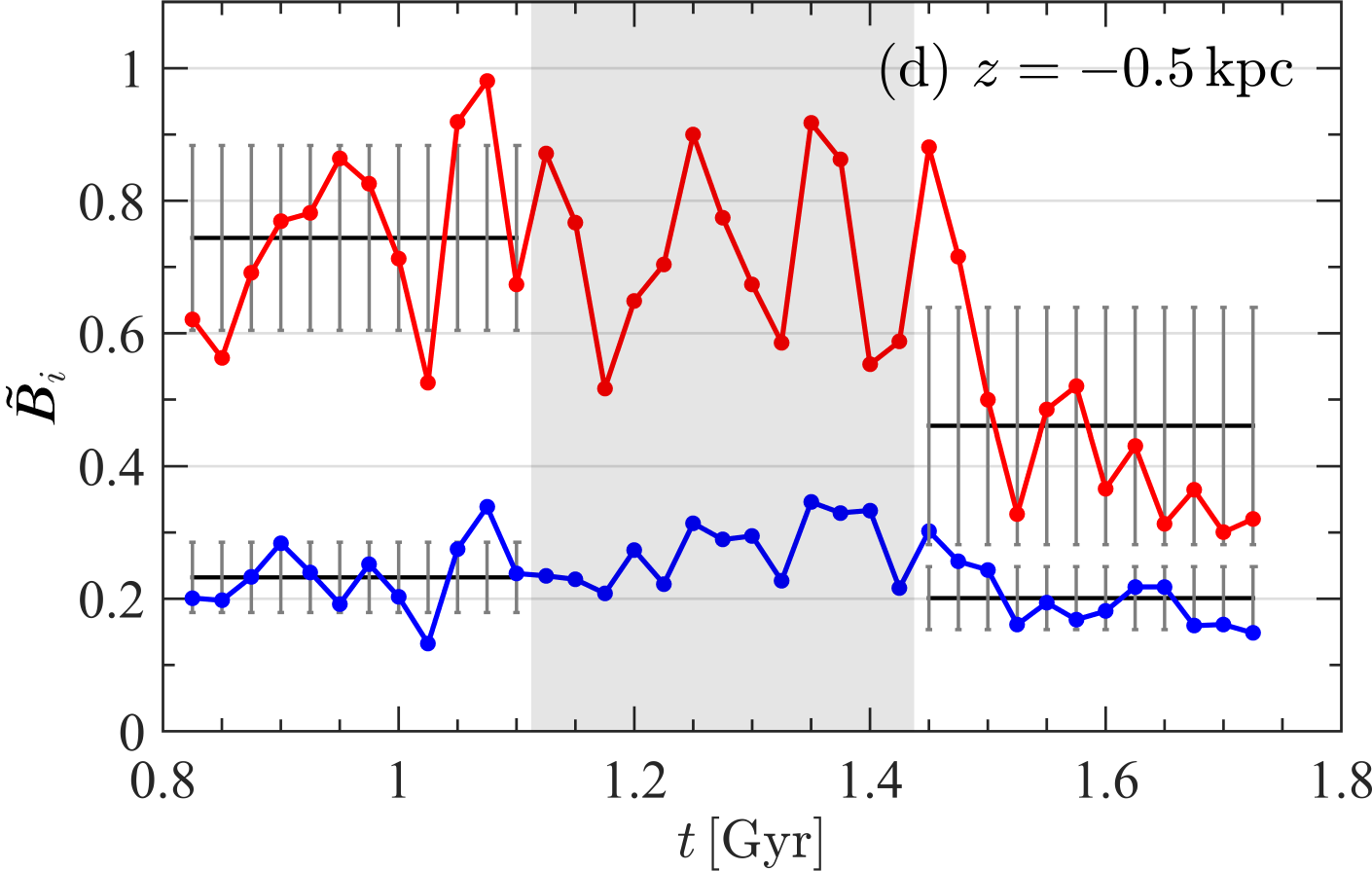} \hspace{6mm}
	\includegraphics[width=0.4\textwidth]{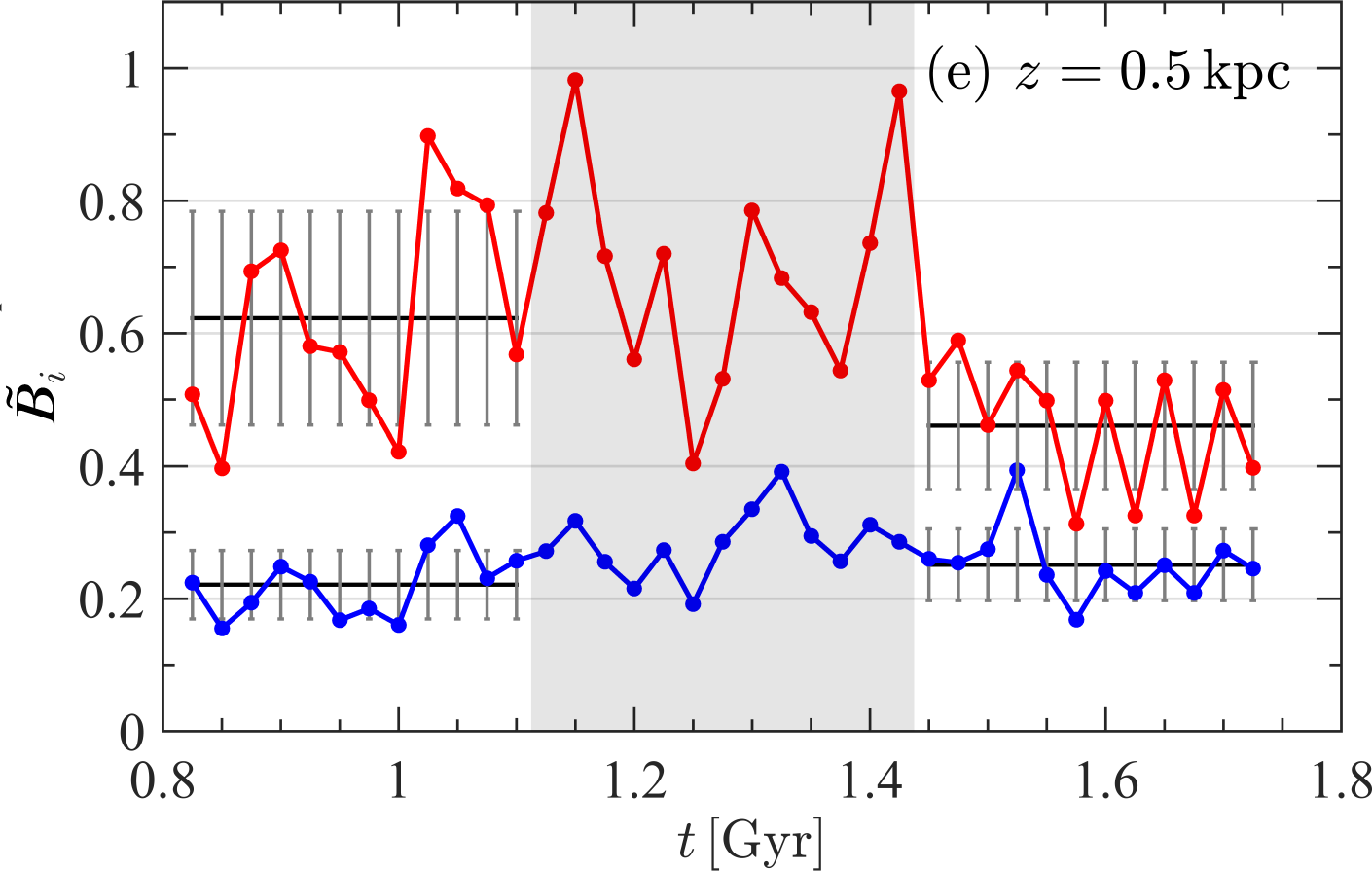}
  \caption{The Betti numbers per correlation area, $\tB_0$
  	(blue) and $\tB_1$ (red), as functions of time at various
  	distances from the mid-plane in the same format as Fig.~\ref{fig:bettiViaTime}.}
  \label{fig:bettiViaTimePerCorLength}
\end{figure*}

The Betti numbers of Table~\ref{tab:bettiPerKpcT} and Fig.~\ref{fig:bettiViaTime},
denoted $\cb_0$ and $\cb_1$ 
are normalised to an area of $1\kpc^2$ whose size has no physical significance.
Meanwhile, the spatial scale of the density fluctuations, controlled by
the correlation length shown in Fig~\ref{fig:corrL}
and Table~\ref{tab:bettiPerL0T}, changes with time and $z$ and differs
between Stages~I and III, especially at small $|z|$. In order to correct for the variation in the
scale of the density fluctuations and isolate changes in their topological
properties, we calculated Betti numbers normalised to the correlation cell area
\begin{equation}\label{tbetti}
\tb_n=\beta_n(l_0/L)^2\,,
\end{equation}
where $L=1.086\kpc$ is the horizontal size of the computational domain.
The corresponding 
values totalled over levels  
are denoted $\tB_n$. 
They are presented in Fig.~\ref{fig:bettiViaTimePerCorLength} and
Table~\ref{tab:bettiPerL0T}.

\begin{table*}
	\centering
	\caption{\label{tab:bettiPerL0T}The correlations lengths and totalled 
		Betti numbers per correlation cell, $\tB_n$
		(with standard deviations within each stage), in Stages~I and III at
		various $z$.}
	\begin{tabular}{ccccccccc}
		\hline
\phantom{$\displaystyle\int$}		& \multicolumn{2}{c}{$l_0$ [pc]} &&
		\multicolumn{2}{c}{$\tB_0\pm\sigma_{\tB_0}$} &&
		\multicolumn{2}{c}{$\tB_1\pm\sigma_{\tB_1}$}\\
		\cline{2-3}\cline{5-6}\cline{8-9}
$z$ [pc] 		&Stage I  	&Stage III 	&&Stage I		&Stage III 		&&Stage I		&Stage III\\
		\hline
\phantom{$-$}0 		&54			&35 		&&$0.26\pm0.04$ &$0.16\pm0.02$ 	&&$0.58\pm0.09$ &$0.34\pm0.03$\\
$-250$   			&45 		&38 		&&$0.22\pm0.03$ &$0.17\pm0.03$ 	&&$0.54\pm0.09$ &$0.33\pm0.07$\\
\phantom{$-$}$250$	&45 		&38 		&&$0.22\pm0.02$ &$0.16\pm0.02$ 	&&$0.52\pm0.08$ &$0.31\pm0.05$\\
$-500$  			&43 		&44			&&$0.23\pm0.05$	&$0.20\pm0.05$ 	&&$0.74\pm0.14$ &$0.46\pm0.18$\\
\phantom{$-$}$500$  &43			&44			&&$0.22\pm0.05$	&$0.25\pm0.05$ 	&&$0.62\pm0.16$ &$0.46\pm0.10$\\
		\hline
	\end{tabular}
\end{table*}

\begin{table}
	\centering
	\caption{\label{chi}
		The Euler characteristic $\chi$ and the Betti number ratios
		between Stages~I and III, $r_0$ and $r_1$ of Eq.~\eqref{r12},
		normalised to the correlation cell area and presented at
		various distances to the mid-plane.}
	\begin{tabular}{lcccc}
		\hline
		$z$ [pc] 			&\multicolumn{2}{c}{$\chi$} &$r_0$ 	&$r_1$ \\
		\cline{2-3}
		&Stage~I	&Stage III  	&		& \\
		\hline
		$\phantom{-25}0$    	& $-0.32\pm 0.10$ 	&$-0.19\pm 0.04$ 		&1.64 	&1.68\\
		$-250$           	& $-0.31\pm 0.09$ 	&$-0.17\pm 0.08$ 		&1.34 	&1.60\\
		$\phantom{-}250$ 	& $-0.30\pm 0.08$ 	&$-0.15\pm 0.05$ 		&1.33 	&1.68\\
		$-500$            	& $-0.51\pm 0.15$ 	&$-0.26\pm 0.19$ 		&1.16 	&1.62\\
		$\phantom{-}500$	& $-0.40\pm 0.17$ 	&$-0.21\pm 0.11$ 		&0.88 	&1.35\\
		\hline
	\end{tabular}
\end{table}

Remarkably, the evolution of the Betti numbers looks different when presented in
terms of $\tB_n$. 
The difference is especially pronounced at small $z$:
while $\cb_n$ 
per kpc$^2$ increase from Stage~I to Stage~III at $z=0$, the
$\tB_n$ 
decrease. Trends at other values of $z$ shown in
Figs.~\ref{fig:bettiViaTime} and \ref{fig:bettiViaTimePerCorLength}
are similar but the change in $\tB_1$ 
between the stages is stronger
in terms of the Betti numbers per correlation cell.
The Betti numbers per correlation cell arguably have a
clearer physical meaning and, conveniently, are dimensionless; we use them to
draw physical conclusions from our results.

In all cases, $\cb_1>\cb_0$ 
independently of the normalisation. This inequality
is consistent with a `spongy' structure, where cavities of hot,
rarefied gas produced by supernovae, contribute to $\beta_1$.
Both Betti numbers per correlation cell decrease with magnetic field strength,
and the change is most pronounced at smaller $|z|$.
The reduction is stronger for the number of holes, $\tB_1$. 
At large values of $|z|$, where the magnetic field is weaker \citep{EGSFB17},
the change in $\tB_n$ 
between Stages~I and III is weaker too.
The only exception is the case of $\tB_0$ 
at $z = 0.5\kpc$ where the change is marginal.

Table~\ref{chi} provides further diagnostics to justify our conclusions.
Negative values of the Euler characteristic $\chi=\tB_0-\tB_1$ 
confirm the `spongy' character of the density distribution
with predominance of cavities. We also present the ratios of the Betti numbers
in Stages~I and III,
\begin{equation}\label{r12}
r_0=\frac{\left.\tB_0\right|_{\rm Stage\  I}}{\left.\tB_0\right|_{\rm Stage\ III}}\,,
\qquad
r_1=\frac{\left.\tB_1\right|_{\rm Stage\  I}}{\left.\tB_1\right|_{\rm Stage\ III}}\,.
\end{equation}
Both ratios exceed unity at almost all $z$ (except for $r_0$ at $z=500\pc$)
quantifying the reduction in the abundance of gas density features in Stage~III
as compared to Stage~I, i.e. the homogenisation of the gas density distribution
by the magnetic field. It is understandable that the ratios are smaller at $|z|=500\pc$
where the magnetic field is weaker. Away from the mid-plane, we observe $r_1>r_0$,
so the effect of the magnetic field
on gas cavities ($r_1$) is somewhat stronger than on density enhancements ($r_0$).

\begin{table*}
	\centering
	\caption{\label{tab:statTestsBetti}The Mann--Whitney--Wilcoxon (MWW) and
		Kolomogorov--Smirnov (KS) statistical tests for the Betti numbers of individual
		snapshots from Stages~I and III: they differ at the significance level of
		95 per cent if the values of $p$ shown are less than 0.05. Entries corresponding
		to a significant difference are shown bold.}
        \begin{tabular}{clllll}
                \hline
                &MWW                    &KS                                                     &&MWW                   &KS\\
                $z$ [pc]                &\multicolumn{2}{c}{$\cb_0$ [kpc$^{-2}$]}       &&\multicolumn{2}{c}{$\tB_0$}           \\
                \cline{2-3}\cline{5-6}
                $\phantom{-}0$  &\bf0.00004     &\bf0.00002                                     &&\bf0.00004    &\bf0.000002\\
                $-250$          &0.49                   &0.79                                           &&\bf0.0002             &\bf0.00002\\
                $\phantom{-}250$&0.54                   &0.79                                           &&\bf0.0002             &\bf0.00002\\
                $-500$                  &0.09                   &0.07                                           &&0.14                  &0.19\\
                $\phantom{-}500$&0.26                   &0.43                                           &&0.16                  &0.19\\\hline
        \end{tabular}
        \hspace{1em}
        \begin{tabular}{clllll}
                \hline
                &MWW                    &KS                                                     &&MWW                   &KS\\
                $z$ [pc] &\multicolumn{2}{c}{$\cb_1$ [kpc$^{-2}$]} &&\multicolumn{2}{c}{$\tB_1$}\\
                \cline{2-3}\cline{5-6}
                $\phantom{-}0$  &\bf0.00004             &\bf0.000002                                    &&\bf0.00004    &\bf0.000002\\
                $-250$          &\bf0.03                &0.07                                           &&\bf0.0001             &\bf0.00002\\
                $\phantom{-}250$&\bf0.009       &\bf0.02                                        &&\bf0.00006    &\bf0.00002\\
                $-500$          &\bf0.0007              &\bf0.0002                                      &&\bf0.001              &\bf0.0002\\
                $\phantom{-}500$&\bf0.007               &\bf0.02                                        &&\bf0.02               &\bf0.02\\
                \hline
        \end{tabular}
\end{table*}

We assess the statistical significance of the variations in Betti numbers
using the Mann--Whitney--Wilcoxon and Kolomogorov--Smirnov tests,
treating the totalled Betti numbers obtained for Stages~I and III as samples from two
distributions. 
Table~\ref{tab:statTestsBetti} shows $p$-values for tests of whether Stages~I and III have the same distributions.
The difference in the distributions is significant at the 95 per cent confidence level
when $p<0.05$. The two tests lead to similar conclusions with just one exception
($\cb_1$ at $z = -250\pc$).
The difference between the Betti numbers of the gas density fluctuations
between the states with and without strong magnetic field is statistically
significant, and the Betti numbers per correlation cell
provide more discriminatory power. The difference between Stages~I and III
is especially strong for the abundance of holes.
The variation of the Betti numbers with the magnetic field strength suggests
that the magnetic field strongly reduces the abundance of gas cavities.

\begin{table*}
	\centering
	\caption{\label{tab:bntableZ0}The mean values and standard deviations of the
		bottleneck distance $D$ (in cm$^{-3}$) between persistence diagrams (PD) for
		$\cb_0$ and $\cb_1$ at $z=0$ for snapshots within Stages~I and III and
		between the stages (the upper part of the table) and at the earliest and latest
		times in Stages~I and III, respectively (the lower part).}
	\begin{tabular}{llcccc}
	\hline
$\phantom{\displaystyle{\int}}$		& &Time interval	&Number of 	&For $\cb_0$ &For $\cb_1$\\ 
		\cline{5-6}
$\phantom{\displaystyle{\int}}$		& &[Gyr]			&PD pairs	&\multicolumn{2}{c}{$\langle D\rangle\pm\sigma_D$ [cm$^{-3}$]}\\
		\hline
		All times& Within Stage I		&0.275			&66  		&$3.5\pm2.3$ 			&$3.8\pm1.7$\\
		& Within Stage III	&0.275			&66			&$1.9\pm1.2$ 			&$2.6\pm0.9$\\
		& Between Stages I and III &
		 &144		&$4.5\pm2.7$ 			&$4.0\pm1.8$\\
\\
		Earliest and& Within Stage I		&0.175 		&28			&$3.5\pm2.1$ 			&$4.2\pm1.8$\\
		latest times& Within Stage III	&0.175		&28 		&$1.8\pm1.0$ 			&$2.5\pm 0.8$\\
		& Between Stages I and III &
		&64			&$5.5\pm2.7$ 			&$4.5\pm 2.0$\\
		\hline
	\end{tabular}
\end{table*}

\begin{table*}
    \centering
    \caption{\label{tab:bntable}As Table~\ref{tab:bntableZ0} but separately for
             various distances to the mid-plane.
             }
             \begin{tabular}{clcccccr}
                \hline
                Quantity &Snapshots &Number of  &\multicolumn{5}{c}{$\langle D\rangle\pm\sigma_D$} \\ 
                \cline{4-8}
                 & & PD pairs & $z = 0$  & $z = -0.25\kpc$ & $z = 0.25\kpc$ & $z = -0.5\kpc$ & $z = 0.5\kpc$\\
                \hline
                $\cb_0$ &Within Stage I & $66$  & $3.5\pm 2.3$ & $2.7\pm 2.1$ & $1.8\pm 1.0$ & $1.4\pm 0.6$ & $1.1\pm 0.4$\\
                                 &Within Stage III  & $66$  & $1.9\pm 1.2$ & $3.0\pm 2.3$ & $2.1\pm 1.2$ & $1.8\pm 1.0$ & $1.9\pm 1.4$\\
                                 &Between Stages I and III & $144$ & $4.5\pm 2.7$ & $3.0\pm 2.1$ & $2.0\pm 1.3$ & $1.6\pm 0.9$ & $1.6\pm 1.2$\\[6pt]
                $\cb_1$ &Within Stage I      & $66$  & $3.8\pm 1.7$ & $2.4\pm 1.8$ & $2.0\pm 0.9$ & $1.8\pm 0.6$ & $1.5\pm 0.5$\\
                                 &Within Stage II      & $66$  & $2.6\pm 0.9$ & $2.8\pm 1.9$ & $2.0\pm 0.8$ & $1.7\pm 0.7$ & $2.1\pm 1.0$\\
                                 &Between Stages I and III & $144$ & $4.0\pm 1.8$ & $2.9\pm 1.5$ & $2.1\pm 0.9$ & $1.7\pm 0.7$ & $1.9\pm 1.0$\\
                \hline
            \end{tabular}
\end{table*}

We calculated the bottleneck distances between each pair of persistence diagrams within
Stages~I and III as well as between the stages, and show
the results in the upper part of Table~\ref{tab:bntableZ0}.
If the bottleneck distance is sensitive to the difference in the gas density
distributions at early and late times, the inter-stage distance should be
larger than the intra-stage distances. The mean values of $D$ conform to this
expectation but the difference is less than one standard deviation.
In an attempt to enhance the inter-stage differences, we treated in a similar manner
the gas density from the earliest and latest periods of the simulation, when the difference
between the magnetic field strengths is larger. The results, presented in the lower part of
Table~\ref{tab:bntableZ0}, show that the mean inter-stage bottleneck distance increases
as expected but remains within one standard deviation despite that fact that
the values of $\sigma_D$ remain similar to those in the larger samples.
There still remains a possibility that the bottleneck distance is more useful at some
distances to the mid-plane even if it is a poor diagnostic for the whole computational domain.
A similar comparison of the intra-stage and inter-stage values of $D$ is presented in
Table~\ref{tab:bntable} for a selection of values of $z$. However, the results
remain marginal.

\begin{table}
	\centering
	\caption{\label{tab:statTests}The Mann--Whitney--Wilcoxon (MWW) and
		Kolomogorov--Smirnov (KS) statistical tests for the bottleneck
		distances: $D$ in Stages~I and III differ at the significance
		level of 95 per cent if the value of the probability shown is less than 0.05.
		Entries corresponding to a significant difference are shown bold.}
	\begin{tabular}{lrrr}
		\hline
&$z$ [pc]			&MWW 			&KS\\
\hline
$\cb_0$ &0    				& $\bf0.0002$  & $\bf0.0002$\\
&$-250$           	& $0.27$       & $0.29$\\
&$\phantom{-}250$	& $0.29$       & $0.54$\\
&$-500$            	& $\bf0.05$    & $0.09$\\
&$\phantom{-}500$  	& $\bf0.02$    & $\bf0.0004$\\
\\

$\cb_1$ &0    				& $\bf0.000003$& $\bf0.0002$\\
&$-250$           	& $\bf0.005$   & $\bf0.00002$\\
&$\phantom{-}250$ 	& $\bf0.006$      & $\bf0.004$\\
&$-500$            	& $0.16$       & $0.20$\\
&$\phantom{-}500$  	& $\bf0.0003$  & $\bf0.006$\\
\hline
\end{tabular}
\end{table}

The Mann--Whitney--Wilcoxon and Kolomogorov--Smirnov tests for the difference between
the probability distributions of $D$ confirm the difference between the intra- and
inter-stage bottleneck distances in some cases (Table~\ref{tab:statTests})
but the results remain mixed and unsystematic, especially for $\cb_0$. As discussed
in Section~\ref{sec:scal}, the effect of the magnetic field on $\cb_1$ is stronger,
and statistical significance of this is reflected in the bottleneck distances for $\cb_1$.
Nonetheless, the bottleneck distance proves to be a poor diagnostic of the difference between
persistence diagrams of a realistic random field, as opposed to
synthetic fields used in theoretical developments of the TDA.
\mbox{\citet{Hend2017arXiv}} show that the bottleneck distance also fails
to distinguish persistence diagrams of non-Gaussian synthetic
random fields.

\section{Discussion and conclusions}\label{sec:con}

We have shown that a magnetic field affects significantly the gas distribution in these ISM simulations.
Topological data analysis of the gas density fluctuations reveals properties of
interstellar turbulence that cannot be obtained from the correlation
analysis. The magnetic field does
not change the form of the autocorrelation function in such a way that it could be used as a diagnostic (e.g., producing local minima or maxima) for regions of strong magnetic fields \citep[cf.][]{HSSFG17}.
Magnetic effects do cause a significant
reduction in the correlation length of the density fluctuations
from $45\text{--}55\pc$ in Stage~I of the simulations, where magnetic
field is dynamically negligible, to $35\text{--}40\pc$ in Stage~III
where magnetic and turbulent energies are comparable. The change
in the correlation length is restricted to the range $|z|\lesssim300\pc$
of distances to the mid-plane where the magnetic field is strong. At
larger values of $z$, the correlation length remains of order $40\pc$
throughout the simulation.

Our analysis focuses on the Betti numbers $\beta_0$ and $\beta_1$
of the gas fluctuations in two-dimensional slices through the
three-dimensional computational domain. These Betti numbers quantify
the number of isolated gas density structures ($\beta_0$) and
holes in the gas distribution ($\beta_1$). Euler's characteristic
in two dimensions is the difference of the two Betti numbers,
$\chi=\beta_0-\beta_1$.
We suggest that the Betti numbers normalised to the size of the
correlation cell, $\tb_0$ and $\tb_1$ defined in Eq.~\eqref{tbetti},
and their total numbers in a filtration, $\tB_0$ and $\tB_1$, equal to the number of points in the
corresponding persistence diagram,
are physically informative and represent a better statistical
diagnostic for topological differences of random fields.

The topological structure of the simulated ISM is characterised
by a persistent inequality $\chi<0$, i.e.
a higher abundance of cavities as compared to isolated gas
clouds.
As a crude estimate, the gas distribution contains
one denser structure per five correlation cells of
$40\text{--}50\pc$ in size (corresponding to $\tB_0\simeq0.2$)
and one cavity per 1--2 correlation cell ($\tB_1 \simeq0.6$) when
the magnetic field is weak (Stage~I). Since $\tB_0<1$, the higher-density
structures represent either large isotropic clouds, which is physically
unlikely, or gas filaments spanning a few correlation cells.
This suggests a spongy and yet filamentary
structure dominated by elongated gas filaments and cavities filled
with rarefied gas. The abundance of dense gas
structures does not change much as the magnetic field grows
but the abundance of cavities reduces to one per three
correlation cells when the magnetic field becomes dynamically
important (Stage~III).

The reduction in both Betti numbers as the magnetic field grows suggests
that it makes the gas distribution more homogeneous.  A magnetic field
of a few $\mu$G in strength can hardly
affect expanding supernova remnants, so it is likely that it reduces
the abundance of the hot, rarefied gas in old remnants or facilitates
their merger with the ambient gas. \citet{EGSFB18} arrive at similar
conclusions from their analysis of the fractional volume of the hot
gas. Remarkably, the modification of the gas distribution by magnetic
field is captured reliably even in a region at $|z|\ga300\pc$
where the magnetic field is weaker and correlation analysis fails to
detect any magnetic effects.
So the topological methods applied here prove to be more 
sensitive to subtle differences between random fields than the
traditional correlation analysis. Perhaps more importantly, topology
reflects features of the random field that cannot be captured by
traditional methods at all.

These methods of topological data analysis could also be applied productively to other ISM data which is expected to be non-Gaussian, both observed and simulated. Structured, often anisotropic gas density fluctuations emerge in simulations
of molecular clouds, calling for methods of analysis more general than 
power spectra and low-order correlation functions. For example,
\citet{B-PML02} note that stronger driving of turbulence produces not only larger fluctuations about the mean density but also more extreme fluctuations (a greater proportion of the gas is at the highest densities). A shift of gas into the extremes in the density field can
dominate the low-order statistical parameters of turbulence such as 
root-mean-square values. Magnetic fields can introduce further 
complications in the form of filamentary and anisotropic density structures.
Such deviations from Gaussian statistical properties
of velocity, density and magnetic field fluctuations are often described as 
intermittency. In such cases, Fourier spectra, with the associated 
random-phase approximation, and comparisons with simple models of 
turbulence (such as Kolmogorov's or Burgers' models) only provide 
necessary but not sufficient evidence of the relevance of the model. 
Moreover, compressibility (e.g., in shock-wave turbulence) and magnetic 
effects imprint nontrivial topological structure on turbulence in both 
molecular clouds and the diffuse interstellar medium. 
New methods of analysis are required that do not rely on the assumption of 
Gaussian statistics or weak deviations from it and are sensitive to  
morphological and topological properties of the random fields. For example, 
the nature of filamentary structures \citep{KKHWBBFL16} prominent in both 
observations and many numerical simulations of interstellar gas requires  
reliable determination of their dimensions 
\citep[see][for examples of such analysis]{MFS15,KKHWBBFL16}. The 
need for new methods of statistical comparison between theory and observations 
is discussed in an insightful review of \citet{Gaa11} and by 
\citet{KWOLR17} who employ genus statistics, a 
topological measure earlier applied to the matter distribution in the 
cosmic web \citep[][and many later papers]{Gott87}.

\section*{Acknowledgements}
We are grateful to Herbert Edelsbrunner, Pratyush Pranav, and
Kandaswamy Subramanian for useful discussions.
Nikolai Makarenko and his group in Pulkovo Observatory (Russia)
helped a lot with understanding some basic topological conceptions
and ideas for comparison of random fields.
Special thanks are due to Arnur Nigmetov and Dmitriy Morozov
who provided a code for fast computation of the bottleneck distance
and helped with its application.
We also thank Can C.~Evirgen  for his help with data visualization.
This work is supported by the Leverhulme Trust Grant RPG-2014-427.
AS, AF and LFSR also acknowledge financial support from the STFC Grant
ST/N000900/1 (Project 2).

\bibliographystyle{mnras}
\bibliography{topsig}
\bsp
\label{lastpage}
\end{document}